\documentclass{aa}  
\usepackage{graphicx}
\usepackage{txfonts}

\usepackage[utf8]{inputenc} 
\usepackage[T1]{fontenc}    
\usepackage[colorlinks = true,
            linkcolor = blue,
            urlcolor  = blue,
            citecolor = blue,
            anchorcolor = blue]{hyperref}
\usepackage{url}            
\usepackage{booktabs}       
\usepackage{amsfonts}       
\usepackage{nicefrac}       
\usepackage{microtype}      
\usepackage{lipsum}
\usepackage{multicol}       
\usepackage{hyperref}       
\hypersetup{colorlinks=true,linkcolor=blue, linktocpage}
\usepackage{textcomp}
\usepackage{gensymb}        
\usepackage{graphicx}       
\usepackage{amsmath}        
\usepackage{amssymb}        
\usepackage{float}          
\usepackage{subcaption}
\usepackage{stfloats}       
\usepackage[export]{adjustbox} 
\usepackage[notquote]{hanging}        
\usepackage[labelfont=bf]{caption}  
\usepackage[font=scriptsize]{caption}   

\usepackage{pdflscape}
\usepackage{afterpage}
\usepackage{capt-of}

\usepackage{times}
\usepackage{tikz}
\usepackage{siunitx} 
\usetikzlibrary{shapes.geometric, arrows}

\begin{document} 

    \title{Characterising X-ray variability in light curves with complex sampling patterns: application to the eROSITA south ecliptic pole survey}

\author{D. Bogensberger \inst{1} \inst{2}, K. Nandra \inst{1}, J. Buchner \inst{1}}
\institute{Max Planck Institute f{\"u}r Extraterrestrische Physik, Gie{\ss}enbachstra{\ss}e 1, 85748 Garching bei München, Germany  
    \and
        Department of Astronomy, The University of Michigan, 1085 South University Avenue, Ann Arbor, Michigan, 48103, USA. 
    \email{\href{mailto:dbogen@umich.edu}{dbogen@umich.edu}}
}

\date{Received <date> / Accepted <date>}

\abstract{
        \textit{\textbf{Aims}}: During its all-sky survey phase, the \emph{eROSITA} X-ray telescope onboard \textit{SRG} scans through the ecliptic poles every 4 hours. This extensive data set of long-duration, frequent, and consistent observations of thousands of X-ray sources is ideal for a detailed long-term X-ray variability analysis. However, individual observations are short, are separated by long but consistent gaps, and have varying exposure times. Therefore, the identification of variable sources, and the characterisation and quantification of their variability requires a unique methodology. We aim to develop and evaluate variability analysis methods for \emph{eROSITA} observations, focusing on sources close to the survey poles. We also aim to detect intrinsically variable sources at any count rate and quantify the variability of low count rate sources. 
        
		\textit{\textbf{Methods}}: We simulate \emph{eROSITA}-like light curves to evaluate and quantify the effect of survey mode observations on the measured periodogram and normalised excess variance. We introduce a new method for estimating the normalised intrinsic variance of a source based on the Bayesian excess variance (bexvar) method.

		\textit{\textbf{Results}}: We determine thresholds for identifying likely variable sources while minimising the false-positive rate, as a function of the number of bins, and the average count rate in the light curve. The bexvar normalised intrinsic variance estimate is significantly more accurate than the normalised excess variance method in the Poisson regime. At high count rates, the two methods are comparable. We quantify the scatter in the intrinsic variance of a stationary pink noise process, and investigate how to reduce it. Finally, we determine a description of the excess noise in a periodogram caused by varying exposure times throughout a light curve. Although most of these methods were developed specifically for analysing variable AGN in the \emph{eROSITA} all-sky survey, they can also be used for the variability analysis of other datasets from other telescopes, with slight modifications.}


\titlerunning{\emph{eROSITA} Variability methodology}
\authorrunning{David Bogensberger}

\maketitle 

\section{Introduction}

The X-ray luminosity of Active Galactic Nuclei (AGNs), X-ray binaries (XRBs), and stars has been observed to vary strongly over a wide range of timescales. Analysing their light curve variability can reveal information about their source properties and indicate various unique phenomena. This variability can be of a persistent or transient nature, depending on the source and the cause of the variability. 

For instance, a measurement of the shortest timescale over which continuous brightness changes are detected can be used to estimate an upper bound on the size of the light-emitting region of a persistently variable source. This is done by equating it to the light-crossing timescale. This kind of variability analysis provided one of the first indications of the true nature of various bright astronomical objects that we now collectively group under the term of an AGN \citep{1963ApJ...138...30M}. Despite the gigantic average luminosities of AGNs, they are observed to vary within hundreds of seconds. This indicates that they are incredibly compact and have radii on a sub-pc scale \citep{1959ApJ...130...38W}. AGN variability studies also revealed a myriad of other intriguing properties. For example, the AGN X-ray variability has been found to anti-correlate with the AGN luminosity \citep{1997ApJ...476...70N, 2016ApJ...831..145Y, 2017ApJ...849..127Z}, and the mass of the supermassive black hole powering the AGN \citep{2001MNRAS.324..653L, 2005MNRAS.358.1405O, 2017MNRAS.471.4398P, 2023arXiv230414228A}. This means that the variability of an AGN can be used as an estimator of the BH mass when other methods are unavailable \citep{2012A&A...542A..83P}. In addition, \citet{2011A&A...536A..84V, 2016A&A...593A..55V} used ensembles of AGNs to determine more correlated X-ray variability scaling relations. 

XRBs have many similar properties as AGNs but vary on much shorter timescales due to the smaller size of the compact object powering their X-ray emission. The X-ray variability in AGNs and XRBs is produced in the innermost part of the accretion disc and the corona \citep{2004astro.ph..9254C}. Therefore, studies of AGN and XRB X-ray variability can determine the physical mechanisms at work there. XRB variability analysis further distinguishes between various states via the variability properties of their light curves \citep{2011BASI...39..409B}. In addition, XRB variability studies often identify particular frequencies that dominate the light curve evolution, known as quasi-periodic oscillations \citep{2019NewAR..8501524I}. These have also been detected in a few AGNs \citep[e.g. ][]{2008Natur.455..369G, 2018NatCo...9.4599Z, 2018ApJ...860L..10S, 2021MNRAS.501.5478A}. 

Many stars also exhibit X-ray variability, typically in the form of flares. These are generated by particle acceleration in the stellar corona due to chromospheric evaporation \citep{1984ApJ...287..917A}. X-ray variability studies of stars can provide constraints on these processes. Young stellar objects are known to be particularly variable X-ray sources \citep{2006A&A...446..155F, 2017ApJ...844..109F}.

The extended ROentgen Survey with an Imaging Telescope Array \citep[\emph{eROSITA};][]{2021A&A...647A...1P, 2021arXiv210413267S} aimed to perform a four-year-long detailed survey of the entire sky in X-rays. \emph{eROSITA} is mounted on the Spectrum-Roentgen-Gamma \citep[\emph{SRG};][]{2021arXiv210413267S} spacecraft, which rotates about itself at a constant angular velocity of $2.5\times10^{-2} ~\mathrm{deg}~ \mathrm{s}^{-1}$, completing one rotation, referred to as an eroday, every 4 hours. Its angular momentum also shifts direction by an average of $10\arcsec$ per eroday, along the ecliptic plane. This observing pattern ensures that the entire sky is observed within six months. Eight \emph{eROSITA} All-Sky Surveys (eRASSs) are planned. 

As the angular momentum vector of \emph{eROSITA}'s rotation about itself always lies in the ecliptic plane throughout all the eRASSs, the South and North Ecliptic Poles (SEP, and NEP, respectively) are observed on every eroday. Therefore, sources lying close to the ecliptic poles are observed consistently every 4 hours for most of the duration of the eRASSs. These observations enable a detailed investigation into the medium to long-term X-ray variability properties of sources in these fields.

However, a variability analysis of \emph{eROSITA} data faces various challenges, including varying exposure times, low count rates, and long gaps between short observations. In this paper, we describe various variability methods that can be used to characterise the intrinsic variability. We also analyse how these could be modified to minimise the detrimental effects of survey mode observations. We also discuss ways to improve upon previous methods, both for \emph{eROSITA}, and for other missions. We particularly focus on AGN variability, and pink noise light curves expected for them in the frequency space probed by \emph{eROSITA}. We also predominantly consider the field near the ecliptic poles, in which there are many more individual observations than for most sources in the sky. However, many results presented here can also be applied to other regions of the sky as observed by \emph{eROSITA}, or variability studies using entirely different instruments.

This paper is structured as follows; in Section \ref{SeceROg}, we discuss the properties of the \emph{eROSITA} observations and the challenges faced by the variability analysis of \emph{eROSITA} data. We describe the variability quantifiers we use in Section \ref{SecMethod}. In that section, we also explain how we simulated light curves for testing and optimising the variability methodology. In Section \ref{Diffvarnonvarsection}, we determine thresholds for selecting variable sources as a function of the count rate and the number of bins. We introduce a new method for estimating the normalised intrinsic variance of a light curve in Section \ref{NEVest}. In Section \ref{Syserr}, we describe the aliasing and red noise leakage, which offset the measured variability from the band-limited power. We also investigate the intrinsic scatter due to the stochastic nature of the variability, and how to reduce it by averaging over multiple segments. Section \ref{Powspecsection} describes how to compute the periodograms of \emph{eROSITA} light curves and calculate the excess noise level due to varying fractional exposures. Finally, in Section \ref{SecDisc}, we discuss our main findings and describe their applicability to \emph{eROSITA} and other variability analyses. 

\section{\emph{eROSITA} light curves} \label{SeceROg}

\subsection{Properties of observations}\label{SeceRO}

While most eRASS sources are observed a handful of times, near the eRASS survey poles, sources receive very dense, more continuous sampling with up to 1080 observations per eRASS. The survey poles nearly coincide with the ecliptic poles, with a variable offset of a few arcminutes. The frequent observations make the regions of the sky close to the ecliptic poles the most interesting for long-term variability analysis. The German \emph{eROSITA} consortium has the rights to the southern ecliptic hemisphere. Therefore, we focused primarily on the properties of the eRASS SEP field. However, this field comes with additional data analysis challenges.

Firstly, the total exposure time drops rapidly with an increasing angle from the ecliptic poles. For example, sources located $5\degree$ away from the poles are observed on an average of 70 erodays per eRASS, but sources within $0.5\degree$ of the poles are observed on 1080 erodays per eRASS. These large exposure gradients affect the source detection and result in a data set featuring a wide range of depths.

Secondly, for most sources, the sensitivity varies in complex patterns. We refer to all the data collected about the brightness of a source within a single eroday period as one observation of the source. \emph{eROSITA} has a large field of view \citep[FoV, $1.03\degree$,][]{2021A&A...647A...1P}, but individual observations of sources are short and have very different effective exposure times. When a source passes through the centre of the FoV, its exposure is 41.2 s, while sources at the border can have exposure times of a fraction of a second. The effective area drops rapidly with off-axis angle, with 80\% at $10'$ and 40\% at $25'$ \citep[for $0.2-1.5~\mathrm{keV}$][Fig. 8]{2021A&A...647A...1P}. Vignetting is even more pronounced at higher energies. Because of survey progression combined with small survey pole variations, the effective exposure time varies strongly over time for each source, with complex patterns.

The fractional exposure ($\epsilon$) quantifies the combination of the source crossing the FoV and experiencing variable vignetting. During an observation of duration $\Delta t$, the effective exposure time $\epsilon\Delta t$ describes the amount of time it would have had to have been observed on-axis to obtain the same exposure depth. By keeping $\Delta t$ constant, $\epsilon$ fully describes the variation in effective exposure. We set $\Delta t=40~\mathrm{s}$, the approximate maximum duration of a single eRASS observation of a source. Within $3\degree$ of the SEP, $\epsilon$ is approximately uniformly distributed between 0.05 and 0.45, with a peak towards 0 and a rapid drop above 0.5.

The \emph{eROSITA} effective area peaks between $0.2-5.0~\mathrm{keV}$ \citep{2021A&A...647A...1P}. There, most count rates range between $10^{-3} - 10^{1} ~\mathrm{cts/s}$ for detected sources. The deep pole exposure enables detecting sources fainter than $10^{-3} ~\mathrm{cts/s}$. However, neither the brightest nor the faintest sources are relevant to this variability analysis. The brightest sources suffer from pileup, and there is too little information on the faintest sources to analyse their variability. This variability analysis instead focuses on the vast majority of sources with sources roughly within the range of $10^{-3} - 10^{1} ~\mathrm{cts/s}$.

We follow previous \emph{eROSITA} analyses \citep{2022A&A...661A..27L, 2022A&A...661A..18B, 2022A&A...661A...8B}, in extracting background count rates from much larger background than source extraction regions. The typical background region near the SEP is 112 times larger than the source region, with typical count rates of $0.71~\mathrm{cts/s}$.

We seek to evaluate the variability of the X-ray source flux. As the effective area varies from observation to observation, the count rate found by dividing the measured number of source counts by the observation duration is not proportional to the source flux in different observations. Instead, we use an exposure-corrected count rate as a proxy of the source flux. \citet{2022A&A...661A...1B} estimated the exposure-corrected count rate of an eRASS source in observation $i$, performed at time $t_i$, as: 

\begin{equation}\label{countratedef}
R_{\mathrm S}(t_i) = \frac{C(t_i)-A(t_i)B(t_i)}{\epsilon(t_i) \Delta t}.
\end{equation}

\noindent
In this equation, $R_{\mathrm S}$ is the exposure-corrected source count rate, $C$ and $B$ are the number of counts measured in the source and background extraction regions, respectively, and $A$ is the background ratio. This equation is applicable to high-count observations, but problematic in low-count observations.

\citet{2022A&A...661A..18B} presented a method for calculating the exposure-corrected count rate that accounts for Poisson uncertainties in both the source and background count rates. A probability density function (PDF) can be constructed based on the inverse incomplete Gamma function, as described by \citet{2014ApJ...790..106K}. We extract the median and $1\sigma$ equivalent quantiles from this PDF, as the exposure-corrected count rate, and its uncertainty. This method is less affected by the varying fractional exposure in the eRASS light curve, and more accurate in the Poisson regime, compared to Eq. \ref{countratedef}. Throughout this paper, we use the term "count rate" to refer to the exposure-corrected source count rate found using this methodology.

\begin{figure}[h]
\resizebox{\hsize}{!}{\includegraphics{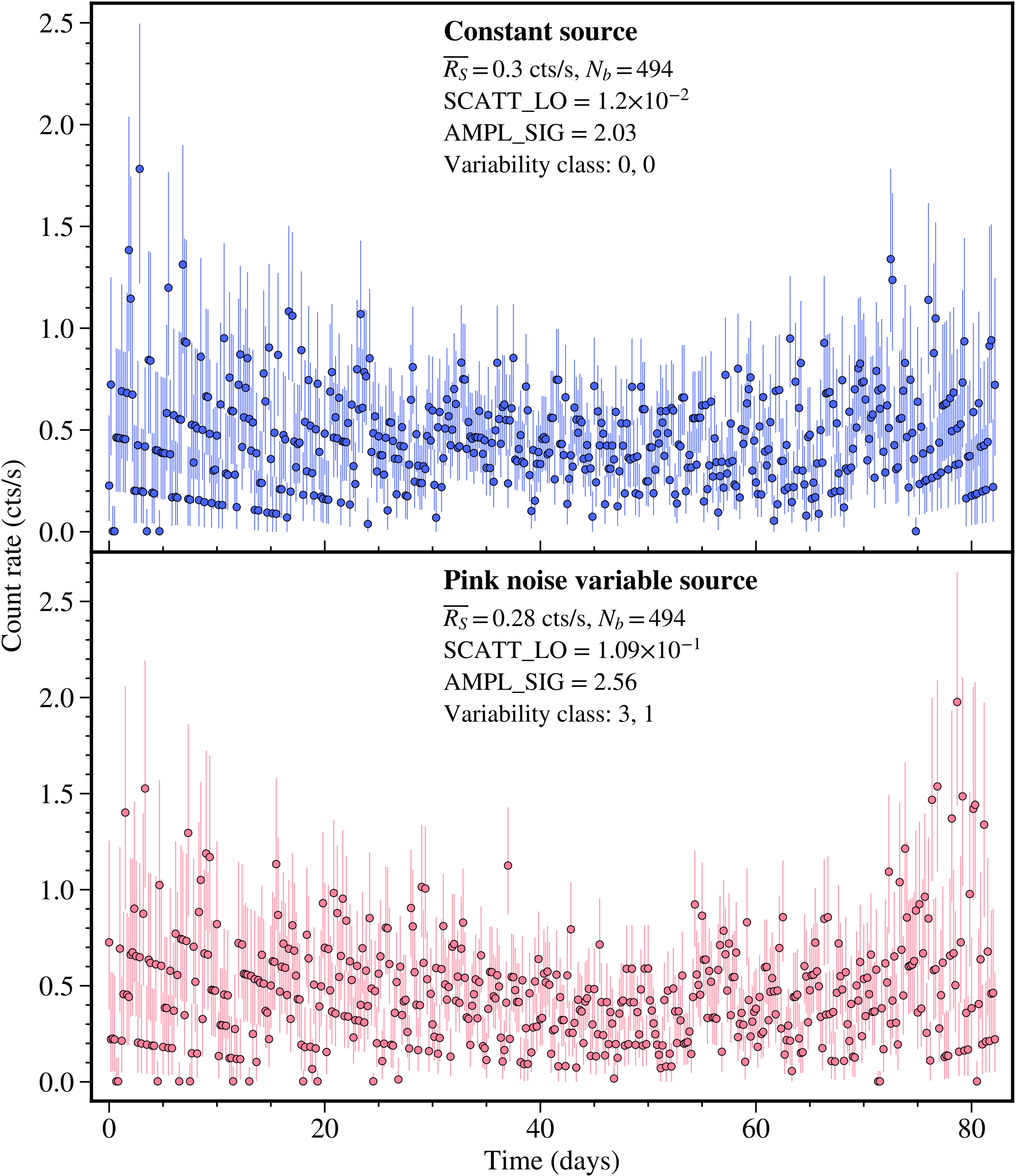}}
\caption{Simulated \emph{eROSITA}-like light curves of a source lying close to either of the two ecliptic poles. These light curves are computed using a typical $\epsilon$ evolution, as observed in actual \emph{eROSITA} light curves. The top panel shows an intrinsically constant source, and the lower panel depicts an intrinsically variable pink noise source, both with an average count rate of $\approx 0.3~\mathrm{cts/s}$. The variability class describes how the value of SCATT\_LO and AMPL\_SIG compares to the variability thresholds computed in Section \ref{Diffvarnonvarsection}. The pink noise variable source is identified to have a SCATT\_LO value above the $3\sigma$ threshold but an AMPL\_SIG value only above the corresponding $1\sigma$ threshold.
 \label{SimLC}}
\end{figure}

\subsection{Challenges for eRASS light curve variability analysis} \label{eRlcchal}

One of the main challenges faced when analysing the variability properties of \emph{eROSITA} light curves, and the key feature that differentiates an \emph{eROSITA} variability analysis from many other variability analyses, is the varying fractional exposure in the light curves. It causes bins to have varying degrees of sensitivity to the source flux, so treating them all identically can bias the analysis.

Visually, the light curve appears to bend upwards when the fractional exposure decreases. Commonly, this appears at the edges of the \emph{eROSITA} observations for sources further than $0.5\degree$ from the poles, leading to `U' shaped light curves, as shown in Fig. \ref{SimLC}. This is illustrated by the two simulated light curves shown in Fig. \ref{SimLC}, one of which is intrinsically constant and the other highly variable.

The `U'-like shape of the light curves can be reduced by coarse rebinning, to several erodays. This is done by summing the counts and exposures from the bins to be merged, and then recomputing the count rate as described above. This eases visual inspection of long term variability, but loses information on eroday to eroday variability. To bridge this gap, we develop and characterize new methods.

Another challenge for analysing \emph{eROSITA} light curves are gaps in the observations. Short observation times (of $41.2 ~ \mathrm{s}$ or less), are separated by long but consistent gaps ($4 ~ \mathrm{h}$). The gaps are at least 350 times longer than the duration of each bin. Sources further than $0.5\degree$ from the poles also have gaps on 6 month time-scales. Variability during the gaps will go undetected.

The SEP \emph{eROSITA} observations are an unparalleled opportunity for variability analysis. Observations of tens of thousands of X-ray sources close to the ecliptic poles occur frequently and consistently over several years. \emph{eROSITA} observations enable a long-term X-ray variability analysis of a large, unbiased sample of AGNs, XRBs, and stars. For this purpose, we first need to decide on the variability quantifiers most useful for an \emph{eROSITA} variability analysis.  

\section{Methods} \label{SecMethod}

\subsection{Variability quantifiers} \label{Varqualsec}

\subsubsection{Normalised excess variance} \label{subsecNEVdes}

A frequently used parameter for quantifying the degree of variability of X-ray light curves of AGNs and XRBs is the normalised excess variance \citep[NEV;][]{1990ApJ...359...86E, 1997ApJ...476...70N}:

\begin{equation} \label{NEVeq}
    \mathrm{NEV} = \frac{\sigma_{\mathrm{obs}}^2 - \overline{\sigma_{\mathrm{err}}^2}}{\overline{R_{\mathrm S}}^2}, 
\end{equation}

\noindent
where the excess variance, $\sigma_{\mathrm{obs}}^2 - \sigma_{\mathrm{err}}^2$, describes how much the observed variance in the source count rates, $\sigma_{\mathrm{obs}}^2$, exceeds the expected variance due to the measurement errors, $\sigma_{\mathrm{err}}^2$. The excess variance is then normalised by the square of the average source count rate, $\overline{R_{\mathrm S}}^2$, to form the NEV. The normalisation removes a scaling with the source flux, and makes the NEV comparable across instruments of different sensitivities.

The NEV measures the excess variance within a time window, from the observed over-dispersion within the set of bins at hand. We refer to the intrinsic variability of the source at the
particular times when it was observed as the NIV (normalised intrinsic
variance) The NEV is an estimator of the NIV, but differs from it due to measurement errors. The NIV is equal to the NEV that would be calculated if the observations were obtained with infinite accuracy and negligible Poisson noise. The NIV is unaffected by Poisson noise, or measurement uncertainties, but does depend on the timing of the observations: the start time, the bin size, and the separation of the bins.

The NIV of variable sources varies stochastically even if the process causing the variability is stationary. The NIV of AGN light curves can, therefore, be described as "weakly non-stationary" \citep{1997ASSL..218...61P}. We define the geometric mean of the NIV of an infinite number of light curves, all covering the same frequency interval and caused by a stationary variability process, as $\mathrm{NIV}_{\infty}$. The NIV of any finite light curve segment varies around the $\mathrm{NIV}_{\infty}$.

For $N_{\mathrm b}$ bins of a variable source at times $t_i$, where $i\in \left\{1, 2, .., N_{\mathrm b}\right\}$, with a measured, background subtracted, exposure-corrected source count rate of $R_{\mathrm S}\left(t_i\right)$, and associated measurement uncertainties of $\sigma_{\mathrm{err}}\left(t_i\right)$, the quantities in Eq. \ref{NEVeq} are defined by:

\begin{align}\label{NEVpar}
\begin{split}
    \overline{R_{\mathrm S}} & = \frac{1}{N_{\mathrm b}} \sum_{i=1}^{N_{\mathrm b}} R_{\mathrm S}\left(t_i\right) \\
    \sigma_{\mathrm{obs}}^2 & = \frac{1}{N_{\mathrm b}-1} \sum_{i=1}^{N_{\mathrm b}} \left(R_{\mathrm S}\left(t_i\right) - \overline{R_{\mathrm S}}\right)^2 \\
    \overline{\sigma_{\mathrm{err}}^2} & = \frac{1}{N_{\mathrm b}} \sum_{i=1}^{N_{\mathrm b}} \sigma_{\mathrm{err}}^2\left(t_i\right).
\end{split}
\end{align}

The lowest NIV that a source can have is 0. This occurs when its flux does not change at all during the observing window, so $\sigma_{\mathrm{obs, NIV}}^2 = 0$. The highest value that the NIV can have occurs when the flux of a variable source is 0 in all bins, except one. For such a light curve, $\overline{R_{\mathrm{S, NIV}}} = R_0/N_{\mathrm b}$, where $R_0$ is the only non-zero count rate of the source. Therefore, the variance of this light curve is $\sigma_{\mathrm{obs, NIV}}^2 = R_0^2/N_{\mathrm b}$, so $\mathrm{NIV}=N_{\mathrm b}$. Hence, the NIV can only have values within the $[0, N_{\mathrm b}]$ range.

In contrast, the NEV can be measured to lie outside of this range. The NEV works well for estimating the NIV of sources with light curves consisting of many bins, each containing many source counts above the background level. Although it can still be used, the NEV faces challenges when applied to light curves of low count rate sources. For instance, it is not unlikely to measure a negative excess variance at low count rates \citep{2017MNRAS.471.4398P}. The NEV is most useful for investigating the variability of light curves consisting of $\gtrsim 20$ bins \citep{1999ApJ...524..667T}.  For more discussion on the NEV, and its uses, see \citet{2003MNRAS.345.1271V, 2004ApJ...611...93P, 2017MNRAS.471.4398P, 2023A&A...673A..68P, 2011A&A...536A..84V, 2016A&A...593A..55V, 2013ApJ...771....9A, 2017ApJ...849..127Z, 2023arXiv230414228A}.

Special care needs to be taken when using the NEV to analyse \emph{eROSITA} light curves.
Firstly, the varying fractional exposures varies the information gained from each bin. However, the parameters of the NEV, as described by Eq. \ref{NEVpar} depend equally on all bins. Therefore, in light curves where the fractional exposures varies, the NEV is biased by, and depends too strongly on the bins with the lowest fractional exposures. To reduce the impact of this, throughout this paper we only keep well-exposed time bins, with $\epsilon>0.1$. Secondly, most bins have low counts. The NEV relies on accurate estimates of the variance in the measured count rates. The measured counts follow a Poisson distribution, which is asymmetric. To account for the resulting asymmetric uncertainty in the inferred net count rates, we chose $\sigma_{\mathrm{err}}$ in Eq. \ref{NEVpar} to be equal to the size of the uncertainty in the direction of the mean. This may, however, cause the NEV be over-, or underestimated.

\subsubsection{Maximum amplitude variation}\label{SecMAD}

The maximum amplitude variation significance \citep[AMPL\_SIG;][]{2016A&A...588A.103B} is another method for detecting and quantifying source variability. The standard definition of the maximum amplitude variation (AMPL\_MAX) uses the bins in which the highest and lowest count rates were measured, which we denote to have occurred at times $t_{\mathrm{max}}$, and $t_{\mathrm{min}}$, respectively. Then AMPL\_MAX, and its significance, AMPL\_SIG, are defined as:

\begin{align}\label{classMADeq}
\begin{split}
    &\mathrm{AMPL\_MAX} = \left[R_{\mathrm S}(t_{\mathrm{max}}) - \sigma_{\mathrm{err}}(t_{\mathrm{max}}) \right] -  \left[R_{\mathrm S}(t_{\mathrm{min}}) + \sigma_{\mathrm{err}}(t_{\mathrm{min}}) \right] \\
    &\mathrm{AMPL\_SIG} = \frac{\mathrm{AMPL\_MAX}}{\sqrt{\sigma_{\mathrm{err}}^2(t_{\mathrm{max}}) + \sigma_{\mathrm{err}}^2(t_{\mathrm{min}})}}. 
\end{split}
\end{align}

\noindent
The advantage of AMPL\_SIG is that it can quickly determine significant differences in the count rate observed within a light curve. AMPL\_MAX only considers the two most extreme points of a light curve, making it suitable for short light curves, and optimal flare detection \citep[see][]{2022A&A...661A..18B}. However, it is less sensitive to milder stochastic variations, especially in the Poisson regime. In \emph{eROSITA}, the highest and lowest count rates measured often occur in bins with the lowest fractional exposures and the largest uncertainties. These cause the AMPL\_SIG to underestimate the actual significance \citep[see calibrated thresholds in][]{2022A&A...661A..18B}.

Therefore, we enhanced equation for investigating light curves featuring varying exposure times. Rather than comparing the two bins with the highest and lowest measured count rates, we instead used the two bins with the highest lower bound and the lowest upper bound confidence interval for the count rate. Therefore, we redefined $t_{\mathrm{max}}$, $t_{\mathrm{min}}$, AMPL\_MAX, and its significance, AMPL\_SIG, as follows: 

\begin{align}\label{amplsigeq}
\begin{split}
    & R_{\mathrm S}(t_{\mathrm{max}})-\sigma_{\mathrm{-err}}(t_{\mathrm{max}}) = \max\left[R_{\mathrm S}(t_i)-\sigma_{\mathrm{-err}}(t_i)\right] \\
    & R_{\mathrm S}(t_{\mathrm{min}})+\sigma_{\mathrm{+err}}(t_{\mathrm{min}}) = \min\left[R_{\mathrm S}(t_i)+\sigma_{\mathrm{+err}}(t_i)\right] \\
    & \mathrm{AMPL\_MAX} = \left[R_{\mathrm S}(t_{\mathrm{max}}) - \sigma_{\mathrm{-err}}(t_{\mathrm{max}}) \right] -  \left[R_{\mathrm S}(t_{\mathrm{min}}) + \sigma_{\mathrm{+err}}(t_{\mathrm{min}}) \right] \\
    & \mathrm{AMPL\_SIG} = \frac{\mathrm{AMPL\_MAX}}{\sqrt{\sigma_{\mathrm{-err}}^2(t_{\mathrm{max}}) + \sigma_{\mathrm{+err}}^2(t_{\mathrm{min}})}}.
\end{split}
\end{align}

\noindent
Where $\sigma_{\mathrm{+err}}$, and $\sigma_{\mathrm{-err}}$ denote the $1\sigma$ errors of the measured count rates in the positive and negative directions, respectively. We used this modified definition of AMPL\_SIG throughout the rest of the paper.  

The AMPL\_SIG can be calculated for all \emph{eROSITA} detected sources, regardless of how often they were observed. However, the more bins there are in a light curve, the more efficient AMPL\_SIG is at detecting variability. Therefore, we did not rebin any light curves for the AMPL\_SIG variability detection and analysis. In Section \ref{Diffvarnonvarsection}, we defined significance thresholds on AMPL\_SIG to identify variable sources.

\subsubsection{Bayesian excess variance}

A third method we used to quantify the variability is the Bayesian excess variance\footnote{\href{https://github.com/JohannesBuchner/bexvar}{https://github.com/JohannesBuchner/bexvar}} \citep[bexvar;][]{2022A&A...661A..18B}. Bexvar uses a hierarchical Bayesian model to determine a posterior probability distribution for the mean and standard deviation of the net count rate, assuming it to follow a log-normal distribution. Background, instrument, and Poisson variability are modelled out. Bexvar models the variability intrinsic to the source, in addition to Poisson variability, background and instrument sensitivity variations with a hierarchical Bayesian model. We refer to samples from the posterior probability distribution of the standard deviation in the log count rate as $\sigma_{\mathrm{b}}$. \citet{2022A&A...661A..18B} also introduced the quantity SCATT\_LO, the 10\% quantile of the distribution of the $\sigma_{\mathrm{b}}$ samples, which is useful for distinguishing between variable and non-variable sources. The standard deviation of the log count rate is estimated by calculating the geometric mean of the samples, which we denote as $\overline{\sigma_{\mathrm{b}}}$. We chose the $15.87\%$ and $84.13\%$ quantiles of the $\sigma_{\mathrm{b}}$ distribution as estimates of the uncertainty in the measurement.

Similar to the NIV, we define the quantity $\sigma_{\mathrm{I}}$ to refer to the standard deviation that would be found by bexvar if all count rates were measured with infinite accuracy. Both the NIV and the $\sigma_{\mathrm{I}}$ are independent quantifiers of the intrinsic variability of a source, unaffected by Poisson noise or measurement uncertainties but dependent on the timing of the observations. Although $\sigma_{\mathrm{I}}$ is not normalised, as the NIV is, it is also invariable to a multiplicative scaling of the flux, by being defined on a logarithmic scale. The NIV describes a variance, and $\sigma_{\mathrm{I}}$ describes a standard deviation. Nevertheless, as the NIV is defined for a linear count rate, and $\sigma_{\mathrm{I}}$ is defined for a logarithmic count rate, the two quantities are not related by a square; $\sigma_{\mathrm{I}} \neq \sqrt{\mathrm{NIV}}$.

The strength of bexvar lies in the self-consistent Bayesian handling, modelling the observed counts with a Poisson distribution and propagating the probability distributions. Unlike the NEV, $\sigma_{\mathrm{b}}$ and $\overline{\sigma_{\mathrm{b}}}$ can never be negative. Bexvar uses the Poisson probability distributions of the measured count rate in each bin, rather than a single value for the uncertainty in the count rate, as used by the NEV and AMPL\_SIG. 

Bexvar estimates the excess variability power on the timescale of the binning, assuming that each bin has an independently drawn count rate. We used a uniform prior within the $[-2, 2]$ interval for $\log\left(\sigma_{\mathrm{b}}\right)$, as this is the range of values we expect to be able to measure for it. Smaller degrees of variability are possible but are unlikely to be distinguished from non-variability in eRASS light curves. 

Unlike the standard NEV methodology, bexvar does not weigh all bins identically. It also uses Poisson probability distributions for the count rate in each bin to determine the degree of variability. These two features enable bexvar to estimate $\sigma_{\mathrm{I}}$ and the error in the measurement accurately, for a log-normal white noise ($P\propto\nu^0$) light curve, with variable fractional exposures, over a wide range of count rates  \citep[][Fig. A.1.]{2022A&A...661A..18B}

Rebinning the light curves of very faint sources, which consist mostly of bins with 0 source counts, can be beneficial for bexvar analysis. Bexvar is more computationally expensive than other variability estimators, and its computation time increases linearly with the number of bins in the light curve. Rebinning should not affect the measured variability as long as most of the variability power contained within the frequency interval of the original light curve is maintained below the Nyquist frequency of the rebinned light curve. Unless a very faint source exhibits brief, large flares, it is difficult to determine a precise or accurate estimate of its variability at the timescale of the separation of the \emph{eROSITA} bins, even with bexvar. Therefore, rebinning the light curves of very faint sources usually does not reduce the ability to investigate their variability, but reduces the computation time. 

Hence, we chose to rebin light curves of faint sources until an average of at least one source count was contained in every two bins, for detecting variability with bexvar. We also required that the rebinned light curve still consisted of at least 20 bins. For flaring sources, it is preferable to use AMPL\_SIG to detect and characterise their variability, but only with the calibrated thresholds on AMPL\_SIG defined by \citet{2022A&A...661A..18B}.

Bexvar is most informative when calculated for light curves consisting of 20 bins or more. It can be used to quantify the variability of all \emph{eROSITA} sources observed for 4 eRASSs or more. Fewer eRASSs of observation are required closer to either of the two ecliptic poles. 

While $\sigma_{\mathrm{b}}$ is much more suitable for quantifying variability in the low count Poisson regime, it may be useful to convert this quantity into one equivalent to the NEV for comparison. We call this quantity $\mathrm{NEV}_{\mathrm{b}}$. In Section \ref{NEVest} we derive an empirical conversion factor between $\sigma_{\mathrm{b}}$ and $\mathrm{NEV}_{\mathrm{b}}$, and evaluate how it compares to the NEV in Appendix \ref{CompNIVest}.

\subsubsection{Power spectral density and periodograms}

The Power Spectral Density \citep[PSD; for a review see][]{1989ASIC..262...27V} describes the distribution of variability power as a function of frequency. A periodogram is an estimate of the PSD of a variable source, between the frequencies $(N_{\mathrm b} \tau)^{-1}$, and $(2\tau)^{-1}$. The quantity $\tau$ represents the separation between one bin and the next. The periodogram is calculated as the modulus square of the Fourier transform. It depends on the time ordering of bins, and describes the correlation of individual measured count rates as a function of their temporal separation. The PSD estimates the variability power for frequencies ranging from the inverse of the duration of the observations, up to the Nyquist frequency, which is half of the sampling frequency.

We normalised periodograms using the fractional rms normalisation \citep{1990A&A...227L..33B}. It has the useful feature that the NEV is equal to the integral of the rms-normalised and noise subtracted periodogram. 

The periodograms of variable sources are often dominated by a power law shape: $P(\nu) \propto \nu^{-\alpha}$, where $\alpha$ is the power law index. A power law PSD with $\alpha = 0$ is known as white noise and corresponds to a light curve in which the count rate in every bin is independent of the count rate in any other bin. A red noise PSD is described by a power law with $\alpha=2$ and is associated with a light curve dominated by long-term trends, in which the count rate in each bin is strongly correlated to that in adjacent bins. A pink noise PSD lies between the two, and has $\alpha = 1$.

AGN and XRB X-ray light curves are typically observed to have periodograms which can be described by a red noise power law of $\alpha \approx 2$ at high frequencies, and a pink noise power law of $\alpha \approx 1$ at lower frequencies \citep{1999ApJ...514..682E, 2002A&A...382L...1P, 2003ApJ...593...96M, 2004MNRAS.348..207P, 2012A&A...544A..80G, 2017ApJ...849..127Z}. The transition from $\alpha \approx 2$ to $\alpha \approx 1$ is usually described by a sharp break, but has also been modelled as a gradual bend in the PSD \citep{2004MNRAS.348..783M, 2012A&A...544A..80G}. This break or bend occurs somewhere between about $10^{-6.4} - 10^{-3.3} ~\mathrm{Hz}$ \citep{2012A&A...544A..80G}. 

Similar power laws have been identified in the periodograms of XRBs, at much lower frequencies. For those, a low frequency break from $\alpha \approx 1$ to $\alpha \approx 0$ can be observed \citep{1990A&A...227L..33B} as well. This has not yet been observed for AGNs. The timescales for XRB variability are on the order of seconds or less. Important features in XRB periodograms typically occur in the $0.01-100~\mathrm{Hz}$ range \citep{1999ApJ...514..939W, 2019NewAR..8501524I}. Therefore, light curves binned into single eroday bins can only detect the long-term evolution of an XRB. Periodograms of such light curves are mainly useful for investigating AGNs. 

To accurately estimate the PSD by computing a periodogram requires more information on the source flux at different times than is required to estimate the NIV, the $\sigma_{\mathrm{I}}$, or for measuring AMPL\_SIG. Therefore, for \emph{eROSITA} observations, this detailed analysis will only be possible for bright variable sources close to either of the two ecliptic poles. 

Periodograms of light curves were computed with the Stingray\footnote{\href{https://docs.stingray.science/}{https://docs.stingray.science/}} timing package \citep{2016ascl.soft08001H}. We particularly used the Stingray \texttt{Powerspectrum} function, which computes periodograms using a fast Fourier transform algorithm.

These four variability measures; the NEV, AMPL\_MAX, $\overline{\sigma_{\mathrm{b}}}$, and the periodogram, each quantify the degree of variability of a source in different ways. In the rest of the paper, we will describe how they are best used for analysing \emph{eROSITA}-like variable light curves. 

\subsubsection{Band-limited power}\label{sec:blp}

We assume a stationary process causing the observed variability. In such a case, this process can be associated with a fixed PSD describing the source variability at all times. We denote the constant value of the integral of such a PSD between two frequencies as the band-limited power, which describes how variable a source is within the selected frequency range. It does not depend on any properties of the observations, and is identical for different sets of observations. 

The band-limited power is similar to the $\mathrm{NIV}_{\infty}$, which is the geometrically averaged NIV over an assumed infinitely many time intervals. The $\mathrm{NIV}_{\infty}$ is also a constant for a stationary process. The difference between them is that the $\mathrm{NIV}_{\infty}$ overestimates the band-limited power due to power leakage \citep[see e.g. ][]{1989ASIC..262...27V, 2003MNRAS.345.1271V}. The properties of the observations can induce variability power at both lower (red noise leak) and higher frequencies (aliasing) than those being investigated, to leak into the $[(N_{\mathrm b} \tau)^{-1}, (2\tau)^{-1}]$ frequency space, and increase the power measured within it. This is discussed in more detail in Section \ref{SecAlias}.

The NIV varies around the $\mathrm{NIV}_{\infty}$ due to the intrinsic scatter of the NIV, caused by the stochastic nature of the variability within a limited set of observations. Finally, the NEV varies around the NIV due to measurement errors.

Estimates of constant quantities, such as the band-limited power, or the $\mathrm{NIV}_{\infty}$ should be used when comparing the variability of different sources, when considering the scaling of the variability with other properties, such as the BH mass or the AGN luminosity, or when investigating whether the variability of an individual source changed over time. Even if the variability process is stationary, the NEV, AMPL\_SIG, $\overline{\sigma_{\mathrm{b}}}$, and periodogram of two different sets of observations are likely to differ by more than their measurement errors would indicate. This happens due to the intrinsic scatter in the variability within a limited window of observations. Without accounting for the intrinsic scatter, differences in variability measurements are likely to be overestimated.

The first step to calculate the band-limited power is to estimate the NIV. This could be done by calculating the NEV, integrating the periodogram, or using $\overline{\sigma_{\mathrm{b}}}$, along with the conversion from $\sigma_{\mathrm{I}}$ to the NIV, that is described in Section \ref{NEVest}. The $\mathrm{NIV}_{\infty}$ can then be estimated by including the intrinsic scatter in the NIV as an additional error. This is discussed in more detail in Sections \ref{SecSysErr} and \ref{SecRedSysErr}. Finally, the $\mathrm{NIV}_{\infty}$ can be converted into an estimate of the band-limited power, by quantifying the strength of the power leakage from higher and lower frequencies into the $[(N_{\mathrm b} \tau)^{-1}, (2\tau)^{-1}]$ frequency space, and subtracting that from the estimate. This final step requires several assumptions about the shape of the source PSD in the frequency space not investigated. That can introduce additional uncertainty in the estimate of the band-limited power. Therefore, unless the strength of the power leakage can be reliably estimated without bias, it can be preferable to compare estimates of the $\mathrm{NIV}_{\infty}$ instead. 

\subsection{Simulations}\label{SecSim}

We performed a variety of different types of simulations throughout this paper, to understand the statistical behaviour of each estimator when applied to \emph{eROSITA} light curves. The true flux of simulated variable sources was determined by selecting the PSD of the source, and then using the \citet{1995A&A...300..707T} method to generate a light curve from it. We typically selected pink noise PSDs for this purpose. The break in AGN PSDs, from an $\alpha \approx 1$ power law, to an $\alpha\approx2$ power law has been observed to often occur within, or above, the frequency range probed by \emph{eROSITA} ($3.96\times 10^{-9} - 3.47\times 10^{-5} ~ \mathrm{Hz}$) \citep{2012A&A...544A..80G}. The aliasing effect often counteracts the steeper power law at high frequencies, and flattens the PSD (see Section \ref{SecAlias}). Therefore, the PSDs of typical \emph{eROSITA} observed AGNs could, to first order, be assumed to approximately follow an $\alpha=1$ power law. The accuracy of this first order assumption is also supported by the periodograms of actual \emph{eROSITA} observations of AGNs, as will be discussed by Bogensberger et al. 2024C. We also simulated power law PSDs with $\alpha=0$ (white noise) and $\alpha=2$ (red noise) for comparison, to determine how strongly the variability methods depend on the PSD shape. For the purpose of identifying significance thresholds for variability detection, we simulated constant light curves, in which the true flux did not vary.

From the simulated light curves of the true flux, we selected intervals of 1050 bins. That length approximately matches the upper limit on the number of observations of a source that can be made per eRASS. To simulate the red noise leak, we selected input PSDs that extended with the same power law to frequencies at least one order of magnitude below the inverse of the total light curve duration. We simulated light curves with at least one order of magnitude more bins than we needed for the analysis, and randomly selected starting points within that interval for the selected portion to be used for further analysis.

The true flux of the simulated light curves was shifted, and scaled such that the mean average source flux matched the desired source count rate at the detector, and there were no bins with a negative true flux. Increasing or decreasing the flux at all points by a constant amount only affects the value of the PSD associated with it at $\nu=0$, which is not relevant to this analysis. Scaling the amplitude of the flux variation affects the NIV. This is how we generated light curves with a wide range of different NIVs. The light curves produced in this way do not contain any background or Poisson noise yet. Therefore, we refer to them as the true light curves of the simulated sources. The NIVs of these light curves were computed directly from the mean and variance of the true light curves, using Eqs. \ref{NEVeq} and \ref{NEVpar}, with $\overline{\sigma_{\mathrm{err}}^2} = 0$.

To investigate the ability to identify variable sources, and the reliability of estimating the NIV from a light curve (Sections \ref{Diffvarnonvarsection} and \ref{NEVest}), we used the true light curves as a basis for generating simulated observed light curves, with properties as similar as possible to those detected by \emph{eROSITA}. For this purpose, we selected a background count rate of $0.71~\mathrm{cts/s}$ and a background area of $0.0089$, which are equal to the mean value of both parameters found for sources detected by \emph{eROSITA} close to the SEP. Next, we randomly selected a fractional exposure for each bin from the distribution observed for sources in the SEP field, assuming $\Delta t = 40~\mathrm{s}$. We cropped this distribution to avoid fractional exposures below 0.1, as is also done for the actual data. The light curves simulated in this way look similar to those in Fig. \ref{SimLC}, except that they feature a random assortment of fractional exposures. These simulated light curves were only used in conjunction with the SCATT\_LO, and AMPL\_SIG methods, which do not depend on the time ordering of bins. 

For each bin of the simulated true light curves, we randomly selected a measured number of source counts from the Poisson distribution with a mean of  $(R_{\mathrm{S,t}}(t_i)+R_{\mathrm{B,t}}(t_i)A(t_i))\epsilon \Delta t$. In this equation, $R_{\mathrm{S,t}}(t_i)$, and $R_{\mathrm{B,t}}(t_i)$ are the true source and background count rates at time $t_i$. We, similarly, selected a measured number of background counts from the Poisson distribution of $R_{\mathrm{B,t}}(t_i)\epsilon \Delta t$. We computed source count rates from the simulated number of source and background counts, the background area, and the fractional exposure, as outlined in Section \ref{SeceRO}. These types of simulated light curves are referred to as being \emph{eROSITA}-like.

We also simulated light curves with fractional exposure distributions that were not based on any observed distribution. This was done to investigate the dependence of the noise level in a periodogram on the mean ($\overline{\epsilon}$) and variance ($\sigma_{\epsilon}^2$) of the fractional exposure distribution in the light curve (Section \ref{Powspecsection}). To span a large parameter space in both parameters, we first selected a minimum and maximum fractional exposure, drawn from a grid of 40 equally spaced points between 0.1 and 1.0, covering all possible combinations. 

We generated scenarios for a maximum, minimum, and intermediate $\sigma_{\epsilon}^2$ for each range of fractional exposures, at a fixed $\overline{\epsilon}$. For the maximum $\sigma_{\epsilon}^2$  scenario, we assigned half of all bins to the maximum fractional exposure, and half to the minimum fractional exposure of the selected range. The ordering of the fractional exposures does not influence the periodogram noise level. For the minimum $\sigma_{\epsilon}^2$  scenario, we assigned one randomly chosen bin with the maximum fractional exposure, one bin with the minimum fractional exposure, and kept the rest at the value halfway between the two. For the intermediate $\sigma_{\epsilon}^2$  scenario, we selected fractional exposures to cover the interval with a constant incremental increase from minimum to maximum. We refer to the simulated true light curves created in this way as the patterned fractional exposure light curves. 

Throughout this paper, we investigated various analytical models, to characterise the use of variability quantifiers for \emph{eROSITA} light curves. These were fitted with the nested sampling Monte Carlo algorithm MLFriends \citep{2014arXiv1407.5459B, 2019PASP..131j8005B} through the
UltraNest\footnote{\href{https://johannesbuchner.github.io/UltraNest/}{https://johannesbuchner.github.io/UltraNest/}} package \citep{2021JOSS....6.3001B}. 

\section{Methods for identifying varying sources}\label{Diffvarnonvarsection}

Out of the millions of X-ray sources detected by \emph{eROSITA} \citep{2021A&A...647A...1P}, we intended to select a much smaller set of sources whose count rate changes significantly throughout the observing interval. These can subsequently be investigated individually in more detail. As described in Section \ref{SeceRO}, eRASS observations of X-ray sources feature a large variety of different properties. We aimed at being able to detect significantly variable sources throughout the observed parameter space, even ones with low count rates. This method can be used to detect variable sources in the eRASS dataset, but may also be applicable to other surveys.

We did not seek to optimally divide the sample into likely variable and likely non-variable sources. Instead, we aimed at being able to select variable sources at a low false positive rate. In addition, we intended the variability thresholds to not be biased towards any particular type of variability, and be able to identify unexpected variability as well.

\citet{2022A&A...661A..18B} investigated the ability of the AMPL\_SIG, NEV, SCATT\_LO, and Bayesian block methods to detect flaring, white noise ($P\propto \nu^0$), and red noise ($P\propto \nu^{-2}$) variability, for the \emph{eROSITA} Final Equatorial-Depth Survey (eFEDS). Of the four methods, \citet{2022A&A...661A..18B} found that SCATT\_LO is almost always the most sensitive to detecting variability, regardless of its type. They, however, also found that AMPL\_SIG is slightly better at detecting flaring sources at high count rates than SCATT\_LO. Both AMPL\_MAX and SCATT\_LO were designed to quantify variability, rather than to distinguish variable from non-variable sources. Nevertheless, they can both be used for that purpose as well. Following the conclusions of \citet{2022A&A...661A..18B}, we decided to use both SCATT\_LO, and AMPL\_SIG to distinguish likely variable from likely non-variable sources in eRASS data sets as well.

To use SCATT\_LO and AMPL\_SIG for variability detection in the eRASSs, we decided to define variability significance thresholds. \citet{2022A&A...661A..18B} defined thresholds on both parameters for detecting variable sources in eFEDS. However, these are not necessarily applicable to eRASS observations, especially not for the regions close to the ecliptic poles, for which there is a large range of different number of bins, and an enhanced sensitivity to detect faint sources. \citet{2022A&A...661A..18B} defined the thresholds as a function of count rate, but did not investigate the dependence on the number of bins. Therefore, we sought to define thresholds as a function of both parameters, and specific to the eRASSs. 

For this purpose, we simulated $4\times 10^5$ \emph{eROSITA}-like light curves of intrinsically non-variable sources, as discussed in Section \ref{SecSim}. We simulated $10^4$ iterations of light curves for 40 sets of combinations of intrinsic source count rates of $\{0.001, 0.003, 0.01, 0.03, 0.1, 0.3, 1.0, 3.0, 10, 30\}$ $\mathrm{cts/s}$, and number of erodays of observation with $\epsilon > 0.1$, of $\{50, 135, 370, 1000\}$. This range of count rates and number of bins was selected to be most useful in selecting variable eRASS sources. For a single eRASS, sources within $\approx7\degree$ of the ecliptic poles are observed between 50-1080 erodays per eRASS. When combining eight eRASSs, all sources in the sky will have been observed on at least 48 different erodays.

\begin{figure}[h]
\resizebox{\hsize}{!}{\includegraphics{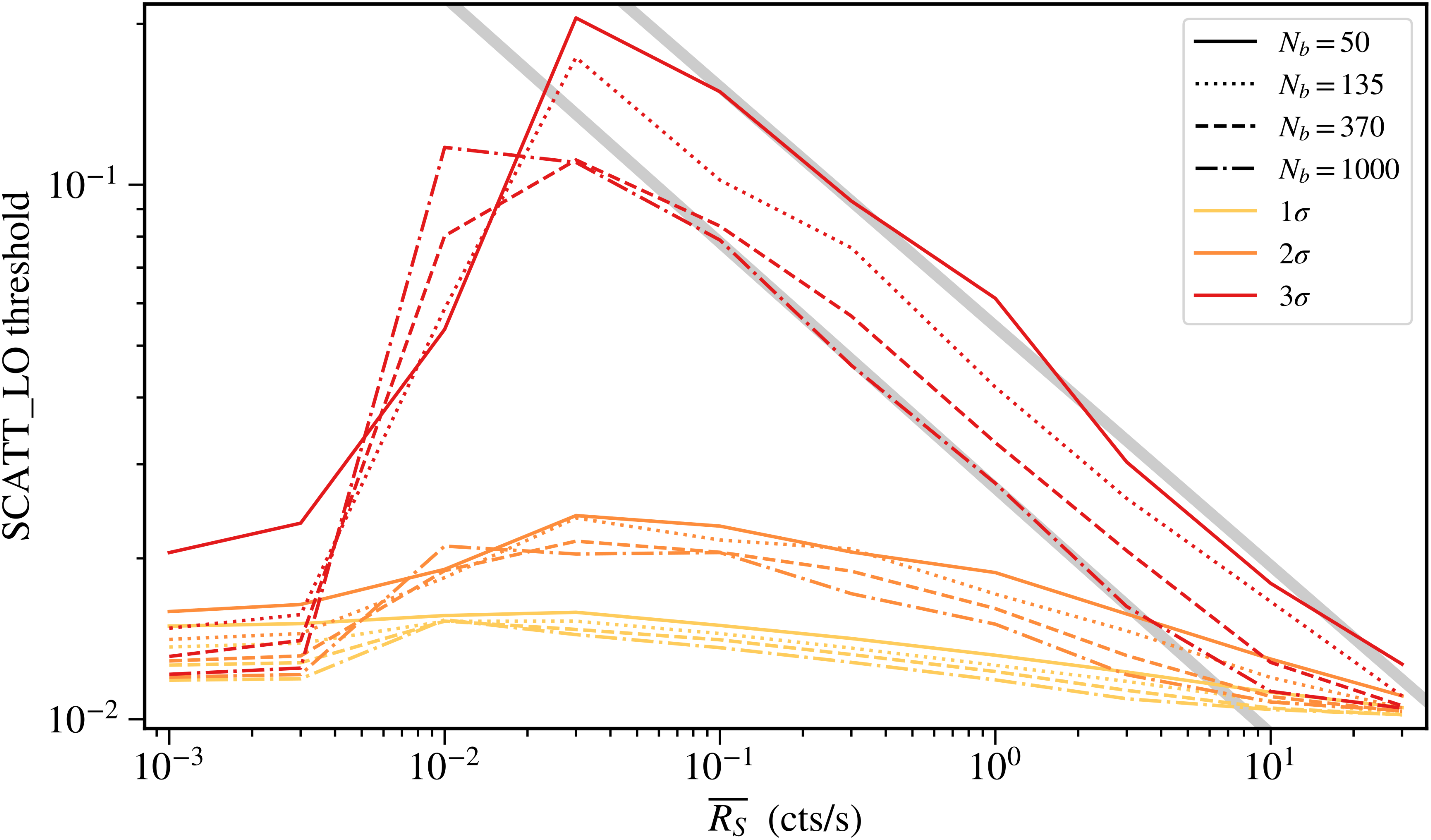}}
\caption{The 1, 2, and 3$\sigma$ thresholds on SCATT\_LO, for identifying variable sources. The thresholds are displayed as a function of the count rate. The dependence on the number of bins is illustrated by using different line styles. The grey lines indicate the best fit power law relationship to the decreasing thresholds with increasing count rate, above the peak. 
 \label{BV2D}}
\end{figure}

\begin{figure}[h]
\resizebox{\hsize}{!}{\includegraphics{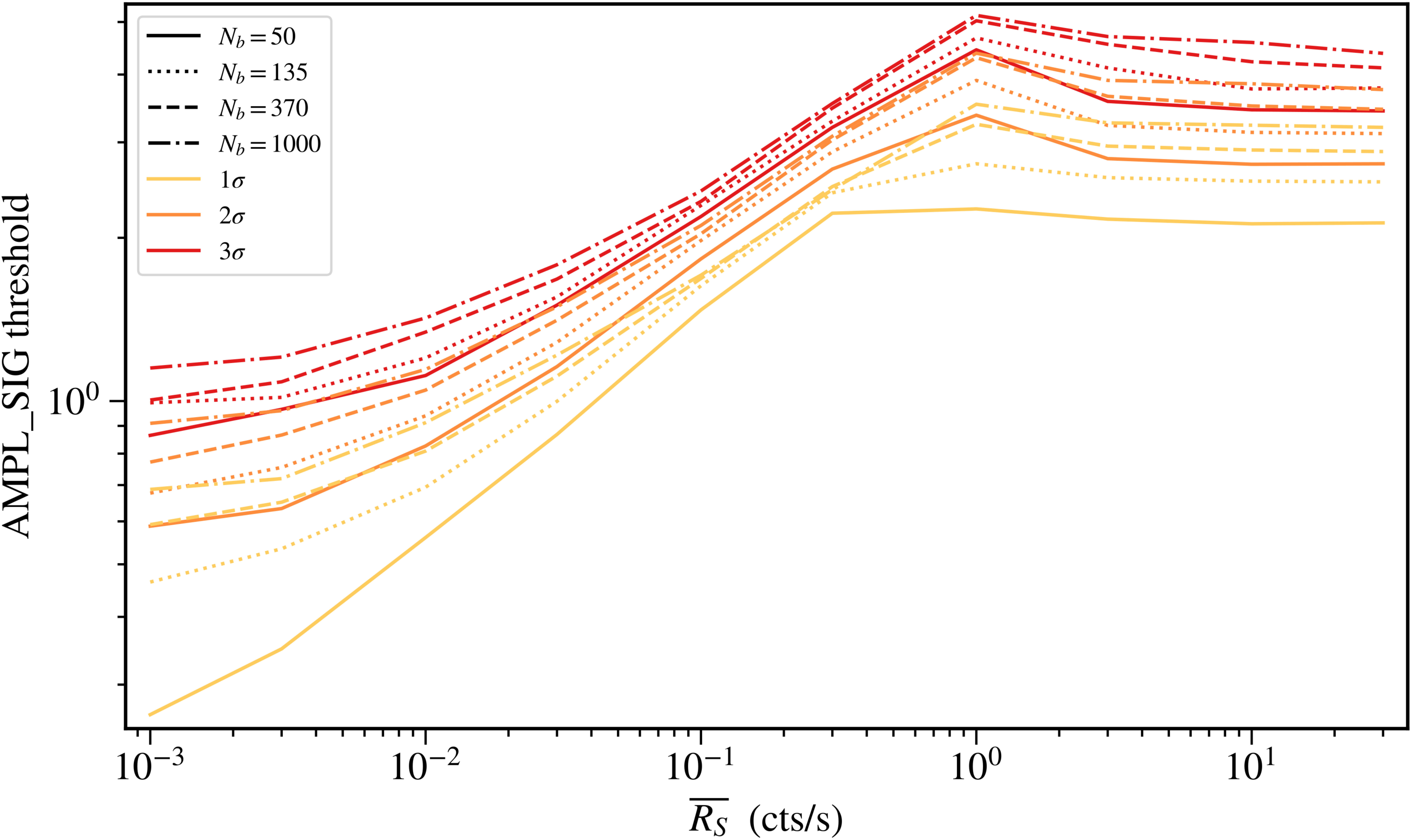}}
\caption{The 1, 2, and 3$\sigma$ thresholds on AMPL\_SIG, for identifying variable sources. The colours and line styles are the same as those in Fig. \ref{BV2D}.
 \label{AS2D}}
\end{figure}

For each simulated light curve, we computed the SCATT\_LO and AMPL\_SIG parameters. From the resulting distribution of values, we determined one-tailed $1\sigma$ (84.13\%), $2\sigma$ (97.72\%), and $3\sigma$ (99.865\%) equivalent quantiles for each input count rate and number of bins. These are displayed in Figs. \ref{BV2D}, and \ref{AS2D}, for SCATT\_LO, and AMPL\_SIG, respectively. They show that the variability significance thresholds of SCATT\_LO, and AMPL\_SIG depend on both the count rate and the number of bins. Variable sources may be selected at different significances, with differing purities, and false positive rates.

The SCATT\_LO thresholds initially rise sharply with increasing count rate. They reach a peak at $0.01-0.03~\mathrm{cts/s}$, before gradually declining over three orders of magnitude in the count rate. The peak is most prominent for the $3\sigma$ threshold, and occurs at a lower count rate for longer light curves. The $1\sigma$ threshold only has a very weak peak and does not change much as a function of the count rate or the number of bins. At high count rates above the peak, the three thresholds converge to one another, and towards $\mathrm{SCATT\_LO} = 10^{-2}$. This is a consequence of the choice of prior for $\log\left(\sigma_{\mathrm b}\right)$, which has a minimum value of -2. 

The decline of the $3\sigma$ thresholds above $\overline{R_{\mathrm S}} = 0.1~\mathrm{cts/s}$ approximately follows a power law of $\mathrm{SCATT\_LO} \propto \overline{R_{\mathrm S}}^{~-0.45}$, which is shown via the grey lines for light curves of 50 and 1000 bins in Fig. \ref{BV2D}. Above the peak, the SCATT\_LO thresholds also show a more simple dependence on the number of bins that can be approximated by $\mathrm{SCATT\_LO} \propto N_{\mathrm b}^{-0.22}$. Both of these trends reflect the increased sensitivity of SCATT\_LO to detect lower degrees of variability when there are a greater number of source counts.

The AMPL\_SIG thresholds have a very different dependence on the count rate and the number of bins. The three thresholds are much closer together than the SCATT\_LO thresholds are. This makes it more challenging to distinguish between sources at different variability significances. The AMPL\_SIG thresholds are dominated by a gradual rise with an increasing count rate until a shallow peak is reached at $1.0~\mathrm{cts/s}$. The thresholds decrease slightly at higher count rates, before plateauing towards the highest count rates we investigated. This general shape of the thresholds can be understood to be a consequence of the accuracy of the assumptions of normal probability distributions on the measured count rate per bin. 

The dependence of the AMPL\_SIG thresholds on the number of bins is the opposite of what was observed for the SCATT\_LO thresholds. The more bins there are, the higher the thresholds are. The reason for this is that it is more likely to find outliers in the count rate of longer light curves. 

These results are qualitatively similar to those of \citet{2022A&A...661A..18B}. Their thresholds were calculated for a different number of bins, and for the eFEDS, rather than the eRASS dataset, so the exact values are not directly comparable. Nevertheless, we found the SCATT\_LO thresholds to have a similar dependence on the count rate. However, whereas \citet{2022A&A...661A..18B} found the AMPL\_SIG thresholds to continuously increase with the count rate, we detected these to reach a plateau above $1.0~\mathrm{cts/s}$. The likely reason for these differences is that the two thresholds use different definitions of AMPL\_SIG. 

We applied the two variability identification methodologies to the intrinsically non-variable and variable simulated \emph{eROSITA}-like light curves of Fig. \ref{SimLC}. Both methods correctly located the non-variable simulated source, shown in the top panel, in the $< 1\sigma$ variability class. In contrast, the intrinsically variable source exhibiting pink noise variability, shown in the lower panel, was identified as variable above the $3\sigma$ threshold by SCATT\_LO. AMPL\_SIG instead identified it between the 1 and $2\sigma$ thresholds. This is a consequence of SCATT\_LO being more sensitive to detecting pink noise variability \citep{2022A&A...661A..18B}.

The higher significance thresholds are based on smaller fractions of simulated light curves. To establish more accurate dependencies of the thresholds on the number of bins and the count rate, would require significantly more simulations.

\section{Intrinsic variance estimation}\label{NEVest}

The thresholds on SCATT\_LO and AMPL\_SIG were set up principally to identify variable sources at a given false positive rate. They do not necessarily indicate the strength of the variability of a given source within a set of observations. In this section, we investigate the correlation between $\sigma_b$, and the NIV, such that a measurement of the former can be used to estimate the latter. 

A successful method for estimating the NIV should be relatively unaffected by the features of eRASS light curves that could create issues for variability analysis (as outlined in Section \ref{eRlcchal}), and remain accurate at both low and high count rates. The method presented in this section was derived for the assumption of \emph{eROSITA}-like pink noise light curves, but is applicable more generally to low count rate light curves with power law PSDs.

The bexvar methodology allows for a more accurate estimate of $\sigma_{\mathrm{I}}$ than the NEV is at estimating the NIV, especially at low count rates and for \emph{eROSITA} light curves (see Appendix \ref{CompNIVest}). However, it is unclear how $\sigma_{\mathrm{I}}$ relates to other measures of the intrinsic variability of a source, such as the NIV or the PSD. 

In contrast, the NIV is more easily interpretable, as it can be associated with the integral of the PSD, and is a measure of the variance of the linear flux distribution. It has been frequently used, and is tied to physical models. There are methods for converting estimates of the NIV into estimates of the constant band-limited power, which is intrinsic to the source, and not dependent on any properties of the observation. However, at the moment, there is no established quantity equivalent to the band-limited power for the $\sigma_{\mathrm{I}}$. Furthermore, the impact of the red noise leak and the aliasing effect on $\sigma_{\mathrm{I}}$ are still unknown. It could be useful to combine the strengths of the NEV and bexvar methodologies, by using the Bayesian framework of bexvar to estimate the NIV.

Therefore, we investigated the possibility of converting the bexvar $\overline{\sigma_{\mathrm{b}}}$ estimate of $\sigma_{\mathrm{I}}$ into an estimate of the NIV of a light curve. If such a function can be found, it could be used in both directions, and allow for variability found by either parameter to be compared with each other. It could also be used to explore the influence that power leakage has on $\sigma_{\mathrm{I}}$, and help determine a stationary parameter equivalent to the band-limited power for $\overline{\sigma_{\mathrm{b}}}$. We denote the NIV estimate based on a $\overline{\sigma_{\mathrm{b}}}$ measurement, using the function relating it to the NIV, as the $\mathrm{NEV}_{\mathrm{b}}$. 

For this purpose, we simulated \emph{eROSITA}-like light curves of intrinsically variable sources exhibiting pink noise variability, as described in Section \ref{SecSim}, in order to investigate the ability of $\overline{\sigma_{\mathrm{b}}}$ to accurately estimate the $\sigma_{\mathrm{I}}$ for these types of sources. We investigated light curves consisting of $\{20, 75, 150, 400, 1050\}$ bins, with mean input count rates of $\{0.015, 0.15, 1.5, 15\}~\mathrm{cts/s}$. This range of count rates and number of bins was chosen to cover the parameter space of eRASS light curves, and is the same as those used to compare different methods of estimating the NIV, in Appendix \ref{CompNIVest}. 

\begin{figure}[h]
\resizebox{\hsize}{!}{\includegraphics{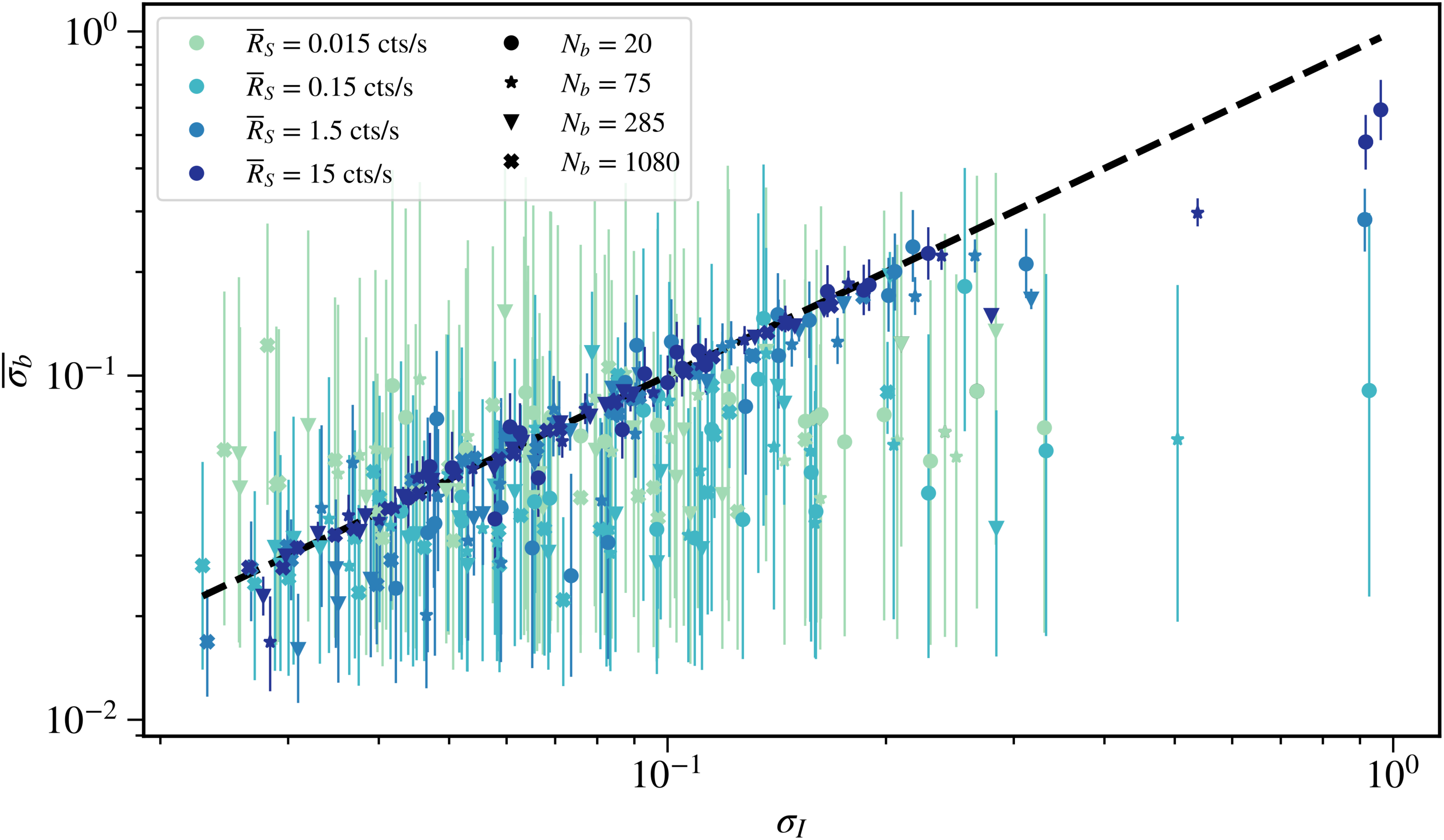}}
\caption{The ability of $\overline{\sigma_{\mathrm{b}}}$ to estimate $\sigma_{\mathrm{I}}$ of a pink noise light curve. Each data point represents a single simulated \emph{eROSITA}-like light curve. The black dashed line indicates the $1:1$ relationship between the two parameters. 
\label{BVtologvar}}
\end{figure}

The results of these simulations are shown in Fig. \ref{BVtologvar}. The $\sigma_{\mathrm{I}}$ was determined from the true light curve, before adding Poisson noise and a background count rate. The $\overline{\sigma_{\mathrm{b}}}$ was calculated for the simulated measured source and background count rates. \citet{2022A&A...661A..18B} showed that $\overline{\sigma_{\mathrm{b}}}$ is an accurate estimator of $\sigma_{\mathrm{I}}$ for light curves with a log-normal distribution of count rates. Fig. \ref{BVtologvar} shows that $\overline{\sigma_{\mathrm{b}}}$ is also mostly accurate at estimating $\sigma_{\mathrm{I}}$ for sources exhibiting pink noise variability. For very low count rate light curves, $\overline{\sigma_{\mathrm{b}}}$ is very uncertain, but bexvar still estimates a reasonable error range within which $\sigma_{\mathrm{I}}$ is most likely to be found. This is very different to the ability of the NEV methodology to estimate the NIV (see Appendix \ref{CompNIVest}). However, this figure also shows that $\overline{\sigma_{\mathrm{b}}}$ appears to systematically underestimate large $\sigma_{\mathrm{I}}$ values for pink noise light curves, even at high count rates. As $\overline{\sigma_{\mathrm{b}}}$ is a mostly accurate estimator of $\sigma_{\mathrm{I}}$, it could potentially also be used to estimate the NIV, when using a function relating the two variability parameters. Therefore, we subsequently investigated the dependence of the measured $\overline{\sigma_{\mathrm{b}}}$ on the NIV of the true light curve.

While a light curve with a larger NIV also tends to have a larger $\sigma_{\mathrm{I}}$, the exact relationship between those two parameters is significantly more complex than the first-order estimate of $\mathrm{NIV} \propto \overline{\sigma_{\mathrm{b}}}^2$. To investigate the nature of the relationship, we simulated $3\times10^4$ \emph{eROSITA}-like pink noise light curves. These consist of 600 light curves for all combinations of the number of bins within the set $\{20, 50, 135, 370, 1000\}$, and count rates of $\{0.001, 0.003, 0.01, 0.03, 0.1, 0.3, 1.0, 3.0, 10.0, 30.0\} ~ \mathrm{cts/s}$ (see Section \ref{SecSim}). We did not investigate the correlation between $\overline{\sigma_{\mathrm{b}}}$ and the NIV for other types of variability, such as red noise.

At the lowest count rates, and the smallest number of bins, the simulations produce so few source counts, that it is not meaningful even to use the measured $\overline{\sigma_{\mathrm{b}}}$ as an estimator of the degree of variability of the source. Nevertheless, we still included these instances, to investigate the ability to use the measured $\sigma_{\mathrm{b}}$ distribution to determine uncertainties and upper limits on an estimate of the NIV.

\begin{figure}[h]
\resizebox{\hsize}{!}{\includegraphics{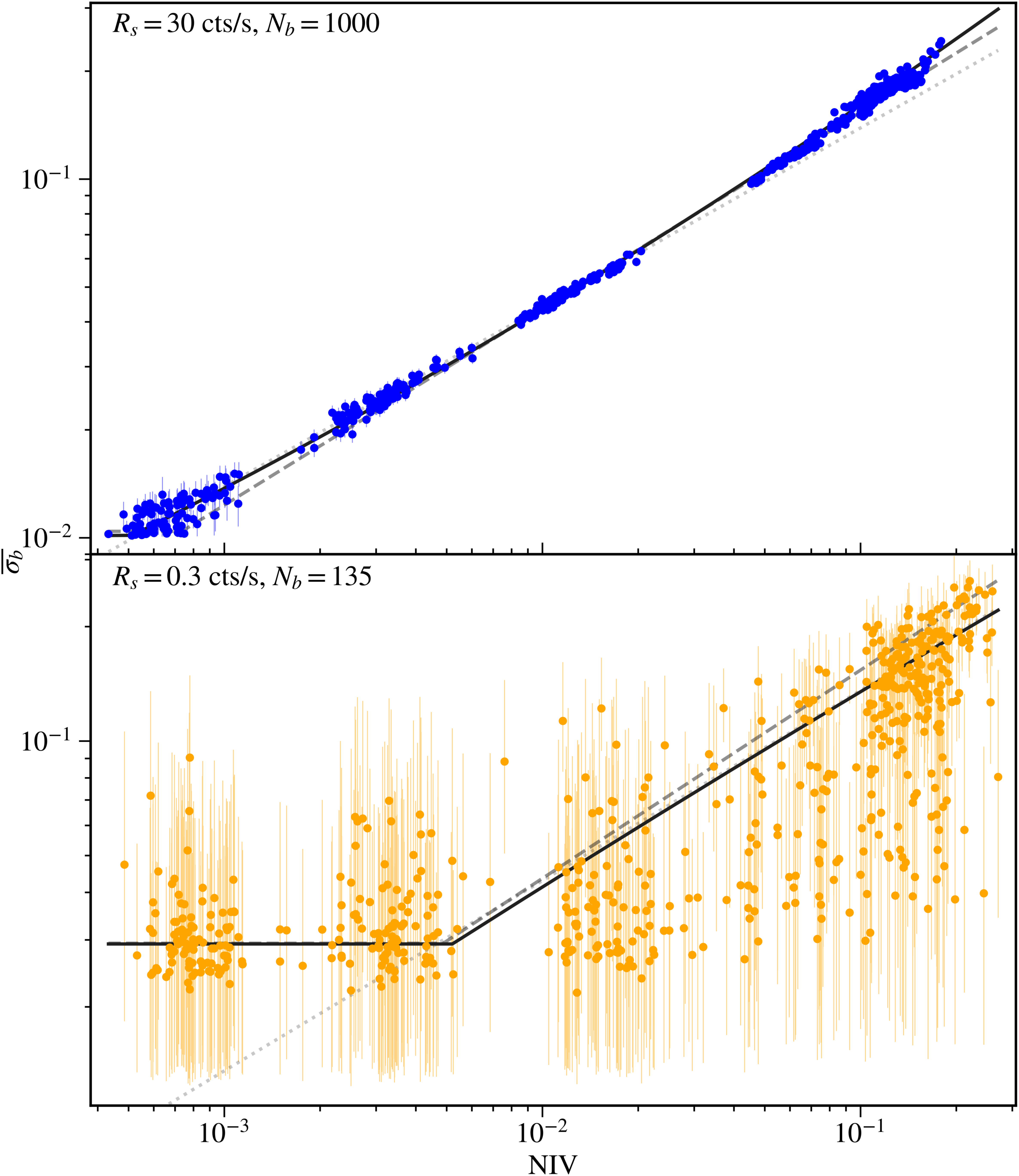}}
\caption{The relationship between $\overline{\sigma_{\mathrm{b}}}$ and the NIV in simulated pink noise \emph{eROSITA}-like light curves. The relationship depends on both the number of bins and the count rate, so this figure showcases two examples. The top panel depicts the relationship for 1000-bin light curves with a mean count rate of $30~\mathrm{cts/s}$. The lower panel shows the relationship for light curves of 135 bins, with a mean count rate of $0.3~\mathrm{cts/s}$. The solid black line depicts the best fit of $\overline{\sigma_{\mathrm{b}}}\left(\mathrm{NEV}, \overline{R_{\mathrm S}}, N_{\mathrm b}\right)$, using Eq. \ref{NEVtoBVfit_quadabc} and the parameters listed in Table \ref{TBVNEVquad(nbcr)}. The dashed line represents using a simpler version of Eq. \ref{NEVtoBVfit_quadabc}, in which the parameters $a$, $b$, and $c$ are constant. The dotted line shows the fit of $\mathrm{NEV}_{\mathrm{b}}\propto \sqrt{\overline{\sigma_{\mathrm{b}}}}$ for simulated values of NIV above the detection limit.
\label{BVtologNEV}}
\end{figure}

Fig. \ref{BVtologNEV} shows the relationship between $\overline{\sigma_{\mathrm{b}}}$ and the NIV in several simulated light curves for two particular average source count rates and number of bins. For bright sources observed in many bins, there is little scatter between $\overline{\sigma_{\mathrm{b}}}$ and the NIV. However, many more variable eRASS sources in the SEP field have count rates and number of bins similar to the values used for the simulations shown in the lower panel ($R_{\mathrm S} = 0.3 ~ \mathrm{cts/s}$, $N_{\mathrm b} = 135$). This indicates that there will still be significant uncertainty in the $\mathrm{NEV}_{\mathrm{b}}$ estimate of the NIV, for most sources. 

At low variabilities, low count rates, and a small number of bins, $\overline{\sigma_{\mathrm{b}}}$ reaches a lower limit plateau, at the minimum value of $\sigma_{\mathrm{I}}$ that bexvar can measure. Whenever this level of $\overline{\sigma_{\mathrm{b}}}$ is measured, it should be treated as an upper limit measurement of the source variability. The minimum variability that can be detected by bexvar depends on both the number of bins, and the average count rate. This is indicated via a horizontal line in Fig. \ref{BVtologNEV}.

At greater degrees of variability, there is an approximately linear relationship between $\log(\overline{\sigma_{\mathrm{b}}})$, and $\log(\mathrm{NIV})$, which has a gradient of approximately 0.5, and does not strongly depend on either the count rate or the number of bins. This is expected from the first order estimate of the relationship between the two parameters. 

However, the top panel of Fig. \ref{BVtologNEV} shows that a linear relationship underestimates $\overline{\sigma_{\mathrm{b}}}$ at both small and large values of the $\mathrm{NIV}$. The gradient between $\log(\overline{\sigma_{\mathrm{b}}})$ and $\log(\mathrm{NIV})$ increases at large values of the NIV for high count rate sources. This effect can also be seen in the underestimation of large values of $\sigma_{\mathrm{I}}$ by $\overline{\sigma_{\mathrm{b}}}$, in Fig. \ref{BVtologvar}. Therefore, as a linear equation is insufficient to describe the relationship between the two parameters, we considered a quadratic equation instead. We found that the fits strongly preferred the parameters of this equation to depend on both the logarithm of the average count rate and number of bins. Therefore, we fitted the relationship between $\overline{\sigma_{\mathrm{b}}}$ and the NIV with:

\begin{align}\label{NEVtoBVfit_quadabc}
\begin{split}
    y_0 & = m_{\mathrm{y}} \log(N_{\mathrm b}) + M_{\mathrm{y}} \log(\overline{R_{\mathrm S}}) + k_{\mathrm{y}} \\
    y_1 & = a \left(\log(\mathrm{NIV})\right)^2 + b \log(\mathrm{NIV}) + c \\
    a & = m_{\mathrm{a}} \log(N_{\mathrm b}) + M_{\mathrm{a}} \log(\overline{R_{\mathrm S}}) + k_{\mathrm{a}} \\
    b & = m_{\mathrm{b}} \log(N_{\mathrm b}) + M_{\mathrm{b}} \log(\overline{R_{\mathrm S}}) + k_{\mathrm{b}} \\
    c & = m_{\mathrm{c}} \log(N_{\mathrm b}) + M_{\mathrm{c}} \log(\overline{R_{\mathrm S}}) + k_{\mathrm{c}} \\
    \log(\overline{\sigma_{\mathrm{b}}}) & = \begin{cases}
        y_1 & \textrm{if } y_1 \geq y_0 \\
        y_0 & \textrm{if } y_1 < y_0.
        \end{cases}
\end{split}
\end{align}

\noindent
where $y_0$ denotes the value of $\log(\overline{\sigma_{\mathrm{b}}})$ at the lower limit plateau, and $y_1$ is the main function relating $\log(\overline{\sigma_{\mathrm{b}}})$ to $\log(\mathrm{NIV})$ above the plateau. We fitted for the best fit values of the parameters $m_{\mathrm{y}}$, $M_{\mathrm{y}}$, $k_{\mathrm{y}}$, $m_{\mathrm{a}}$, $M_{\mathrm{a}}$, $k_{\mathrm{a}}$, $m_{\mathrm{b}}$, $M_{\mathrm{b}}$, $k_{\mathrm{b}}$, $m_{\mathrm{c}}$, $M_{\mathrm{c}}$, and $k_{\mathrm{c}}$. We defined the lower limit plateau as a level in $\log(\overline{\sigma_{\mathrm{b}}})$, rather than as a function of $\log(\mathrm{NIV})$, to reduce the degeneracy of the fit parameters. 

Table \ref{TBVNEVquad(nbcr)} lists the best parameters when using Eq. \ref{NEVtoBVfit_quadabc} to fit the relationship between $\log(\overline{\sigma_{\mathrm{b}}})$ and $\log(\mathrm{NIV})$. To reduce degeneracy and improve the fit, we rescaled the NIV as: $\log(\mathrm{NIV}') = \log(\mathrm{NIV}) - \overline{\log(\mathrm{NIV})}$, where $\overline{\log(\mathrm{NIV})}$ is the average NIV over all simulations. The parameter values presented in the table have been rescaled back, to describe the dependence of $\log(\overline{\sigma_{\mathrm{b}}})$ on $\log(\mathrm{NIV})$. 

\begin{table}[pt]
\centering
\setlength{\tabcolsep}{4pt}
\def\arraystretch{1.1}
\begin{tabular}{c|c}
    \textbf{Parameter} & \textbf{Value} \\ \hline
    \textbf{$m_{\mathrm{y}}$} & $-0.1034\pm0.0047$ \\
    \textbf{$M_{\mathrm{y}}$} & $-0.1850\pm0.0021$ \\
    \textbf{$k_{\mathrm{y}}$} & $-1.410\pm0.0011$ \\
    \textbf{$m_{\mathrm{a}}$} & $-0.0296\pm0.0019$ \\
    \textbf{$M_{\mathrm{a}}$} & $0.0363\pm0.0025$ \\
    \textbf{$k_{\mathrm{a}}$} & $0.0698\pm0.0055$ \\
    \textbf{$m_{\mathrm{b}}$} & $-0.1137\pm0.0066$ \\
    \textbf{$M_{\mathrm{b}}$} & $0.1474\pm0.0088$ \\
    \textbf{$k_{\mathrm{b}}$} & $0.796\pm0.019$ \\
    \textbf{$m_{\mathrm{c}}$} & $-0.1036\pm0.0061$ \\
    \textbf{$M_{\mathrm{c}}$} & $0.1580\pm0.0083$ \\
    \textbf{$k_{\mathrm{c}}$} & $-0.078\pm0.018$ \\
\end{tabular}
\caption{Table of the best fitting parameters of $\log(\overline{\sigma_{\mathrm{b}}})$ as a function of $\log(\mathrm{NIV})$, using Eq. \ref{NEVtoBVfit_quadabc}.
\label{TBVNEVquad(nbcr)}}
\end{table}

The amplitude of the quadratic term, $a$, is close to zero, and even changes sign within the parameter space we investigated. It is positive for light curves with a large source count rate, and few bins, and negative for low count rates and many bins. The linear and the constant terms, whose strength is defined by parameters $b$ and $c$, respectively, depend more strongly on both the count rate and the number of bins. These three parameters all decrease with an increasing number of bins and increase with an increasing count rate. There is a similar dependence of parameters $b$ and $c$ on both the count rate and the number of bins.

Fig. \ref{BV(NEV)quadabc} in Appendix \ref{AppCorPlot} is the corner plot of the best fit of Eq. \ref{NEVtoBVfit_quadabc} to all the simulated data relating $\overline{\sigma_{\mathrm{b}}}$ to the rescaled NIV in the light curve, $\log(\mathrm{NIV}')$. There are some degeneracies, most notably between $m_y$ and $k_y$ and between other $m$ and $k$ parameters. This is probably because the number of bins does not change as much as the count rate within the sample of simulated light curves, so the $m$ parameters can often act similarly to the constant $k$ terms. There is also a slight negative degeneracy between $M_a$ and $M_{\mathrm{b}}$, as well as between some parameters of $a$ and $c$. Nevertheless, the fitting parameters are otherwise well constrained. We tried to keep the degeneracies between parameters as minimal as possible, and only maintained parameters necessary to the fit. There might, however, be the possibility of simplifying Eq. \ref{NEVtoBVfit_quadabc} by setting $m_{\mathrm{b}} = m_{\mathrm{c}}$ and $k_{\mathrm{b}} = k_{\mathrm{c}}$. 

The ability of Eq. \ref{NEVtoBVfit_quadabc} to fit the relationship between $\overline{\sigma_{\mathrm{b}}}$ and the NIV within the parameter space we investigated, is shown via the solid black lines in Fig. \ref{BVtologNEV}. The Eq. \ref{NEVtoBVfit_quadabc} can be rearranged to determine the bexvar estimate of the NIV, $\mathrm{NEV}_{\mathrm{b}}$, from the measurement of $\overline{\sigma_{\mathrm{b}}}$ as follows: 

\begin{align}\label{NEVbeq}
\log(\mathrm{NEV}_{\mathrm{b}}) & = \frac{-b + \sqrt{b^2 - 4a(c-\log(\overline{\sigma_{\mathrm{b}}}))}}{2a} \\
\log(\overline{\sigma_{\mathrm{b,l}}}) & = m_{\mathrm{y}} \log(N_{\mathrm b}) + M_{\mathrm{y}} \log(\overline{R_{\mathrm S}}) + k_{\mathrm{y}},
\end{align}

\noindent
where $\overline{\sigma_{\mathrm{b,l}}}$ is the approximate value of the lower limit of $\overline{\sigma_{\mathrm{b}}}$ that is measurable at that particular count rate and number of bins. The equation for $\mathrm{NEV}_{\mathrm{b}}$ should only be considered as an upper limit estimate of the NIV if $\overline{\sigma_{\mathrm{b}}} \approx \sigma_{\mathrm{b,l}}$. 

The above function can fail in two particular instances. Firstly, it is undefined when $a = 0$. In that case, Eq. \ref{NEVtoBVfit_quadabc} is solved as a simple linear equation. Secondly, it fails when $\log(\overline{\sigma_{\mathrm{b}}}) < c - (b^2 / 4a)$ if $a > 0$, or $\log(\overline{\sigma_{\mathrm{b}}}) > c - (b^2 / 4a)$ if $a < 0$. However, within the parameter space we investigated, that would require incredibly small, or incredibly large degrees of variability, neither of which is likely to be found for a pink noise light curve with a significant degree of variability, detected using the methods of Section \ref{Diffvarnonvarsection}.

Finally, we tested how well this methodology would allow us to estimate the NIV with $\mathrm{NEV}_{\mathrm{b}}$. Just as for Fig. \ref{BVtologvar}, we simulated \emph{eROSITA}-like light curves of pink noise variable sources, consisting of $\{20, 75, 150, 400, 1050\}$ bins, and average count rates of $\{0.015, 0.15, 1.5, 15\}~\mathrm{cts/s}$. We specifically selected different values for the number of bins and the average count rate than in the simulations used for determining the relationship between $\overline{\sigma_{\mathrm{b}}}$, and the NIV. In this way, we could independently verify the usefulness of this method, for other parameters not previously investigated. However, we used light curves of 20 bins both for defining the method and testing it, as that is the lower limit we chose for this type of variability analysis.  

\begin{figure}[h]
\resizebox{\hsize}{!}{\includegraphics{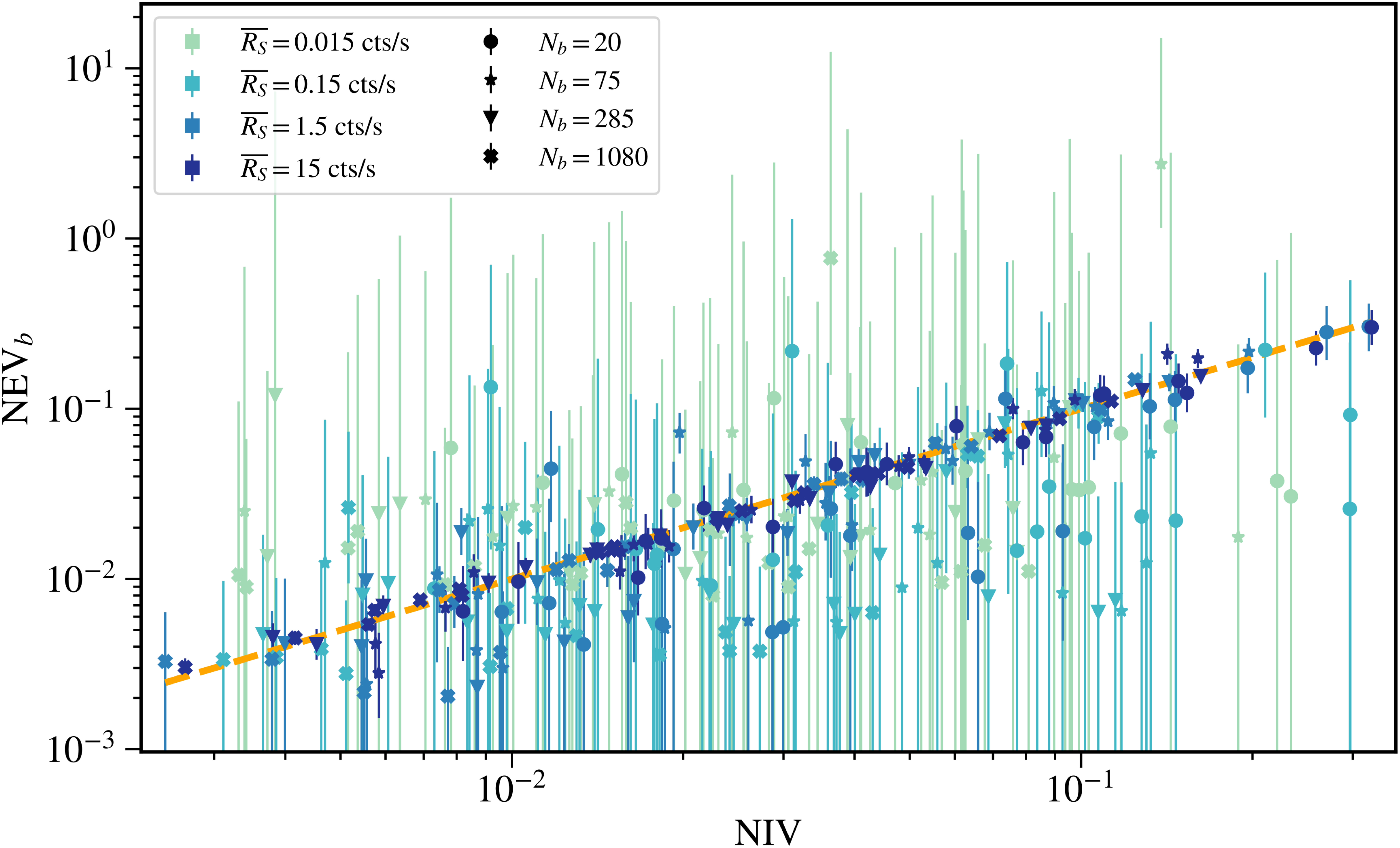}}
\caption{Comparison of the $\mathrm{NEV}_{\mathrm{b}}$ measurements with the NIV of the light curve that it seeks to estimate, for pink noise light curves, using Eq. \ref{NEVtoBVfit_quadabc}, and the values listed in Table \ref{TBVNEVquad(nbcr)}. The orange dashed line indicates the $1:1$ relationship between $\mathrm{NEV}_{\mathrm{b}}$, and the NIV. 
 \label{bvNEVfig}}
\end{figure}

Fig. \ref{bvNEVfig} shows that $\mathrm{NEV}_{\mathrm{b}}$ can accurately estimate the NIV for this parameter space. We note that there is no apparent discrepancy between the two variability measures at high degrees of variability, as there was for $\overline{\sigma_{\mathrm{b}}}$ and $\sigma_\mathrm{I}$. The uncertainties of $\mathrm{NEV}_{\mathrm{b}}$ depicted in this figure are determined by converting the upper and lower bounds of the $1\sigma$ confidence interval of $\overline{\sigma_{\mathrm{b}}}$ into $\mathrm{NEV}_{\mathrm{b}}$ values, using Eq. \ref{NEVtoBVfit_quadabc}, and the parameters of Table \ref{TBVNEVquad(nbcr)}. Whenever the lower bound error on $\overline{\sigma_{\mathrm{b}}}$ extends below $10^{y_0}$, the lower bound error on $\mathrm{NEV}_{\mathrm{b}}$ is extended to a value of 0. This is due to the inability to determine lower degrees of variability for those particular light curves. At very low count rates, when there is insufficient information available to properly constrain the NIV of a light curve, $\mathrm{NEV}_{\mathrm{b}}$ is still able to provide an accurate confidence interval within which the NIV of the light curve is likely to be.

In Appendix \ref{CompNIVest}, we compared the $\mathrm{NEV}_{\mathrm{b}}$ against other estimators of the NIV. We also investigated to what extent the $\mathrm{NEV}_{\mathrm{b}}$ estimate of the NIV remains accurate for PSDs that differ from the assumed pink noise shape. We found that $\mathrm{NEV}_{\mathrm{b}}$ provides the most accurate estimate of the NIV and the most accurate boundaries on the estimate, for \emph{eROSITA}-like light curves with count rates below $1.5~\mathrm{cts/s}$, for pink, white, and red noise variable sources. At a count rate of $0.015~\mathrm{cts/s}$, the average difference ratio between the estimated value and the NIV, is two orders of magnitude lower for $\mathrm{NEV}_{\mathrm{b}}$, than it is for the NEV. Above $1.5~\mathrm{cts/s}$, the $\mathrm{NEV}_{\mathrm{b}}$ is comparable to other methods. The $\mathrm{NEV}_{\mathrm{b}}$ was defined for \emph{eROSITA}-like light curves, but can be used to estimate the NIV of any source observed with $N_{\mathrm b} \leq 1000$, and a power law PSD with indices between $0<\alpha<2$. Due to its higher accuracy as compared to other methods, we decided to subsequently only use $\mathrm{NEV}_{\mathrm{b}}$ for estimating the NIV. 

\section{Power leakage and the intrinsic scatter in the NIV}\label{Syserr}

We have developed and evaluated methods for accurately estimating the NIV of variable sources. However, the NIV is non-stationary, and varies stochastically, as outlined in Sections \ref{subsecNEVdes} and \ref{sec:blp}. In this section, we discuss these effects, and introduce a new way to estimate the size of the intrinsic scatter of the NIV of a single, or multiple segments of a light curve. The results presented in this section are not specific to \emph{eROSITA}, but apply to any variability analysis of approximately pink-noise light curves. 

\subsection{Aliasing and the red noise leak} \label{SecAlias}

All the variability power contained at frequencies above the inverse of the bin duration is integrated out within each bin. If the bins of a light curve are adjacent, having no gaps in between, the average count rate in each bin is not affected by power above the sampling frequency, so there is no aliasing effect. However, for light curves consisting of gaps of a constant duration between bins, power at frequencies between the inverse of the separation of bins, $\tau^{-1}$, and the inverse of the bin duration, $\Delta t^{-1}$, increases the flux difference from one bin to the next. This affects both the NIV and the periodogram. It is known as the aliasing effect \citep[see][]{1989ASIC..262...27V, 2005PhRvE..71f6110K}, and it is particularly strong for \emph{eROSITA} light curves, as the gaps between observations are at least 360 times as long as the observations themselves. 

If individual bins are substantially shorter than the gaps between them, and if there is negligible power at frequencies larger than the inverse of the duration of each bin, the timing of the observations can be mathematically described as delta functions. In that case, which applies to \emph{eROSITA} light curves, we can simplify the mathematical description of the impact that aliasing has on a PSD, to:

\begin{equation}\label{aliasfn}
P_{\mathrm a}(\nu) = P(\nu) + \sum_{k=1}^\infty \left( P(\nu - k \nu_{\mathrm s}) + P(\nu + k \nu_{\mathrm s}) \right),
\end{equation}
\noindent
where $P_{\mathrm a}(\nu)$ is the observed periodogram affected by aliasing, $P(\nu)$ is the true PSD that we want to determine accurately, and $\nu_{\mathrm s}$ is the sampling frequency. For a derivation of this equation, see \citet{2005PhRvE..71f6110K}.

\begin{figure}[h]
\resizebox{\hsize}{!}{\includegraphics{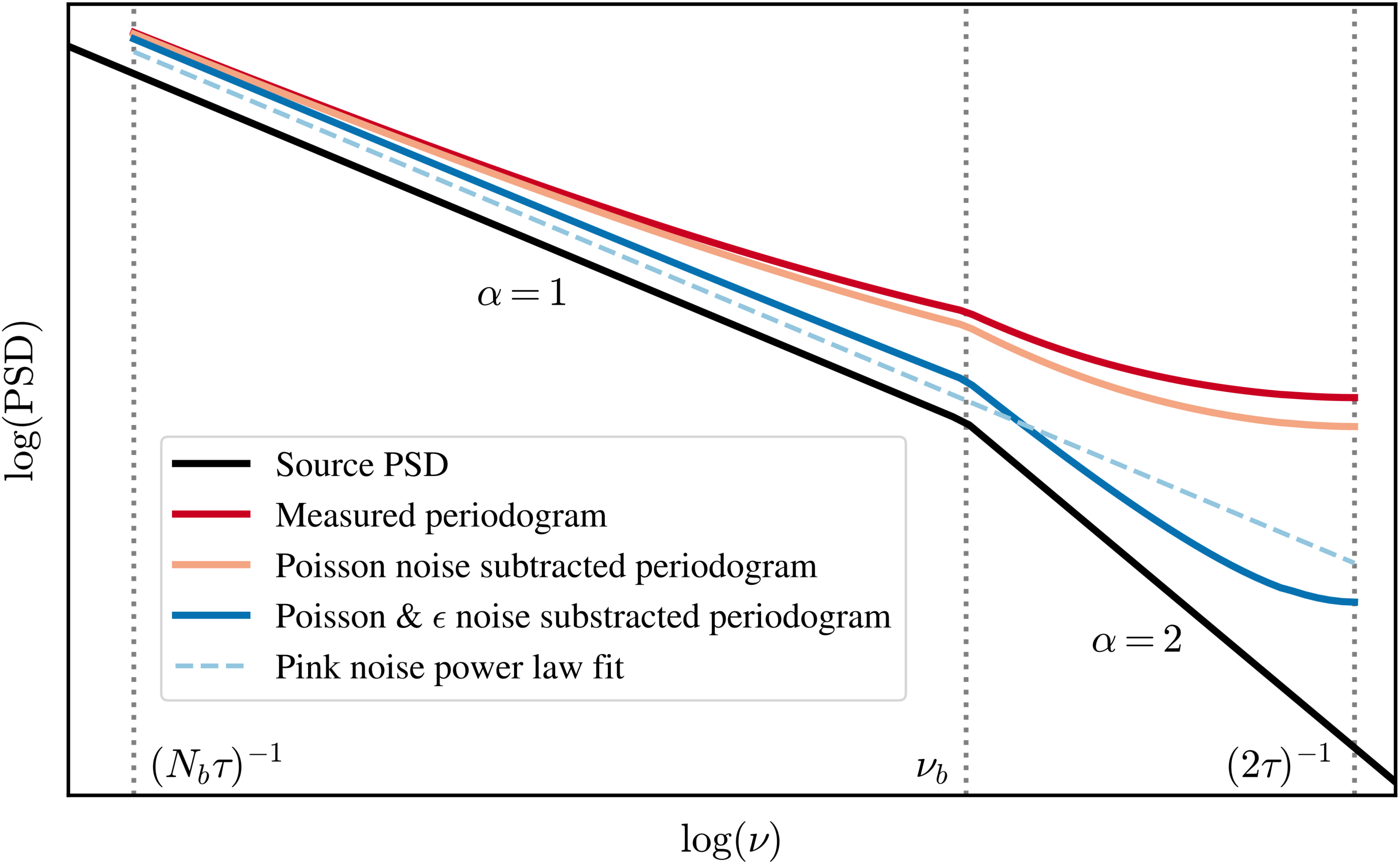}}
\caption{A comparison of a typical AGN PSD, and how a measured periodogram differs from it, within the frequency space that can be explored by \emph{eROSITA} for sources lying close to the ecliptic poles. The intrinsic PSD is depicted as following a broken power law of $P\propto \nu^{-1}$ at $\nu < \nu_{\mathrm{b}}$, and $P\propto \nu^{-2}$ at $\nu \geq \nu_{\mathrm{b}}$, and continues beyond the frequency range probed by the observations. The measured periodogram is offset from the intrinsic PSD by the Poisson noise, the $\epsilon$ noise, aliasing, and the red noise leak. The $\epsilon$ noise (Eq. \ref{excPoisNois}) is often significantly larger than the Poisson noise (Eq. \ref{PoisNoisL_meanFE}). The flattening of the Poisson and $\epsilon$ subtracted periodogram at the highest frequencies is due to aliasing. The red noise leak increases the amplitude of the power laws. The relative strength of the Poisson, fractional exposure noise, aliasing, and the red noise leak depicted in this figure are illustrative, and depend on the shape of the PSD, the average count rate, and the mean, and variance of the fractional exposure. Periodogram noise is not depicted. 
 \label{TypPSD}}
\end{figure}

For AGN PSDs described by power laws, in which variability power decreases with increasing frequency, aliasing predominantly affects the periodogram shape close to the Nyquist frequency. It causes the power law slope to flatten at the highest frequencies observed. Fig. \ref{TypPSD} depicts various effects that afflict a measured periodogram and offset it from the source PSD. 

Aliased periodograms can be fitted using Eq. \ref{aliasfn} to determine an estimate of the intrinsic shape of the PSD of the variable source, if there were no aliasing. However, this requires some assumptions about the shape of the PSD above the Nyquist frequency, for which no data are available. Fortunately, we can rely on previous analyses of AGN periodograms at higher frequencies to inform this assumption \citep{2002A&A...382L...1P, 2012A&A...544A..80G}. 

Another effect causes variability power at lower frequencies to leak into the observed frequency space, enhancing the measured degree of variability. This is known as the red noise leak. However, whereas the aliasing of $\alpha > 0$ power laws predominantly affects the variability measured at the highest frequencies, the red noise leak increases the power at all frequencies. For power law PSDs, the red noise leak mainly increases their normalisation, but does not affect the power law slopes \citep{2016ApJ...825...56Z}. 

Even without removing the impact that the red noise leak and aliasing have on the periodograms of light curves of a set of variable sources, it is still possible to compare their estimates of $\mathrm{NIV}_{\infty}$, if it can be assumed that the combination of the two effects has a similar impact on the sources being compared. However, to compare $\mathrm{NIV}_{\infty}$ estimates, it is first necessary to determine the intrinsic scatter in the NIV.

\subsection{The intrinsic scatter of the NIV}\label{SecSysErr}

The NIV is the intrinsic source variability at the time of the observations. It is unaffected by Poisson statistics and measurement errors. Nevertheless, it varies over time in a stochastic way, even if the process causing the variability is stationary. This has previously been discussed by \citet{2003MNRAS.345.1271V} and \citet{2013ApJ...771....9A}.  

\begin{figure}[h]
\resizebox{\hsize}{!}{\includegraphics{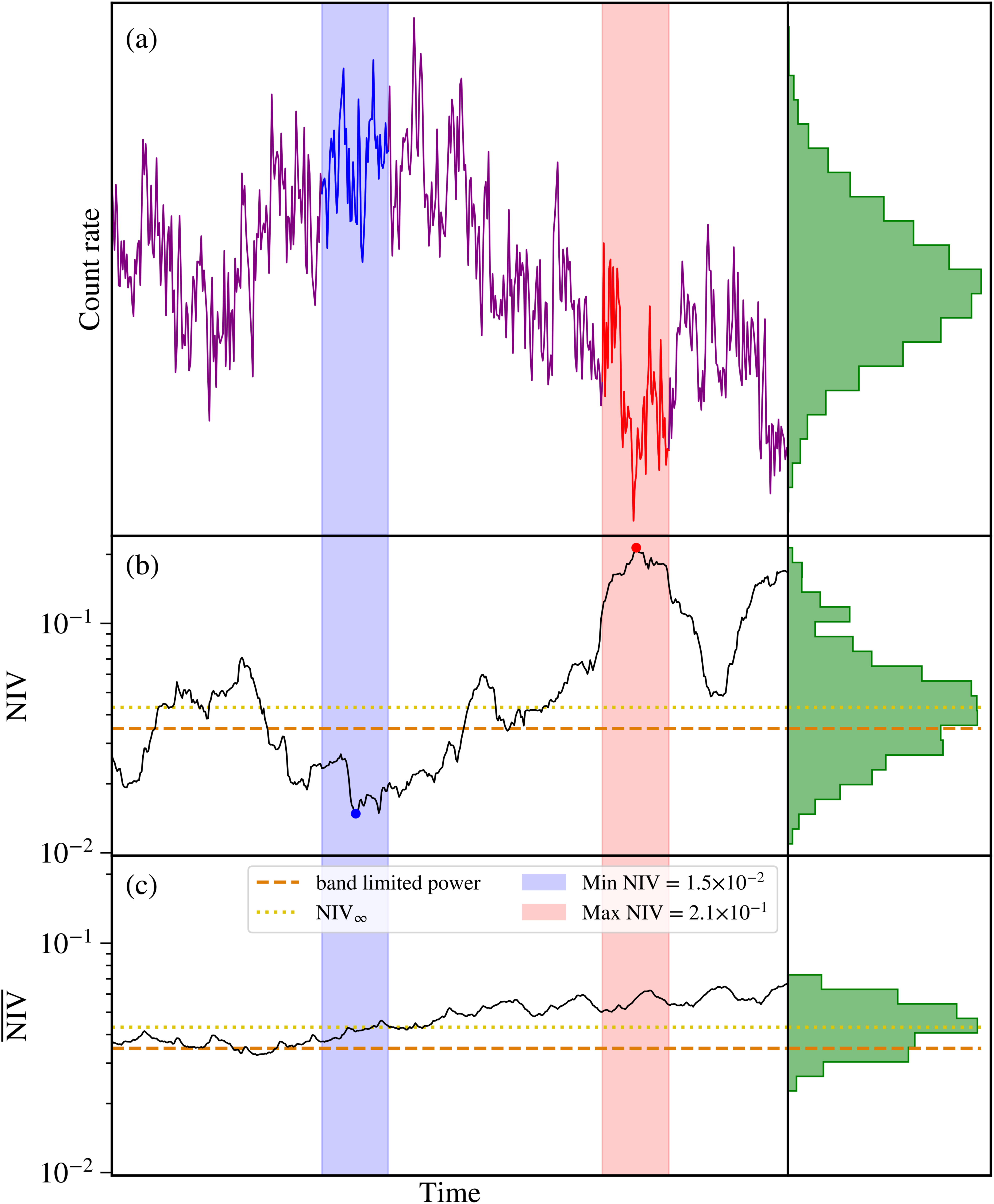}}
\caption{The NIV of a variable source can change over time, even if the mechanism generating the variability, the PSD, and the band-limited power remain unchanged. Panel (a) shows a 500-bin segment of a simulated pink noise light curve without measurement errors or Poisson noise. Panel (b) shows the NIV of a sliding window interval of 50 bins from the light curve in panel (a). Each NIV data point is placed at the time of centre of the selected interval. The maximum and minimum NIV of this interval is indicated, and the parts of the light curve for which they were obtained are highlighted. Panel (c) shows the average NIV obtained by computing the geometric mean NIV of 10 adjoined segments, each consisting of 50 bins. Each value is placed at the center of the segments used. Panels (b) and (c) use data from the light curve outside of the interval shown in panel (a). The panels on the right show the distribution of these parameters, throughout the entire $5\times10^3$ bin simulated light curve. 
 \label{NEVrange}}
\end{figure}

Fig. \ref{NEVrange} demonstrates how strongly the NIV can vary between light curves observed at different times, even if they are generated by the same PSD with the same band-limited power. In this particular example, a $5\times10^3$ bin pink noise light curve was simulated, of which a 500 bin segment is shown. The NIV was computed for a sliding window interval of 50 bins each. The NIV was spread out over more than one order of magnitude between its maximal ($0.21$) and minimal ($0.015$) value in just this 500 bin segment. For steeper power laws, the scatter of the NIV is even more pronounced \citep{2003MNRAS.345.1271V}. 

In this simulation, the band-limited power, for the frequency range corresponding to a 50 bin light curve, has a value of $0.035$. The $\mathrm{NIV}_{\infty}$ is offset from the band-limited power by the red noise leak, to a value of $0.043$. Aliasing was not included in this simulation. If it was, it would have caused an even larger difference between $\mathrm{NIV}_{\infty}$ and the band-limited power. Fig. \ref{NEVrange} also shows that $\log(\mathrm{NIV})$ follows a normal distribution with a mean of $\mathrm{NIV}_{\infty}$. 

\citet{2003MNRAS.345.1271V} provided a table of the scatter to expect in the log variance of the count rate in light curves for a few different PSD slopes, for a few different number of bins. \citet{2013ApJ...771....9A} further investigated the mean, variance, and skewness of the distribution of the NEV for a sample of 5000 simulated light curves, with continuous, or sparse sampling, and different PSD slopes.

In this section, we instead focused on determining a more general description of the scatter in the NIV, rather than the variance, of a stationary pink noise process. We aimed at being able to provide a means to estimate the intrinsic scatter in the NIV for any number of bins in the light curve. We also investigated the dependence of the scatter on other parameters besides the number of bins. The following results are based on 20 times more simulated light curves for each selected number of bins, than those used by \citet{2003MNRAS.345.1271V}. We also investigated more than three times as many different number of bins. 

The intrinsic scatter found in this way can be directly used alongside any NIV estimate, analogously to a sampling error, to estimate how much the NIV can be expected to vary between two different sets of observations. In so doing, the $\mathrm{NEV}_{\mathrm{b}}$ (or similar) measurement can also be used as an estimator of $\mathrm{NIV}_{\infty}$. None of this section is specific to \emph{eROSITA}, and the results can be used for any approximately pink noise light curve.

We label the intrinsic scatter of the NIV in log-space as $\Delta_{\mathrm s}$. It is equivalent to the standard deviation of the distribution of $\log(\mathrm{NIV})$, which can be seen in Fig. \ref{NEVrange}. 

We simulated $3.2\times10^5$ pink noise light curves like the one shown in Fig. \ref{NEVrange} to determine the standard deviation of the distribution of NIVs for light curves consisting of $\{5, 10, 20, 30, 40, 50, 60, 75, 90, 100, 120, 140, 170, 200, 500, 1000\}$ bins. We mainly focused on the range of bins most useful for an eRASS variability analysis. We extended the range somewhat, to better determine the dependence of the intrinsic scatter of the NIV on the number of bins of the light curve. We simulated $10^5$ bins for each instance of a pink noise light curve, to ensure that the NIV in each randomly selected segment was part of the same distribution about a fixed $\mathrm{NIV}_{\infty}$. Unlike previous simulations, we did not include Poisson noise, a background count rate, or a fractional exposure, as this analysis seeks to only determine the intrinsic scatter of the NIV. 

We randomly selected starting positions within the simulated light curves and calculated the NIV for each selected interval. We investigated different degrees of intrinsic variability by scaling the range of fluxes in the simulated light curves, while keeping the mean flux constant. A lower range at a constant mean, results in a lower $\mathrm{NIV}_{\infty}$. For each set of 2000 such simulated light curve intervals, which all have the same number of bins, and were generated from the same PSD, with the same $\mathrm{NIV}_{\infty}$, we determined the standard deviation of the $\log(\mathrm{NIV})$ distribution ($\Delta_{\mathrm s}$). We subsequently investigated the dependence of $\Delta_{\mathrm s}$ on various parameters.

We found that the intrinsic scatter depends on both the number of bins, and $\mathrm{NIV}_{\infty}$. Previously, \citet{2003MNRAS.345.1271V} and \citet{2013ApJ...771....9A} investigated the dependence of it on the slope of the PSD, and the number of bins in the light curve, but not the dependence on $\mathrm{NIV}_{\infty}$. The parameters $\mathrm{NIV}_{\infty}$ and $N_{\mathrm b}$ are correlated with one another, for a stationary process, if $\alpha > 0$. Nevertheless, we still treat them as mostly independent, as the $\mathrm{NIV}_{\infty}$ can also vary across a wide range at a fixed number of bins of the light curve. 

\begin{figure}[h]
\resizebox{\hsize}{!}{\includegraphics{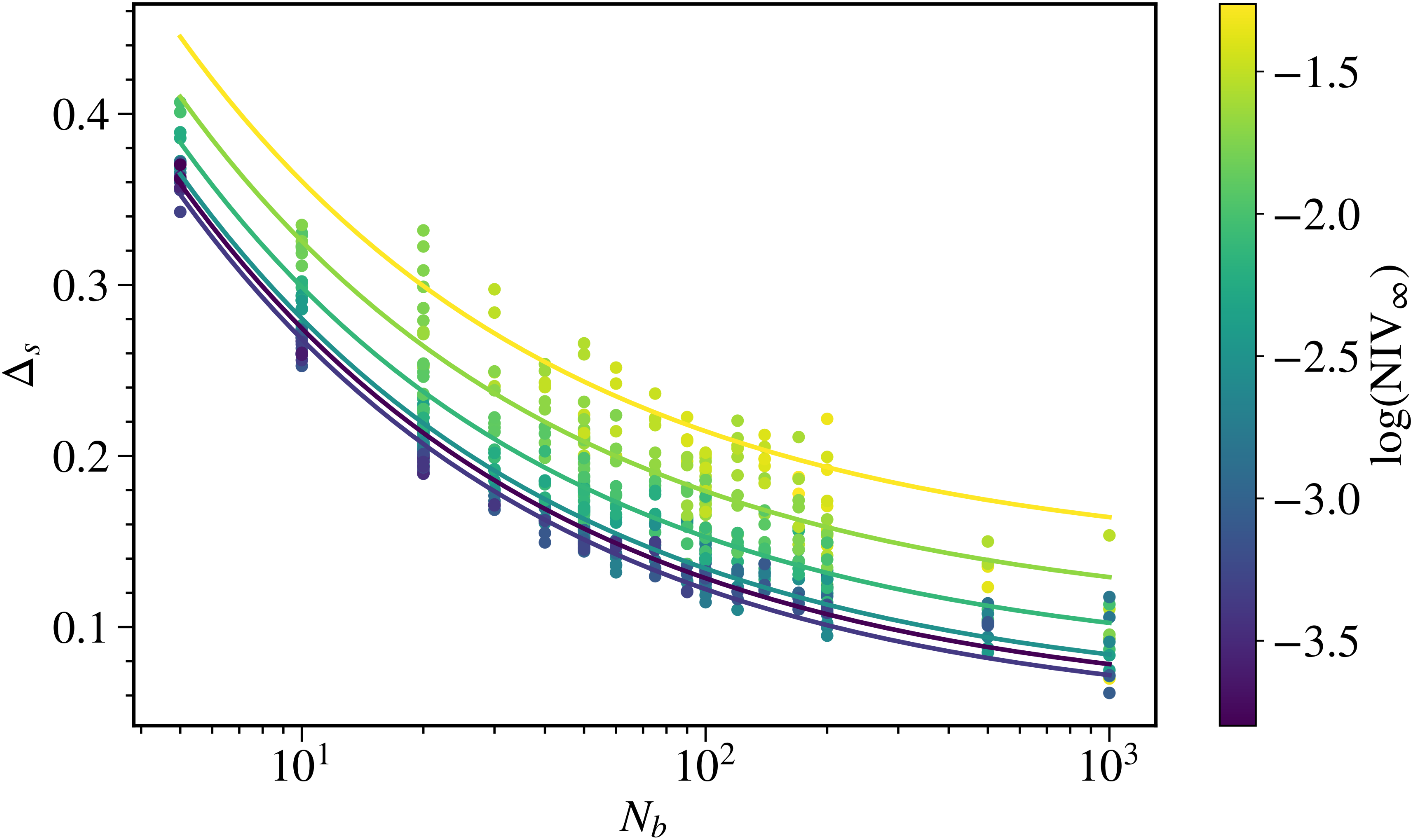}}
\caption{The dependence of $\Delta_{\mathrm s}$ on the number of bins, and the $\mathrm{NIV}_{\infty}$. Each data point represents the intrinsic scatter in the NIV, determined from 2000 light curves, randomly selected from a $10^5$ bin simulated pink noise light curve produced from a constant input PSD. The curves show the best fit relation of $\Delta_{\mathrm s}(N_{\mathrm b})$, at fixed levels of $\mathrm{NIV}_{\infty}$, using Eq. \ref{Syserreq} with parameter values as contained in Table \ref{TBFsyserr}. There is a significant scatter among the data points due to the probabilistic nature of this estimate. 
 \label{3dSyserr}}
\end{figure}

Fig. \ref{3dSyserr} depicts the dependence of $\Delta_{\mathrm s}$ on the number of bins and the $\mathrm{NIV}_{\infty}$. For a fixed $N_{\mathrm b}$, the intrinsic scatter depends on the $\mathrm{NIV}_{\infty}$ with an approximately quadratic relationship:  $\Delta_{\mathrm s}(\mathrm{NIV}_{\infty}) = a\left(\log(\mathrm{NIV}_{\infty})\right)^2 + b\log(\mathrm{NIV}_{\infty}) + \gamma$. We further analysed the dependence of $a$, $b$, and $\gamma$ on the number of bins of the light curve. Only $\gamma$ had a dependence on it, which could be approximately described by $\gamma(N_{\mathrm b}) = (c/N_{\mathrm b})^d + f$. Therefore, we fit the results of our simulations using:

\begin{equation}\label{Syserreq}
    \Delta_{\mathrm s} = a\left(\log(\mathrm{NIV}_{\infty})\right)^2 + b \log(\mathrm{NIV}_{\infty}) + c N_{\mathrm b}^{-d} + f.
\end{equation}

\noindent
As Fig. \ref{3dSyserr} shows, the range of $\Delta_{\mathrm s}$ values at a fixed number of bins does not change significantly with an increasing number of bins in the light curve. That is why we modeled the dependence of $\Delta_{\mathrm s}$ on $\mathrm{NIV}_{\infty}$ and $N_{\mathrm b}$ as additive, rather than multiplicative. 

The largest value of  $\mathrm{NIV}_{\infty}$ is limited by the number of bins of the light curve. Very short light curves cannot have the same maximum value of $\mathrm{NIV}_{\infty}$ as significantly longer light curves do. This prevents the low $N_{\mathrm b}$, high $\log(\mathrm{NIV}_{\infty})$ corner of this figure from being populated with data points. The intrinsic scatter depends more strongly on the number of bins of the light curve, but the dependence on $\mathrm{NIV}_{\infty}$ cannot be ignored, especially not for large variabilities of $\mathrm{NIV}_{\infty} > 10^{-2}$. 

To reduce the degeneracy between the two parameters $a$ and $b$ in the fit, the NIV was rescaled, so that Eq. \ref{Syserreq} was instead fitted as a function of $\log(\mathrm{NIV}_{\infty}')=\log(\mathrm{NIV}_{\infty})-\overline{\log(\mathrm{NIV}_{\infty})}$, where $\overline{\log(\mathrm{NIV}_{\infty})}$ denotes the average $\log(\mathrm{NIV}_{\infty})$ over all simulations. When fitting Eq. \ref{Syserreq} to all the values of $\Delta_{\mathrm s}(\mathrm{NIV}_{\infty}', N_{\mathrm b})$ that had been obtained from the simulations, we obtained the corner plot for the best fit parameters $a'$, $b'$, $c$, $d$, and $f'$, that is shown in Fig. \ref{cplotsyserr} in Appendix \ref{ACpSysErr}. Table \ref{TBFsyserr} lists the non-rescaled best fit parameters of $\Delta_{\mathrm s}(\mathrm{NIV}_{\infty}, N_{\mathrm b})$. 

\begin{table}[h]
\centering
\setlength{\tabcolsep}{4pt}
\def\arraystretch{1.1}
\begin{tabular}{c|c}
    \textbf{Parameter} & \textbf{Value} \\ \hline
    \textbf{$a$} & $0.0232\pm0.0013$ \\
    \textbf{$b$} & $0.1513\pm0.0061$ \\
    \textbf{$c$} & $0.648\pm0.010$ \\
    \textbf{$d$} & $0.464\pm0.012$ \\
    \textbf{$f$} & $0.2920\pm0.0079$ \\
\end{tabular}
\caption{Table of the best fitting parameters of $\Delta_{\mathrm s}$, as a function of $N_{\mathrm b}$, and $\log(\mathrm{NIV}_{\infty})$, described by Eq. \ref{Syserreq}.
\label{TBFsyserr}}
\end{table}

We had expected $\Delta_{\mathrm s}$ to approximately depend on $1/\sqrt{N_{\mathrm b}}$, and indeed we found $d$ to be close to 0.5. There is some degeneracy between $c$ and $d$, and between $d$ and $f$. Nevertheless, all five parameters were essential to fit the dependence of $\Delta_{\mathrm s}$ on $N_{\mathrm b}$ and $\log(\mathrm{NIV}_{\infty})$. 

The best estimate of $\mathrm{NIV}_{\infty}$ is $\mathrm{NEV}_{\mathrm{b}}$, but with according upper, and lower limits. The $1\sigma$ upper limit on the $\mathrm{NIV}_{\infty}$ estimate based on a single measurement of the $\mathrm{NEV}_{\mathrm{b}}$, is finally found using:

\begin{align}\label{SampErrUL}
\begin{split}
    \log(\mathrm{NIV}_{\mathrm{\infty, u}}) & = \log(\mathrm{NEV}_{\mathrm{b}}) + \Delta_{\mathrm s} \\
    \log(\mathrm{NIV}_{\mathrm{\infty, u}}) & = \frac{1-b-\sqrt{(b-1)^2-4a(g + \log(\mathrm{NEV}_{\mathrm{b}}))}}{2a} \\
    g & = c N_{\mathrm b}^{-d} + f. \\
\end{split}
\end{align}

\noindent
Similarly, the lower limit is calculated by: 

\begin{align}\label{SampErrLL}
\begin{split}
    \log(\mathrm{NIV}_{\mathrm{\infty, l}}) & = \log(\mathrm{NEV}_{\mathrm{b}}) - \Delta_{\mathrm s} \\
    \log(\mathrm{NIV}_{\mathrm{\infty, l}}) & = \frac{-1-b+\sqrt{(b+1)^2-4a(g - \log(\mathrm{NEV}_{\mathrm{b}}))}}{2a}.
\end{split}
\end{align}

\noindent 

The content of the square root in these two equations is not likely to ever be negative. We only recommend using these equations for pink noise light curves with fewer than 1000 bins, and  $-4 < \log(\mathrm{NIV}_{\infty, l}) < -1$. The full error of the estimate of $\mathrm{NIV}_{\infty}$ is found by combining the measurement error of $\mathrm{NEV}_{\mathrm b}$ with the intrinsic scatter found using Eqs. \ref{SampErrUL}, and \ref{SampErrLL}.

Our results are comparable to those of \citet{2003MNRAS.345.1271V}, for the five different number of bins in the light curve discussed by them. When converting their $90\%$ interval on the scatter in the variance for pink noise light curves to a $1\sigma$ range in the NIV, we found values within the range of $\Delta_{\mathrm s}$ values we found, for differing $\mathrm{NIV}_{\infty}$. We found a similar dependence on the number of bins as can be deduced from the results provided by \citet{2003MNRAS.345.1271V}. In contrast, Eqs. \ref{SampErrUL} and \ref{SampErrLL}, provide the intrinsic scatter in the NIV for any number of bins between 5 and 1000, and are based on 67 times more simulations. We also investigated, and quantified the dependence of $\Delta_{\mathrm s}$ on $\mathrm{NIV}_{\infty}$, and found it to be non-negligible. Our results are also comparable to those of \citet{2013ApJ...771....9A}, for $\alpha=1$.

\subsection{Reducing the intrinsic scatter in the NIV by averaging over multiple segments}\label{SecRedSysErr}

The intrinsic scatter is best reduced by observing the source again at a different time. Without additional observations, the scatter can instead be reduced by splitting the total observed light curve into several segments. Instead of computing one NIV estimate for the entire frequency range, a more precise estimate over a smaller frequency range can be found by computing the geometric mean of the NIV estimates found in $N_{\mathrm{seg}}$ segments of $N_{\mathrm b}$ bins each. In choosing how to split a light curve into individual segments, a balance needs to be found between reducing the intrinsic scatter, by increasing the number of segments, and maximising the frequency space investigated, by having as many bins in each segment as possible.

\citet{2003MNRAS.345.1271V}, and \citet{2013ApJ...771....9A} also describe the use of such an ensemble average to reduce the intrinsic scatter of the NIV. In contrast to our approach, they use an arithmetic, rather than a geometric ensemble average of the NEV. If the distribution of the NIV is log-normal, a geometric mean provides a more accurate estimate of the $\mathrm{NIV}_{\infty}$. Furthermore, a linear ensemble average NEV will tend to overestimate the $\mathrm{NIV}_{\infty}$, if it does not follow a linear distribution. This difference in methodology makes our results not directly comparable to those of \citet{2003MNRAS.345.1271V} and \citet{2013ApJ...771....9A} in regards to ensemble averages.

Calculating a geometric mean NEV is not possible if the NEV of any segment is negative. As the $\mathrm{NEV}_{\mathrm{b}}$ is never negative, its geometric mean, $\overline{\mathrm{NEV}_{\mathrm{b}}}$, can always be computed, regardless of the mean count rate, or degree of variability in different segments.

The intrinsic scatter is affected by the placement of the segments in relation to each other. If all the segments used to compute $\overline{\mathrm{NEV}_{\mathrm{b}}}$ are spaced randomly, such that there are long, and inconsistent gaps between each of them, then the NIV in each segment is independent of the NIV of any of the other segments. In that case, the logarithmic intrinsic variance of $\overline{\mathrm{NIV}}$, which we refer to as $\Delta_{\mathrm{s, n}}$, is:

\begin{equation}\label{Syserr_rand}
    \Delta_{\mathrm{s, n}} = \frac{\Delta_{\mathrm s}}{\sqrt{N_{\mathrm{seg}}}}.
\end{equation}

\noindent
In this equation, $\Delta_{\mathrm s}$ is described by Eq. \ref{Syserreq}, with parameter values as listed in Table \ref{TBFsyserr}. The parameter $N_{\mathrm b}$ in Eq. \ref{Syserreq} now refers to the number of bins in each of the segments of the light curve. The best estimate of $\mathrm{NIV}_{\infty}$ is now $\overline{\mathrm{NEV}_{\mathrm b}}$. The upper and lower limits of that estimate are still found by Eqs. \ref{SampErrUL}, and \ref{SampErrLL}, but using $a_{\mathrm n}=a/\sqrt{N_{\mathrm{seg}}}$ instead of $a$, and $b_{\mathrm n}$, $c_{\mathrm n}$, and $f_{\mathrm n}$, that are all adjusted from their values in Table \ref{TBFsyserr} in the same way. This is also found by \citet{2003MNRAS.345.1271V} and \citet{2013ApJ...771....9A}.

\begin{figure*}[h]
\resizebox{\hsize}{!}{\includegraphics{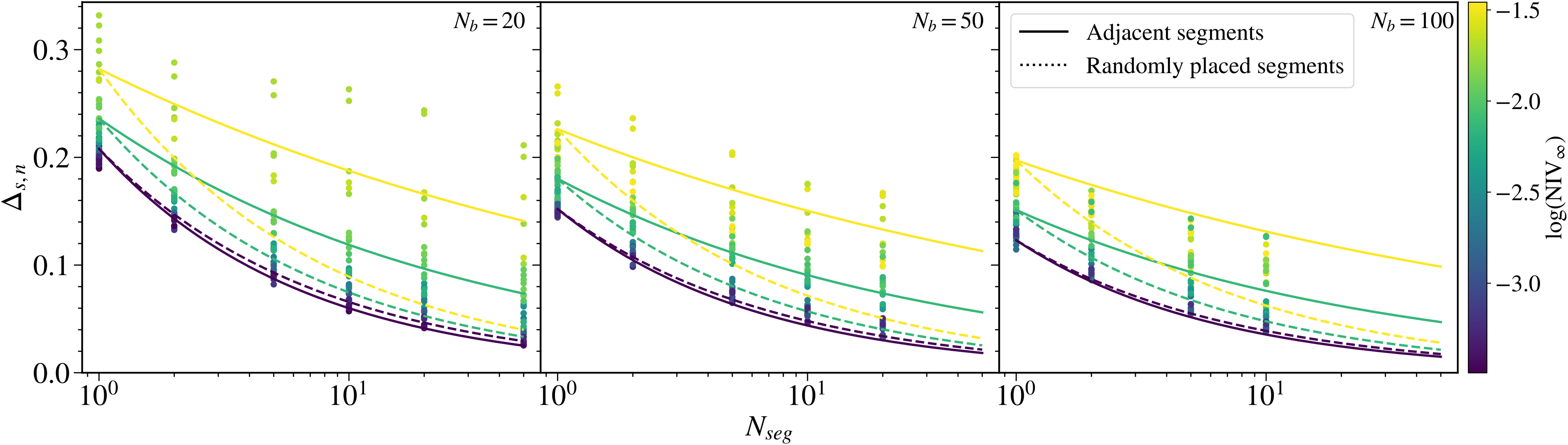}}
\caption{Reducing the intrinsic scatter, by computing the geometric mean NIV for $N_{\mathrm{seg}}$ distant, or adjoined segments of $N_{\mathrm b}$ bins bins each. The data points indicate the results of individual simulations for adjoined segments. The lines depict the best fits to the simulations, using Eqs. \ref{Nsegadjeq} and \ref{Syserreq}, and the  parameter values in Tables \ref{TBFsyserr}, and \ref{TBFsyserradj}. The solid lines show the dependence of $\Delta_{\mathrm{s, n}}$ on $N_{\mathrm{seg}}$ for adjoined segments, and the dotted lines for distant, randomly placed segments. 
 \label{3dSyserr_nseg}}
\end{figure*}

Most cases are, however, more complex. For example, the NIV in one segment is not independent of the NIV of directly neighbouring segments. In such cases, the intrinsic scatter of the NIV cannot be reduced linearly following Eq. \ref{Syserr_rand}, but decreases more gradually with an increasing number of segments. The geometric mean NIV of ten adjoined segments consisting of 50 bins each is illustrated in Fig. \ref{NEVrange}(c). The range of values of the average NIV, computed over ten adjoined segments of 50 bins, is smaller than the range of the NIV in 50 bins, but not by a factor of $1/\sqrt{10}$. 

For \emph{eROSITA} observations of sources spanning multiple eRASSs, there is a natural way to separate the light curve into segments during which the source was observed every eroday, separated by intervals with no observation. The gaps between the segments are not random, but are, in most cases, considerably longer than the segments, so Eq. \ref{Syserr_rand} is applicable. However, this does not apply to sources close to the ecliptic poles. 

The reduction of the intrinsic scatter of the NIV in adjoined segments is studied with more simulations, identical in number and methodology as those described in Section \ref{SecSysErr}. Fig \ref{3dSyserr_nseg} plots the trade-off between the number of bins per segment and the number of segments, restricted to combinations of $N_{\mathrm b}\times N_{\mathrm{seg}} \leq 1000$. The dashed curves in Fig. \ref{3dSyserr_nseg} decline steeper than the solid curves, demonstrating that the intrinsic scatter is reduced more quickly for randomly positioned segments, than adjacent ones. At the lowest $\mathrm{NIV}_{\infty}$ values we simulated, $\Delta_{\mathrm{s,n}}$ approximately follows Eq. \ref{Syserr_rand}. For higher variabilities, the decline is more gradual.  There is a significant scatter in the data points. Increasing the precision would have required a significantly greater number of simulations than were used.

Based on the form of Eq. \ref{Syserr_rand}, we investigated a power law dependence of $\Delta_{\mathrm{s, n}}$ on $N_{\mathrm{seg}}$. Visual inspection suggests that the exponent of $N_{\mathrm{seg}}$ changes linearly with $\log(\mathrm{NIV}_{\infty})$, and is independent of $N_{\mathrm b}$:

\begin{equation}\label{Nsegadjeq}
    \Delta_{\mathrm{s, n}} = \Delta_{\mathrm s} N_{\mathrm{seg}}^{m  \log(\mathrm{NIV}_{\infty})+k}. 
\end{equation}

We fitted this equation and obtained the parameters listed in Table \ref{TBFsyserradj}. Since $\mathrm{NIV}_{\infty}$ is not known, it can be approximated by $\overline{\mathrm{NEV}_{\mathrm{b}}}$.

\begin{table}[h]
\centering
\setlength{\tabcolsep}{4pt}
\def\arraystretch{1.1}
\begin{tabular}{c|c}
    \textbf{Parameter} & \textbf{Value} \\ \hline
    \textbf{$m$} & $0.1782\pm0.0027$ \\
    \textbf{$k$} & $0.0824\pm0.0063$ \\
\end{tabular}
\caption{Table of the best fitting parameters describing the intrinsic scatter of the NIV, as a function of $N_{\mathrm{seg}}$ and $\mathrm{NIV}_{\infty}$, if the segments are adjoined. They relate to Eq. \ref{Nsegadjeq}. These parameter values can be used alongside those listed in Table \ref{TBFsyserr} to determine the intrinsic scatter, when combing Eqs. \ref{Syserreq}, and \ref{Nsegadjeq}.
\label{TBFsyserradj}}
\end{table}

This relationship has only been derived for simulated pink noise light curves. Eq. \ref{Nsegadjeq} only applies to the interval $5.75\times10^{-4} < \overline{\mathrm{NEV}_{\mathrm{b}}} < 0.263$, for which the exponent in Eq. \ref{Nsegadjeq} is between $0$ and $-0.5$. Pink noise light curves of significantly variable sources with $N_{\mathrm b} \leq 1000$, should have $\log\left(\overline{\mathrm{NEV}_{\mathrm{b}}}\right)$ values well within these limits.

\section{Periodogram analysis} \label{Powspecsection}

The shape of the power spectrum is a diagnostic of the variability process. Estimating the source PSD based on a periodogram requires understanding the biases introduced by the characteristics of the observation. For eRASS, the long but consistent gaps between observations lead to aliasing, described in Section \ref{SecAlias}. In addition, Poisson noise and the varying fractional exposures generate additional noise that affect the periodogram. These effects are discussed in turn in this section. The results presented here apply to all variability analyses, and are not specific to \emph{eROSITA}.

Poisson noise in the count rate measurements affects the observed periodogram by increasing all powers by a constant value of $N_{\mathrm P} = 2 / \overline{R_{\mathrm S}}$, in the fractional rms normalisation \citep{1990A&A...227L..33B}. However, while the \emph{eROSITA} count rates are estimated from bins of effective duration $\epsilon\Delta t$, the PSD frequencies are based on the bin separation, $\tau$. Therefore, the Poisson noise level for light curves with regularly spaced gaps between bins is:

\begin{equation}\label{PoisNoisL_meanFE}
N_{\mathrm P} = \frac{2\tau}{\overline{\epsilon R_{\mathrm S}}\Delta t}.    
\end{equation}

The varying fractional exposures in a light curve generate additional noise, which acts similar to the Poisson noise, by increasing the PSD power at all frequencies. The excess noise has the same dependence on the count rate, but also depends on the mean ($\overline{\epsilon}$), and variance ($\sigma_\epsilon^2$) of the fractional exposure. Therefore, we describe the total Poisson and fractional exposure noise in the PSD as:

\begin{equation}\label{excPoisNois}
    N_{\mathrm{P, \epsilon}} = \frac{2 \tau}{\overline{\epsilon R_{\mathrm S}} \Delta t} \left(1+f(\overline{\epsilon}, \sigma_\epsilon^2)\right).
\end{equation}

\noindent
For other periodogram normalisations, the above equation is modified by replacing $2/\overline{R_{\mathrm S}}$ with the corresponding term. 

In the case of a constant fractional exposure, $\sigma_\epsilon^2=0$, we define $f=0$, and the equation simplifies to Eq. \ref{PoisNoisL_meanFE}. Besides that, we quantified $f$ with simulations. We did not find a dependence of $f$ on the count rate, the number of bins, or the time-ordering of the fractional exposures.

\begin{figure}[h]
\resizebox{\hsize}{!}{\includegraphics{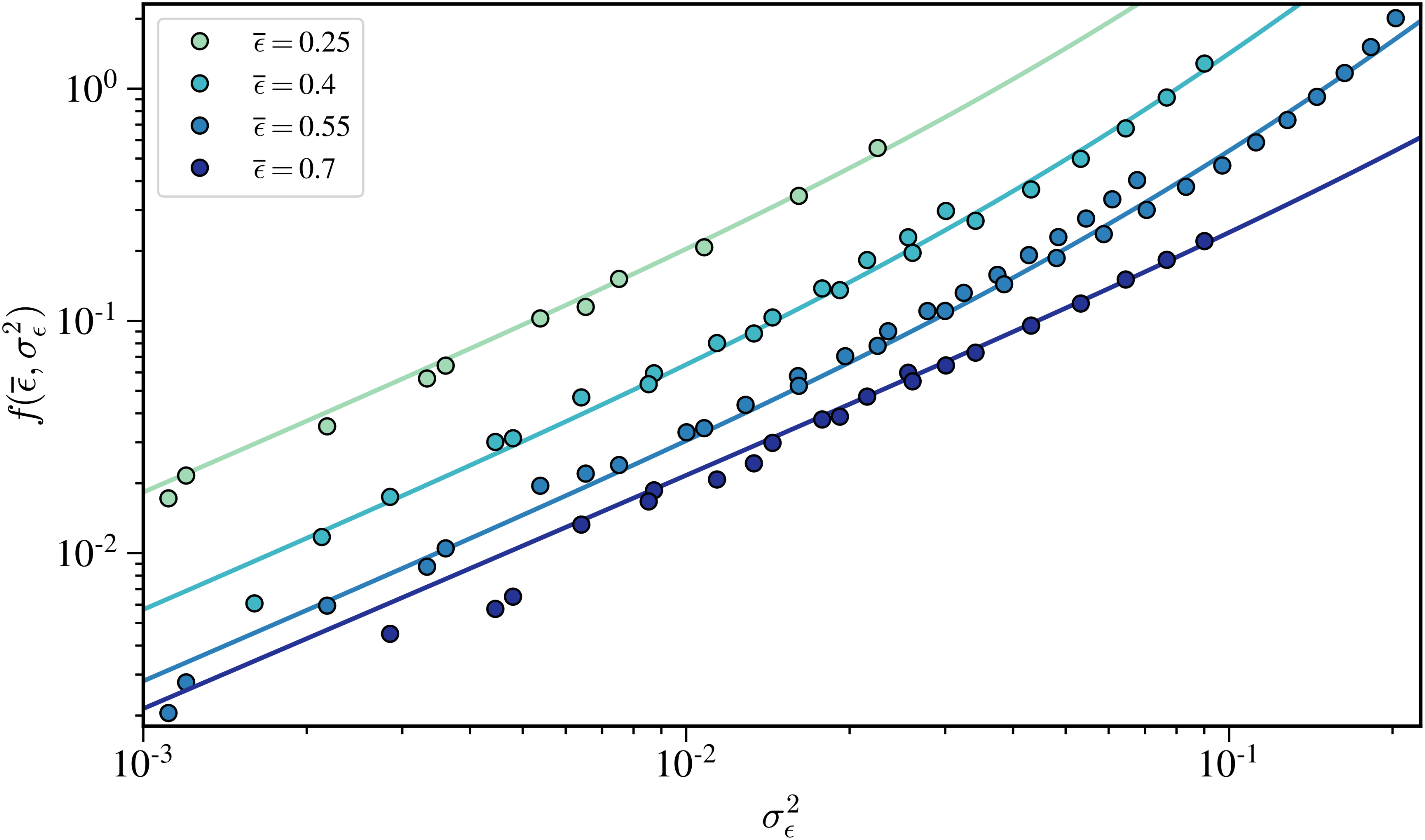}}
\caption{The dependence of the excess noise in a periodogram caused by varying fractional exposures, as a function of $\overline{\epsilon}$, and $\sigma_{\epsilon}^2$. The data points indicate a subset of the results of the simulations used to determine $f(\overline{\epsilon}, \sigma_{\epsilon}^2)$ (Eqs. \ref{excPoisNois} and \ref{fplexeq}). The lines indicate the best fit to all data, for four particular values of $\overline{\epsilon}$, using Eq. \ref{fplexeq}, and the parameter values in Table \ref{Tabbfpar_exPLfit}.
 \label{ExcNoiseFEg}}
\end{figure}

We simulated $2\times10^6$ instances of 1000 bin, high count rate ($30~\mathrm{cts/s}$) patterned fractional exposure light curves (see Section \ref{SecSim}), of intrinsically constant sources with a negligible background. This corresponds to 1000 simulated light curves for each combination of $\overline{\epsilon}$ and $\sigma_\epsilon^2)$. We determined the fractional exposure noise level, and averaged it over the simulations. The results of this are shown in Fig. \ref{ExcNoiseFEg}. The range of $\overline{\epsilon}$ and $\sigma_\epsilon^2)$ values we investigated encompasses the entire parameter space these parameters can have. Therefore, the results found here apply to any light curve with variable fractional exposures, observed by any instrument, and are not merely applicable to \emph{eROSITA}.

We approximated the trends depicted in Fig. \ref{ExcNoiseFEg} with a functional form. At a fixed $\overline{\epsilon}$, we adopted the quadratic equation $f(\sigma_\epsilon^2) = a \sigma_\epsilon^2 + b \sigma_\epsilon^4$, fulfilling $f(\overline{\epsilon}, 0) = 0$. At fixed $\sigma_\epsilon^2$, we found that $f(\overline{\epsilon})$ can be fitted with an exponential and a constant. Therefore, we chose the following fitting function for the results of these simulations:

\begin{equation}\label{fplexeq}
    f(\overline{\epsilon}, \sigma_\epsilon^2) = \left(c_a e^{-d_a \overline{\epsilon}} + g_a\right) \sigma_\epsilon^2 + \left(c_{\mathrm{b}} e^{-d_{\mathrm{b}} \overline{\epsilon}} + g_{\mathrm{b}}\right) \sigma_\epsilon^4.
\end{equation}

To reduce the degeneracy between the fitting parameters, we fitted $f$ as a function of $\sigma_\epsilon^2$, and $\overline{\epsilon}' = \overline{\epsilon} - \sum_{i}(\overline{\epsilon}_i)$, where $\overline{\epsilon}_i$ is the mean fractional exposure of simulation $i$. The best fitting parameter values are described in Table \ref{Tabbfpar_exPLfit}, for the non-rescaled parameters $\overline{\epsilon}$, and $\sigma_\epsilon^2$. Fig. \ref{exPLfitcorner} in Appendix \ref{ACpExcessNoise} depicts the correlation between the parameters in the fit. There is a degeneracy between $c_a$ and $d_a$, and between $c_{\mathrm{b}}$ and $d_{\mathrm{b}}$, as these pairs of parameters increase the effect of the exponential functions involved.

\begin{table}[h]
\centering
\setlength{\tabcolsep}{4pt}
\def\arraystretch{1.1}
\begin{tabular}{c|c}
    \textbf{Parameter} & \textbf{Value} \\ \hline
    \textbf{$c_a$} & $188.48\pm0.96$ \\
    \textbf{$d_a$} & $9.836\pm0.019$ \\
    \textbf{$g_a$} & $1.9451\pm0.0057$ \\
    \textbf{$c_{\mathrm{b}}$} & $1159.9\pm9.7$ \\
    \textbf{$d_{\mathrm{b}}$} & $6.174\pm0.022$ \\
    \textbf{$g_{\mathrm{b}}$} & $-12.74\pm0.18$ \\
\end{tabular}
\caption{Table of the best fitting parameters of the fit from Eq. \ref{fplexeq} to the excess periodogram noise level induced by the variable fractional exposure.
\label{Tabbfpar_exPLfit}}
\end{table}

The fit slightly underestimates the excess fractional exposure noise at the highest values of $\sigma_\epsilon^2$, and overestimates it at the lowest values. However, neither of these regimes is relevant for eRASS light curves. Therefore, we consider the accuracy of this approximation as sufficient for analysing \emph{eROSITA} periodograms. 

Importantly, the varying fractional exposures noise can be much larger than the standard Poisson noise. This is illustrated in Fig. \ref{TypPSD} for typical values of $\overline{\epsilon}$, and $\sigma_\epsilon^2$ in \emph{eROSITA} observations. Finally, estimating the NIV by integrating a periodogram of an eRASS source, after removing the Poisson and varying fractional exposure noise, is validated in Appendix \ref{SecNEVi}. There, we demonstrate the impact of the excess noise and the accuracy of our analytic estimate.

\section{Discussion}\label{SecDisc}

For optimal variability analysis of eRASS observations, we recommend first reducing the extensive data set to a more manageable size of significantly variable sources, using the methods detailed in Section \ref{Diffvarnonvarsection}. After that, we recommend using the $\mathrm{NEV}_{\mathrm{b}}$ methodology to estimate the NIV of all significantly variable sources. The light curve might be split into individual segments to reduce the intrinsic scatter of the NIV. For comparison with other sources, the $\mathrm{NIV}_{\infty}$ is estimated by computing the geometric mean $\mathrm{NEV}_{\mathrm{b}}$ of all segments. The error in the $\mathrm{NIV}_{\infty}$ estimate is found by adding the measurement error with the intrinsic scatter in quadrature.

The thresholds for identifying variable sources using the two variability quantifiers SCATT\_LO and AMPL\_SIG were determined specifically for applicability to \emph{eROSITA} observations of sources close to the SEP. They apply to light curves of between 50 and 1000 bins, for average count rates of between $0.001$ and $30~\mathrm{cts/s}$. As neither of the two variability quantifiers uses the timing of the observations, it is possible to combine data collected over multiple eRASSs. The greater the number of bins of the light curve, at a constant bin duration, the easier it is to identify intrinsically variable sources. Therefore, it is recommended to combine as many measures of the source flux as possible. For \emph{eROSITA}, this means combining all observations of the same source obtained over multiple eRASSs. After eRASS:8, all X-ray sources in the sky will have been observed on $\gtrsim48$ erodays. Therefore, the thresholds described in Section \ref{Diffvarnonvarsection} can be used to identify likely variable sources throughout the entire sky, as observed by \emph{eROSITA}. A caveat is that the thresholds might need to be adjusted for regions of the sky with significantly stronger background count rates, especially for faint sources. 

Selecting variable sources in observations by other instruments may require these thresholds to be modified. They will be impacted by the properties of the observation, the fractional exposure distribution of the light curve bins, and the strength of the background count rate, which varies across the sky. It is unclear to what extent a variable background could affect these results. However, the \emph{eROSITA} background is very stable \citep{2021A&A...647A...1P}, so we do not expect it to have a significant impact on any of the results discussed here. The thresholds for variability detection are based on the specific fractional exposure distribution of \emph{eROSITA} light curves. Variability thresholds for other surveys might differ to some extent from the ones presented here, and will likely require separate simulations.

The $\mathrm{NEV}_{\mathrm{b}}$ method is the most accurate way to estimate the NIV of a variable source. We encourage its use in future variability analyses of light curves with at least a few bins in the Poisson regime. This is of particular relevance for \emph{eROSITA} variability analysis. However, the NEV is still the preferred method for high count rate sources. The conversion from $\overline{\sigma_{\mathrm{b}}}$ to $\mathrm{NEV}_{\mathrm{b}}$ (Section \ref{NEVest}), was made with the assumption of a pink noise PSD, and for typical source and background count rates, and fractional exposures of eRASS observations of SEP sources. However, the the function relating the two parameters can also be used for light curves obtained by other instruments. We only expect potential slight differences in the lower limit of $\overline{\sigma_b}$ for different background count rates. A similar set of simulations can also be performed to improve the accuracy of a similar $\mathrm{NEV}_{\mathrm{b}}$ estimate for different types of PSDs generating the variability. However, as discussed in Appendix \ref{CompNIVest}, $\mathrm{NEV}_{\mathrm{b}}$ also works well for white and pink noise variability. Even without any modification to the function relating $\overline{\sigma_{\mathrm{b}}}$ to the NIV, the $\mathrm{NEV}_{\mathrm{b}}$ provides the most accurate estimate of the NIV of a variable source for these types of variability. We did not investigate non-power law PSDs, for which this might not be true. The $\mathrm{NEV}_{\mathrm{b}}$ has been tested to work for all \emph{eROSITA} light curves with between 20 and 1000 counsecutive bins. That corresponds to sources within $\approx 17 \degree$ of the ecliptic poles. We recommend using AMPL\_MAX as a variability measure for shorter light curves instead. 

The $\mathrm{NEV}_{\mathrm{b}}$ estimate of the NIV can also be calculated for light curves of other instruments. To do so, the count rates can be converted to an average number of source counts per bin. The conversion is: $\overline{C_{\mathrm s}} = 12.177 \overline{R_{\mathrm S}} $, where $\overline{C_{\mathrm s}}$ is the average number of source counts per bin. This is a consequence of the choice of bin size of $40 ~\mathrm{s}$, and the mean of the fractional exposure distribution above 0.1, of $\approx 0.3044$ (see Bogensberger et al. 2024C).

In order to compare the NIV estimate of different sources, the relevant light curves should cover the same frequency interval. This can be achieved by cropping longer light curves to be the same length as shorter ones. Alternatively, all light curves can be split into smaller segments of the same length. Then the geometric mean $\mathrm{NEV}_{\mathrm{b}}$ can be computed for all segments of the same source, and these quantities can then be compared. This method also helps to reduce the intrinsic scatter of the NIV, which corresponds to a smaller error on the estimate of the $\mathrm{NIV}_{\infty}$.   

The equations that were determined for the intrinsic scatter of the NIV in Sections \ref{SecSysErr} and \ref{SecRedSysErr}, apply to light curves of sources with approximately pink noise PSDs. They are not specific to \emph{eROSITA} and can be used to analyse pink light curves found by any instrument.

The description of the excess noise in the periodogram in Section \ref{Powspecsection}, applies to all light curves from any single instrument or collection of instruments. It is not specific to \emph{eROSITA}, or the fields closest to the ecliptic poles. 

Periodograms can only reliably be computed for the brightest variable sources observed for the longest continuous intervals by \emph{eROSITA}. In order to analyse the periodograms of \emph{eROSITA} sources, it is first necessary to subtract the Poisson and fractional exposure noise from them (Section \ref{Powspecsection}), and consider the effects of aliasing (Section \ref{SecAlias}). 

\section{Summary and Conclusions} \label{Secconc}

In this paper, we investigated and described how to best perform an eRASS light curve variability analysis. We also improved previously determined variability methodologies, particularly for low count rate sources, and light curves with varying exposure times. 

We established a method for selecting significantly variable eRASS sources from a large sample with a wide range of different count rates and exposure times. We also determined reliable thresholds on the two variability quantifiers SCATT\_LO and AMPL\_SIG. These thresholds determine the significance of the detected variability and minimise the rate of false positives. They are defined as functions of both the count rate and the number of bins of \emph{eROSITA}-like light curves.

The NIV quantifies the intrinsic degree of variability of a source within the duration of the observations and is usually estimated by the NEV. However, the NEV methodology faces challenges at low count rates, and for varying fractional exposures. We presented a new method for estimating the NIV that remains accurate even in the Poisson regime, and is not biased by varying fractional exposures. It uses the Bayesian excess variance (bexvar) and a function that converts that into an estimate of the NIV. This method is based on simulations of a pink noise PSD but remains more accurate than the NEV for both red, and pink noise PSDs at low count rates. Above $1.5~\mathrm{cts/s}$, the methods are comparably accurate. 

The NIV of one light curve can substantially differ from the NIV in another light curve covering the same frequency interval, even if the process generating the variability is stationary. Therefore, the intrinsic scatter of the NIV must be determined, to compare the variability of different sources or the same source at different times. The best way to reduce the scatter is to split a light curve into shorter segments of equal length, and compute the geometric mean of the $\mathrm{NEV}_{\mathrm{b}}$ measurements of individual segments. We found analytic expressions for the size of the intrinsic scatter as a function of the number of bins in each segment, the number of segments, and the intrinsic variability of the source, for pink noise variability. The intrinsic scatter also depends on whether the individual segments are adjacent, or randomly placed, and far apart. It decreases more rapidly with an increasing number of segments in the latter case, especially for strongly variable sources. 

Additionally, we investigated how to analyse the periodograms of variable sources observed by \emph{eROSITA}. If a regularly sampled light curve consists of bins with differing exposure times, its periodogram will feature an additional noise component above the Poisson noise. We determined the dependence of this fractional exposure noise on its mean and variance throughout the light curve. 

The code that we developed and used throughout this paper is available at: \newline\noindent \href{https://github.com/DavidBogensberger/eROSITA_SEP_Variability}{https://github.com/DavidBogensberger/eROSITA\_SEP\_Variability}. It contains further instructions detailing when and how these methods can be used, how to interpret the results, and when to consider modifying them for use in other areas of the sky, or other instruments.

The methods presented here are used to analyse the variability of X-ray sources detected within $3\degree$ of the SEP in the first three eRASSs, which will be discussed in Bogensberger et al. 2024C. 

\section{Acknowledgements}

This work is based on data from \emph{eROSITA}, the soft X-ray instrument aboard SRG, a joint Russian-German science mission supported by the Russian Space Agency (Roskosmos), in the interests of the Russian Academy of Sciences represented by its Space Research Institute (IKI), and the Deutsches Zentrum für Luft- und Raumfahrt (DLR). The SRG spacecraft was built by Lavochkin Association (NPOL) and its subcontractors and is operated by NPOL with support from the Max Planck Institute for Extraterrestrial Physics (MPE).

The development and construction of the \emph{eROSITA} X-ray instrument was led by MPE, with contributions from the Dr. Karl Remeis Observatory Bamberg \& ECAP (FAU Erlangen-Nuernberg), the University of Hamburg Observatory, the Leibniz Institute for Astrophysics Potsdam (AIP), and the Institute for Astronomy and Astrophysics of the University of Tübingen, with the support of DLR and the Max Planck Society. The Argelander Institute for Astronomy of the University of Bonn and the Ludwig Maximilians Universität Munich also participated in the science preparation for \emph{eROSITA}.

This work made use of the software packages astropy \citep[\href{https://www.astropy.org/}{https://www.astropy.org/};][]{2013A&A...558A..33A, 2018AJ....156..123A}, bexvar \citep[\href{https://github.com/JohannesBuchner/bexvar}{https://github.com/JohannesBuchner/bexvar};][]{2022A&A...661A..18B}, numpy \citep[\href{https://www.numpy.org/}{https://www.numpy.org/};][]{harris2020array}, matplotlib \citep[\href{https://matplotlib.org/}{https://matplotlib.org/};][]{Hunter:2007}, scipy \citep[\href{https://scipy.org/}{https://scipy.org/};][]{2020SciPy-NMeth}, Stingray \citep[\href{https://docs.stingray.science/}{https://docs.stingray.science/};][]{2016ascl.soft08001H}, and Ultranest \citep[\href{https://johannesbuchner.github.io/UltraNest/}{https://johannesbuchner.github.io/UltraNest/};][]{2021JOSS....6.3001B}.

We thank the anonymous referee for their helpful comments.

\bibliographystyle{aa} 
\bibliography{bibliography.bib}      

\appendix

\section{Comparing methods for NIV estimation}\label{CompNIVest}

In this section, we considered two other methods that could be used to estimate the NIV of the light curve of a variable source, and compared them against the $\mathrm{NEV}_{\mathrm{b}}$ method. The first of these methods estimates the NIV by integrating the periodogram of the light curve, and we label it $\mathrm{NEV}_{\mathrm{i}}$. The second methodology is based on using Eqs. \ref{NEVeq} and \ref{NEVpar}, but with an adjustment for reducing the effect of the lowest exposure bins on the estimate. We denote this estimate of the NIV as $\mathrm{NEV}_{\mathrm{eq}}$. Finally, we evaluated the strengths and weaknesses of these three methods. 

\subsection{$\mathrm{NEV}_{\mathrm{i}}$}\label{SecNEVi}

Parseval's theorem states that in the high count rate limit, when the assumption of a normal probability distribution of the measured count rate in every bin is true, the definition of the NEV as shown in Eq. \ref{NEVeq}, corresponds to the integral of the Poisson noise subtracted, rms normalised periodogram \citep{1989ASIC..262...27V}:

\begin{equation}
    \mathrm{NEV}_{\mathrm{i}} = \int_{(N_{\mathrm b}\tau)^{-1}}^{(2\tau)^{-1}} (P(\nu)-N_{\mathrm{P, \epsilon}}) d\nu.
\end{equation}

\noindent
Therefore, it is also possible to estimate the NIV of the light curve of a variable source in this way. The Poisson noise of a light curve with variable fractional exposures is discussed in more detail in Section \ref{Powspecsection}.

This method is more computationally expensive than calculating $\mathrm{NEV}_{\mathrm{eq}}$. Nevertheless, it can be useful for estimating the NIV for a different frequency range. 

\begin{figure}[h]
\resizebox{\hsize}{!}{\includegraphics{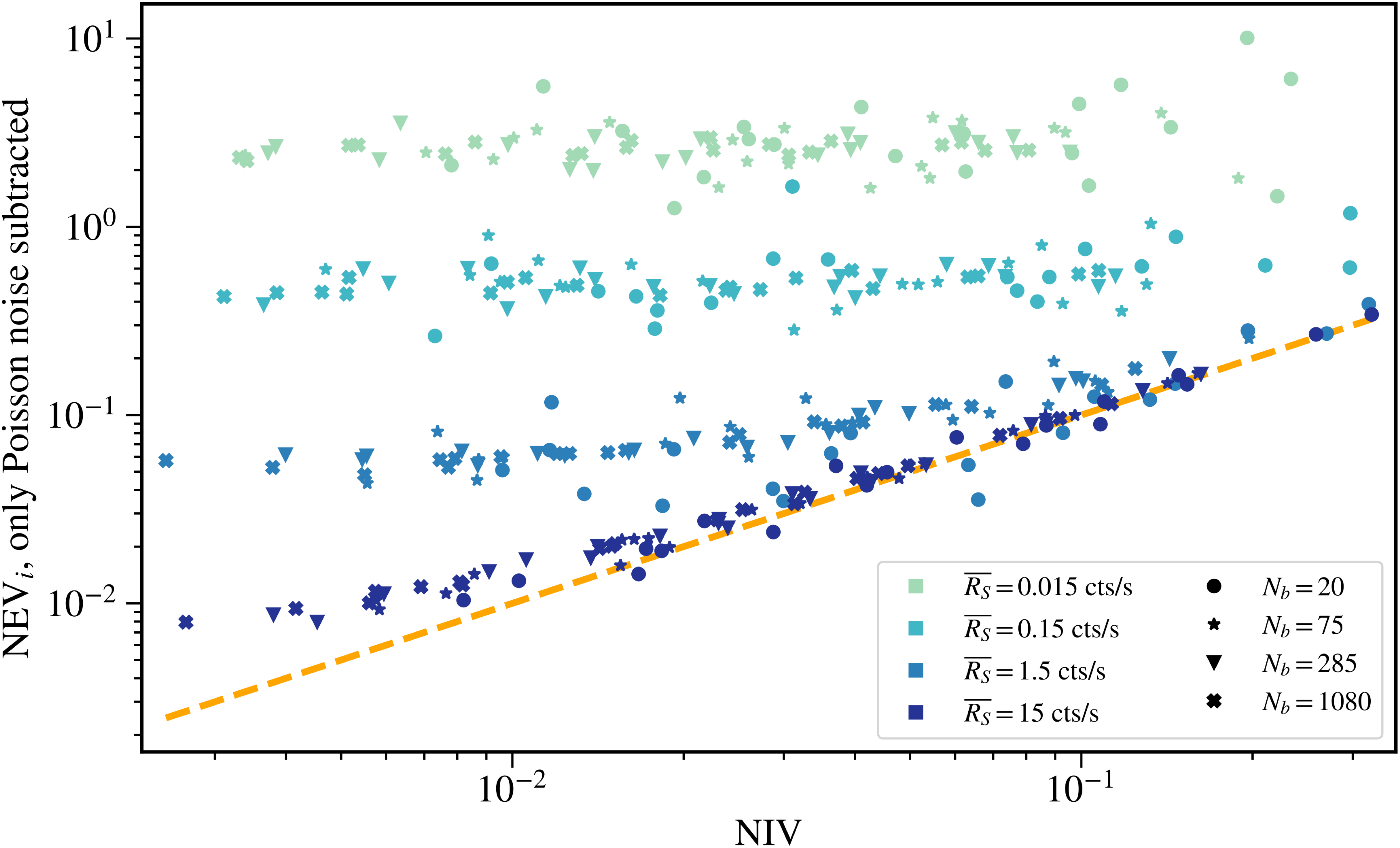}}
\caption{The ability to estimate the NIV by integrating the periodograms of simulated \emph{eROSITA}-like pink noise light curves. Each data point indicates the result of a single simulation. The closer a data point lies to the orange dashed 1:1 line, the better the estimate of the NIV is. 
For this plot, only the standard Poisson noise (Eq. \ref{PoisNoisL_meanFE}) was subtracted from the periodograms.
 \label{iNEVfig}}
\end{figure}

Fig. \ref{iNEVfig} shows the ability of $\mathrm{NEV}_{\mathrm{i}}$ to estimate the NIV in simulated pink noise light curves, when subtracting the standard Poisson noise (Eq. \ref{PoisNoisL_meanFE}) from the periodogram. It can be compared against Fig. \ref{bvNEVfig}, to see the different accuracies of the two methods. This shows that $\mathrm{NEV}_{\mathrm{i}}$ is only a good estimator of the NIV at the highest count rates and the greatest NIVs. At lower count rates and NIVs, $\mathrm{NEV}_{\mathrm{i}}$ levels off, diverging from the 1:1 line that the data points should ideally be distributed around. The minimum level of $\mathrm{NEV}_{\mathrm{i}}$ decreases with an increasing count rate. However, it seems largely independent of the number of bins in the light curve. At the lowest count rates considered for this figure, this plateau level even lies above the maximum NIV of the simulated light curves. The plateaus are caused by varying fractional exposures, and the excess noise they cause in the periodograms, which has not been subtracted in this example. This shows how important it is to subtract the fractional exposure noise as well, as discussed in Section \ref{Powspecsection}. The properties of the plateaus helped motivate the expression of the total Poisson and fractional exposure noise of Eq. \ref{excPoisNois}. 

\begin{figure}[h]
\resizebox{\hsize}{!}{\includegraphics{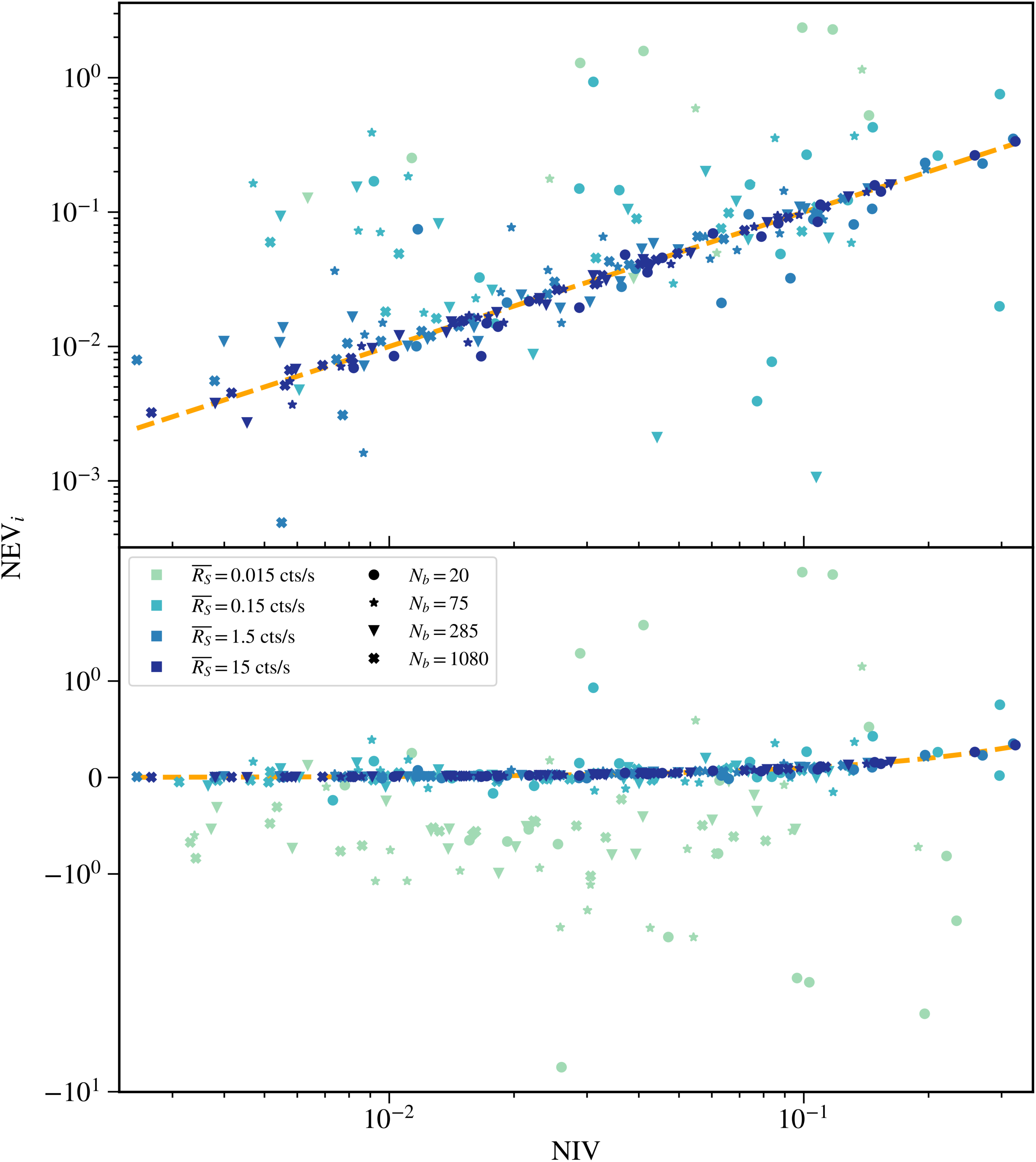}}
\caption{The accuracy of estimating the NIV by integrating periodograms of simulated \emph{eROSITA}-like pink noise light curves. Unlike in Fig. \ref{iNEVfig}, the NIV is estimated here by subtracting both the Poisson and fractional exposure noise, using Eqs. \ref{excPoisNois}, and \ref{fplexeq}, with parameters listed in Table \ref{Tabbfpar_exPLfit} from the periodograms. This figure shows the relationship on both a logarithmic (top panel) and a symmetric logarithmic (lower panel) scale, to showcase how close the individual $\mathrm{NEV}_{\mathrm{i}}$ measurements are to the NIV, as well as the complete range of values measured for $\mathrm{NEV}_{\mathrm{i}}$.
 \label{iNEVfig_corrPN}}
\end{figure}

When subtracting the combined Poisson and fractional exposure noise from the periodograms of the simulated light curves, using Eqs. \ref{excPoisNois}, and \ref{fplexeq}, as well as the parameters listed in Table \ref{Tabbfpar_exPLfit} in Section \ref{Powspecsection}, the plateaus disappear, and $\mathrm{NEV}_{\mathrm{i}}$ becomes a much more accurate estimator of the NIV, as is shown in Fig. \ref{iNEVfig_corrPN}. Henceforth, we will use $\mathrm{NEV}_{\mathrm{i}}$ to refer to the Poisson and fractional exposure noise subtracted integral of the periodogram.

At high count rates ($\overline{R_{\mathrm S}}\gtrsim1.5~\mathrm{cts/s}$), $\mathrm{NEV}_{\mathrm{i}}$ is an accurate estimator of the NIV, for the entire range of values of the NIV, and number of bins in the light curve, that we investigated. However, its accuracy decreases rapidly at lower count rates. At $\overline{R_{\mathrm S}} \leq 0.15~\mathrm{cts/s}$, the $\mathrm{NEV}_{\mathrm{i}}$ often differs significantly from the NIV, and is frequently underestimated. 87\% of the light curves we simulated at a count rate of $0.015~\mathrm{cts/s}$ had a negative $\mathrm{NEV}_{\mathrm{i}}$. At $0.15~\mathrm{cts/s}$, 34\% of the light curves are still found at a negative $\mathrm{NEV}_{\mathrm{i}}$, and even at $1.5~\mathrm{cts/s}$, 7\% of light curves were found to have a negative $\mathrm{NEV}_{\mathrm{i}}$. Possible reasons for this is that the periodogram weighs all bins equally, regardless of their fractional exposure, and the challenges of computing a periodogram from a very small number of counts.

\subsection{$\mathrm{NEV}_{\mathrm{eq}}$}\label{SecNEVeq}

The standard NEV, as defined in Eqs. \ref{NEVeq} and \ref{NEVpar} works well for normal probability distributions for the count rate in each bin. When the components used to calculate the NEV (Eq. \ref{NEVeq}) are defined as in Eq. \ref{NEVpar}, the NEV treats every bin identically, regardless of their fractional exposure. This might cause a significant portion of the excess variance detected in an \emph{eROSITA}-like light curve to originate from the bins with the smallest exposure times. 

One way of reducing the effect that the lowest exposure bins have on the measured NEV, is to weight each of the terms of the NEV by the fractional exposure itself. Rather than using the definitions described in Eq. \ref{NEVpar}, we would use: 

\begin{align} \label{FEweighNEVpar}
\begin{split}
    \overline{R_{\mathrm S}} & = \frac{\sum_{i=1}^{N_{\mathrm b}} \epsilon\left(t_i\right) R_{\mathrm S}\left(t_i\right)}{\sum_{i=1}^{N_{\mathrm b}} \epsilon\left(t_i\right)} \\
    \sigma_{\mathrm{obs}}^2 & = \frac{N_{\mathrm b}}{N_{\mathrm b}-1} \frac{\sum_{i=1}^{N_{\mathrm b}} \epsilon^2\left(t_i\right) \left(R_{\mathrm S}\left(t_i\right) - \overline{R_{\mathrm S}}\right)^2}{\sum_{i=1}^{N_{\mathrm b}} \epsilon^2\left(t_i\right)} \\
    \overline{\sigma_{\mathrm{err}}^2} & = \frac{\sum_{i=1}^{N_{\mathrm b}} \epsilon^2\left(t_i\right) \sigma_{\mathrm{err}}^2\left(t_i\right)}{\sum_{i=1}^{N_{\mathrm b}} \epsilon^2\left(t_i\right)}.
\end{split}
\end{align}

\noindent
If the background area and count rate are constant, then the first equation of Eq. \ref{FEweighNEVpar} is identical to Eq. \ref{countratedef}. These equations are applicable to instances of varying fractional exposures, unlike those of Eq. \ref{NEVpar}. By weighting the terms, the equations effectively deal with the individually measured source counts, rather than count rates. 

Since the uncertainties in the measured count rates are asymmetric, it is worth considering which error to use for $\sigma_{\mathrm{err}}^2\left(t_i\right)$. The excess variance compares the difference between the contribution of each measured count rate to the total variance, with the error of that measurement. Therefore, we decided to use the error in the direction of the mean count rate, similar to the definition of the modified AMPL\_SIG in Section \ref{SecMAD}. So if $R_{\mathrm S}\left(t_i\right) \leq \overline{R_{\mathrm S}}$, we set $\sigma_{\mathrm{err}}\left(t_i\right) = \sigma_{\mathrm{+err}}\left(t_i\right)$, and if $R_{\mathrm S}\left(t_i\right) > \overline{R_{\mathrm S}}$, we set $\sigma_{\mathrm{err}}\left(t_i\right) = \sigma_{\mathrm{-err}}\left(t_i\right)$ in Eq. \ref{FEweighNEVpar}. We label the NIV estimate found by using these modification to the NEV method, as the $\mathrm{NEV}_{\mathrm{eq}}$. The value of $\mathrm{NEV}_{\mathrm{eq}}$ is still found from Eq. \ref{NEVeq}, but with the updated parameters of Eq. \ref{FEweighNEVpar}. 

These modifications allow $\mathrm{NEV}_{\mathrm{eq}}$ to not be as biased by varying fractional exposures. However, they do not deal with the inherent problems with these equations in Poisson regime, as discussed in Section \ref{subsecNEVdes}.

\begin{figure}[h]
\resizebox{\hsize}{!}{\includegraphics{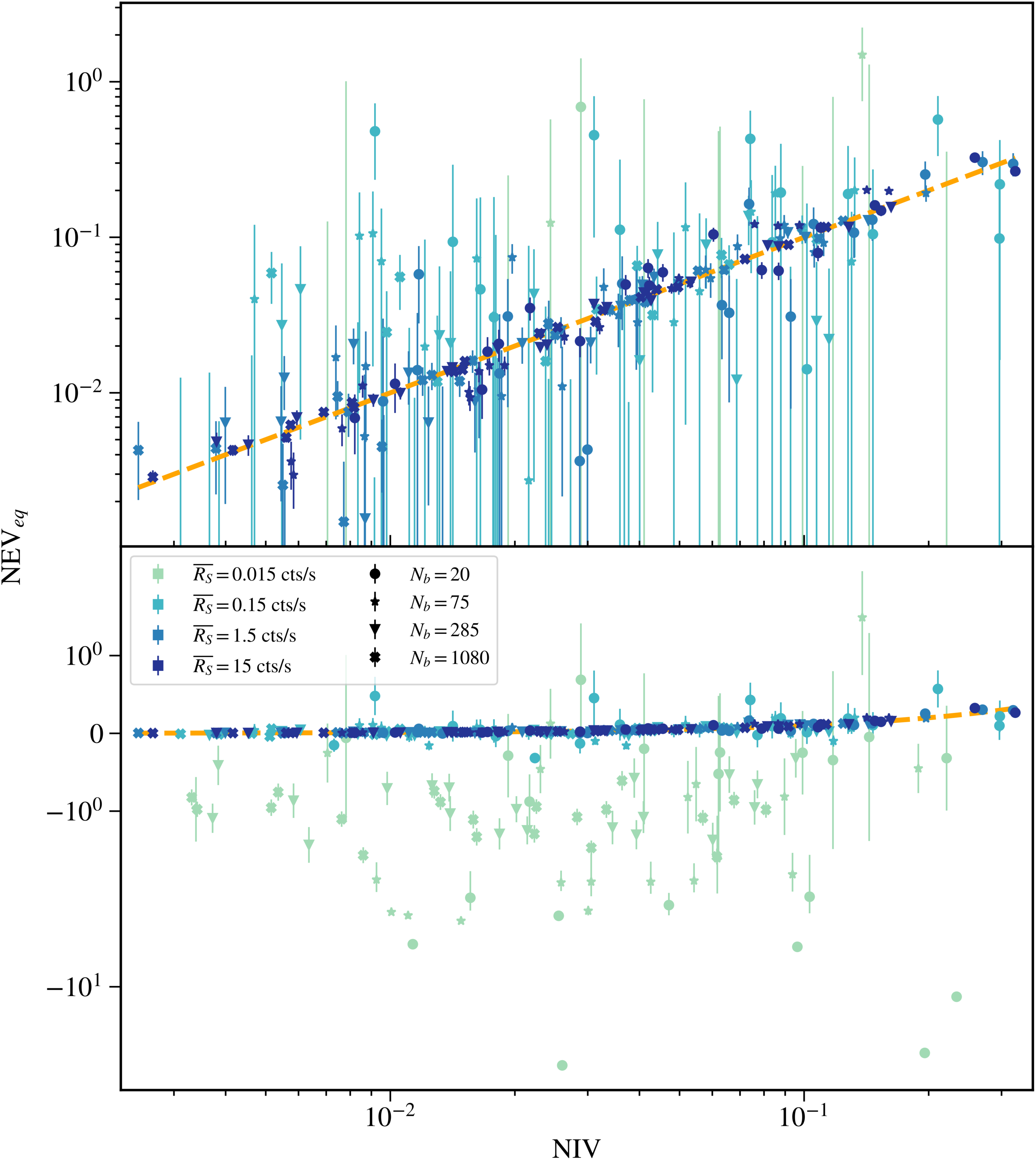}}
\caption{Comparison of the $\mathrm{NEV}_{\mathrm{eq}}$ obtained by using the weighted parameters of Eq. \ref{FEweighNEVpar} in Eq. \ref{NEVeq}, with the NIV. This figure uses the same structure as Fig. \ref{iNEVfig_corrPN}.
 \label{eqwNEVfigl}}
\end{figure}

Fig. \ref{eqwNEVfigl} shows the performance of using $\mathrm{NEV}_{\mathrm{eq}}$ as an estimator of the NIV, using Eq. \ref{NEVeq}, and the weighted parameters of Eq. \ref{FEweighNEVpar}. This method works well at high count rates. However, 95\% of the simulated light curves at a count rate of $0.015~\mathrm{cts/s}$, and 29\% of the light curves with a count rate of $0.15~\mathrm{cts/s}$ were found to have a negative $\mathrm{NEV}_{\mathrm{eq}}$, some at extremely negative values. At a count rate of $1.5~\mathrm{cts/s}$, 3\% of simulated light curves still have a negative $\mathrm{NEV}_{\mathrm{eq}}$. Negative $\mathrm{NEV}_{\mathrm{eq}}$ values are not necessarily a problem, given the proper statistical approaches, but can indicate a systematic offset from the NIV, and will prevent a geometric mean NEV to be calculated, for estimating $\mathrm{NIV}_{\infty}$. At low count rates, $\mathrm{NEV}_{\mathrm{eq}}$ is less likely to significantly overestimate the NIV than $\mathrm{NEV}_{\mathrm{i}}$, but is more likely to significantly underestimate it.

We used the results of \citet{2003MNRAS.345.1271V}, their Equation 11, to determine the uncertainties in the individual measurements of the $\mathrm{NEV}_{\mathrm{eq}}$, which we plot in Fig. \ref{eqwNEVfigl}. However, this equation might not be accurate for light curves with varying fractional exposures or for the modifications to the standard method discussed above. It also does not reliably estimate the uncertainty in the measured $\mathrm{NEV}_{\mathrm{eq}}$ when it is less than 0, as can be seen by the lower panel in Fig. \ref{eqwNEVfigl}. Accurate error bars should extend to above 0 in most instances. 

Figs. \ref{BVtologvar}, \ref{iNEVfig}, \ref{iNEVfig_corrPN}, and \ref{eqwNEVfigl}, which depict the ability of various methods to estimate the NIV, are all based on the same set of simulated light curves. These were generated to have $\{20, 75, 150, 400, 1050\}$ bins, and mean input count rates of $\{0.015, 0.15, 1.5, 15\}~\mathrm{cts/s}$. Comparing Fig. \ref{BVtologvar} to Figs. \ref{iNEVfig_corrPN} and \ref{eqwNEVfigl}, shows that $\mathrm{NEV}_{\mathrm{b}}$ is significantly more accurate at estimating the NIV in the light curve at low count rates, than $\mathrm{NEV}_{\mathrm{i}}$ or $\mathrm{NEV}_{\mathrm{eq}}$ are.

\subsection{Comparison}

\begin{figure*}[h]
\begin{subfigure}{\textwidth}
    \resizebox{\hsize}{!}{\includegraphics{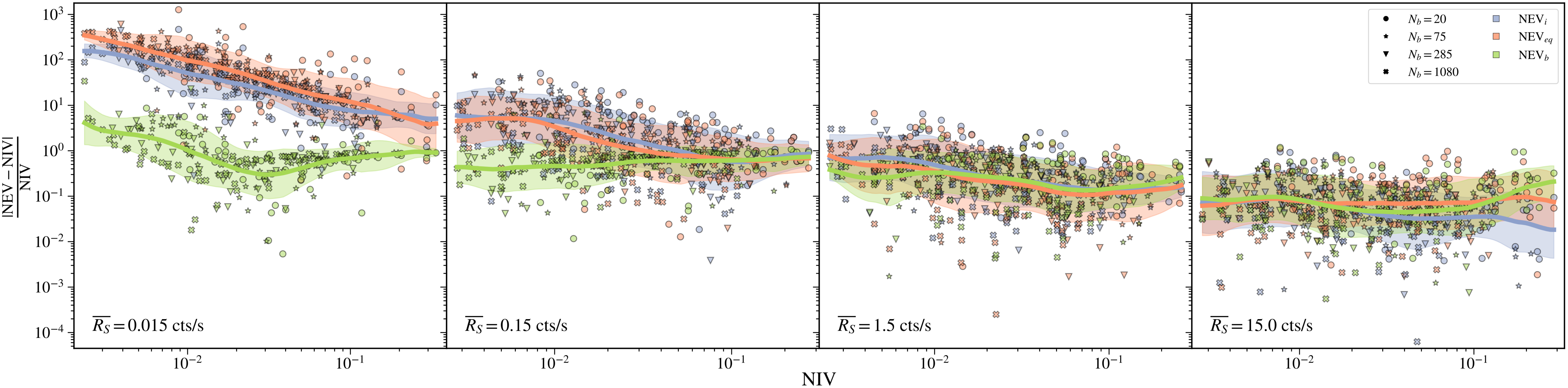}}
    \caption{For light curves simulated with white noise PSDs of $P\propto \nu^0$}
    \label{MetCompNEV0}
\end{subfigure}
\begin{subfigure}{\textwidth}
    \resizebox{\hsize}{!}{\includegraphics{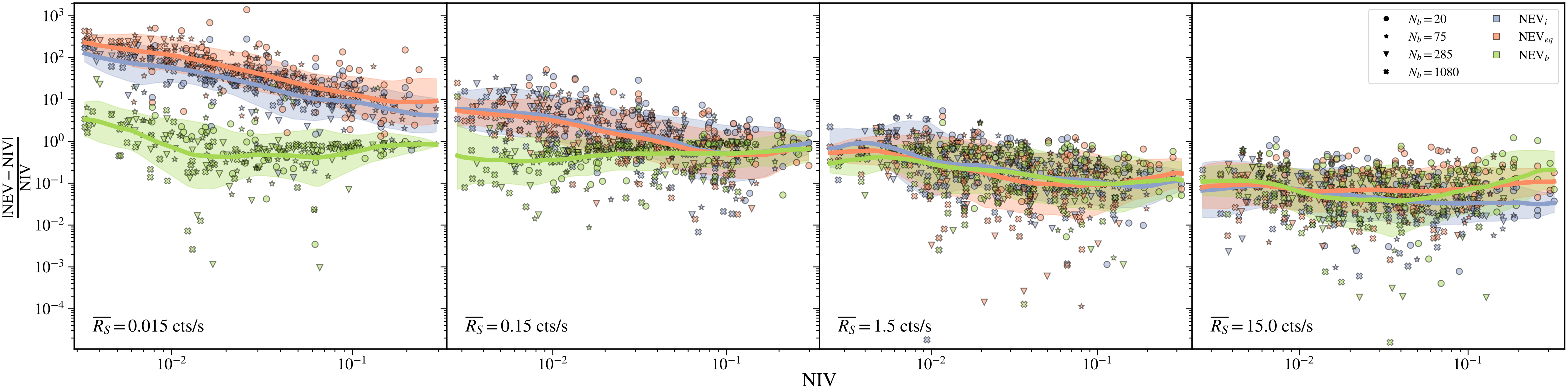}}
    \caption{For light curves simulated with pink PSDs of $P\propto \nu^{-1}$}
    \label{MetCompNEV-1}
\end{subfigure}
\begin{subfigure}{\textwidth}
    \resizebox{\hsize}{!}{\includegraphics{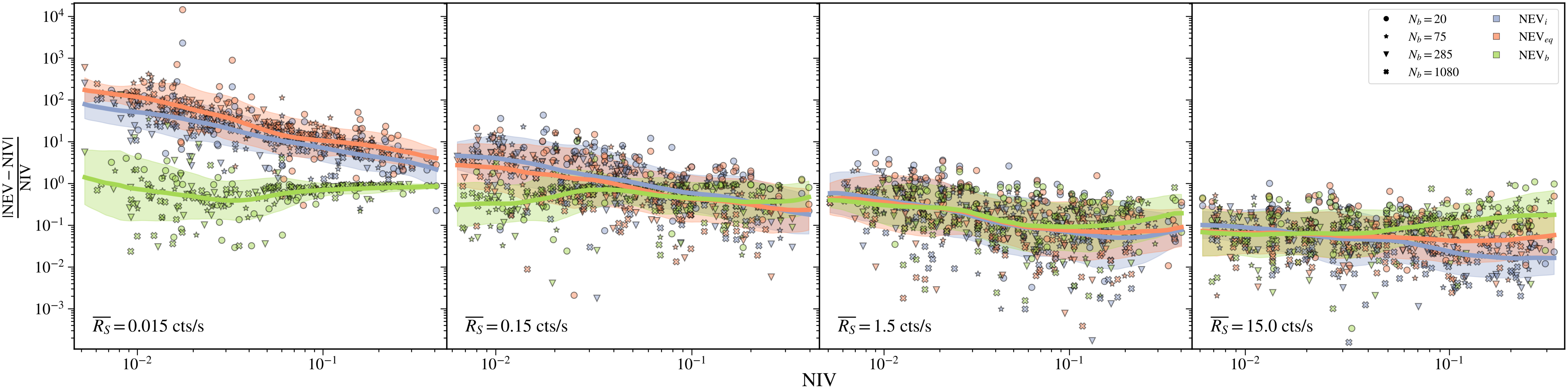}}
    \caption{For light curves simulated with red PSDs of $P\propto \nu^{-2}$}
    \label{MetCompNEV-2}
\end{subfigure}
\caption{Comparison of the ability of the integral of the periodogram ($\mathrm{NEV}_{\mathrm{i}}$), the modified NEV equation ($\mathrm{NEV}_{\mathrm{eq}}$), and the bexvar NIV estimate ($\mathrm{NEV}_{\mathrm{b}}$), to accurately estimate the NIV of \emph{eROSITA}-like light curves. This figure plots the absolute value of the difference between the measured value, $\mathrm{NEV}_{\mathrm m}$ (which refers to either $\mathrm{NEV}_{\mathrm{i}}$, $\mathrm{NEV}_{\mathrm{eq}}$, or $\mathrm{NEV}_{\mathrm{b}}$), and the NIV ($\mathrm{NIV}$), normalised by the NIV, as a function of the NIV. The lower the value of this parameter, the more accurate the estimate of the NIV is. Each data point represents one instance of a simulated light curve. The same light curves were used for computing $\mathrm{NEV}_{\mathrm{i}}$, $\mathrm{NEV}_{\mathrm{eq}}$, and $\mathrm{NEV}_{\mathrm{b}}$. Each panel represents a different count rate. The dependence on the number of bins is showcased via different symbols. Fig (a) depicts the accuracy of the NIV estimates for white noise light curves. Fig (b) does the same, but for pink noise light curves, and in (c) for red noise light curves. Solid lines are plotted on top of the distribution of points, corresponding to the geometric mean of the distribution of $\log\left(|\mathrm{NEV}_{\mathrm m} - \mathrm{NIV}| / \mathrm{NIV} \right)$ as a function of $\log(\mathrm{NIV})$. Transparent areas indicate the part of the distribution corresponding to one standard deviation from the mean, in either direction.}
\label{MetCompNEV}
\end{figure*}

In this section, we compare the ability of the three methods previously discussed ($\mathrm{NEV}_{\mathrm{b}}$, $\mathrm{NEV}_{\mathrm{i}}$, and $\mathrm{NEV}_{\mathrm{eq}}$), to provide an accurate estimate of the NIV in the light curve. We compared the accuracies as a function of the count rate, the number of bins, the degree of variability, and the type of variability observed. Fig. \ref{MetCompNEV} depicts the absolute magnitude of the difference between the measured NIV estimate (which we label as $\mathrm{NEV}_{\mathrm m}$ to refer to the three methods), and the actual NIV, normalised by the NIV of the simulated light curves: $\left|\mathrm{NEV}_{\mathrm m} - \mathrm{NIV}\right|/\mathrm{NIV} = \left|D_{\mathrm m}\right|$. The figure shows the dependence of $\left|D_{\mathrm m}\right|$ on the NIV in each plot, as the ability to estimate the NIV strongly depends on its value. The three methods were applied to the same simulated light curves for a fair comparison. Three different types of variability were considered; white noise ($P\propto \nu^0$), pink noise ($P\propto \nu^{-1}$), and red noise ($P\propto \nu^{-2}$). We again selected mean count rates of $\{0.015, 0.15, 1.5, 15.0\}$ cts/s, and a number of bins of $\{20, 75, 150, 400, 1050\}$ for the construction of the simulated light curves used in this analysis.

The results of these simulations are shown in Fig. \ref{MetCompNEV}. As expected, the ability to estimate the NIV improves with an increasing number of bins, for all three methods. Fig. \ref{MetCompNEV} shows that $\mathrm{NEV}_{\mathrm{b}}$ is more accurate at estimating the NIV than $\mathrm{NEV}_{\mathrm{i}}$ and $\mathrm{NEV}_{\mathrm{eq}}$ at low count rates, and low degrees of variability. Even though the conversion from $\overline{\sigma_{\mathrm{b}}}$ to $\mathrm{NEV}_{\mathrm{b}}$ was determined specifically for pink noise PSDs, the conversion is still very accurate for light curves generated from white and red noise PSDs. Even for those types of variability, $\mathrm{NEV}_{\mathrm{b}}$ is significantly more accurate than $\mathrm{NEV}_{\mathrm{i}}$, or $\mathrm{NEV}_{\mathrm{eq}}$ at low count rates and low degrees of variability. At the lowest count rates we investigated, of $0.015~\mathrm{cts/s}$, $\mathrm{NEV}_{\mathrm{b}}$ is always the most accurate NIV estimate, for the three types of PSDs, for all values of the NIV, and the entire range of the number of bins that we investigated. At this count rate, $\mathrm{NEV}_{\mathrm{b}}$ is often 2 orders of magnitude more accurate than the other two methods. 

At $0.15~\mathrm{cts/s}$, $\mathrm{NEV}_{\mathrm{b}}$ is more accurate at low variabilities than either $\mathrm{NEV}_{\mathrm{i}}$ or $\mathrm{NEV}_{\mathrm{eq}}$, but all three methods are approximately equally accurate at high degrees of variability. For count rates of $1.5~\mathrm{cts/s}$ and $15~\mathrm{cts/s}$, the measured number of source counts per bin can reasonably accurately be assumed to have a normal probability distribution for most, or all erodays. As a result, all three methods have similar accuracies at these count rates, across the entire range of simulated NIVs. However, $\mathrm{NEV}_{\mathrm{b}}$ provides a slightly less accurate estimate of the NIV of high count rate sources ($15~\mathrm{cts/s}$) exhibiting red noise variability of $\mathrm{NIV} \gtrsim 5\times 10^{-2}$), as compared to $\mathrm{NEV}_{\mathrm{i}}$ or $\mathrm{NEV}_{\mathrm{eq}}$. 

\begin{figure*}[h]
\begin{subfigure}{\textwidth}
    \resizebox{\hsize}{!}{\includegraphics{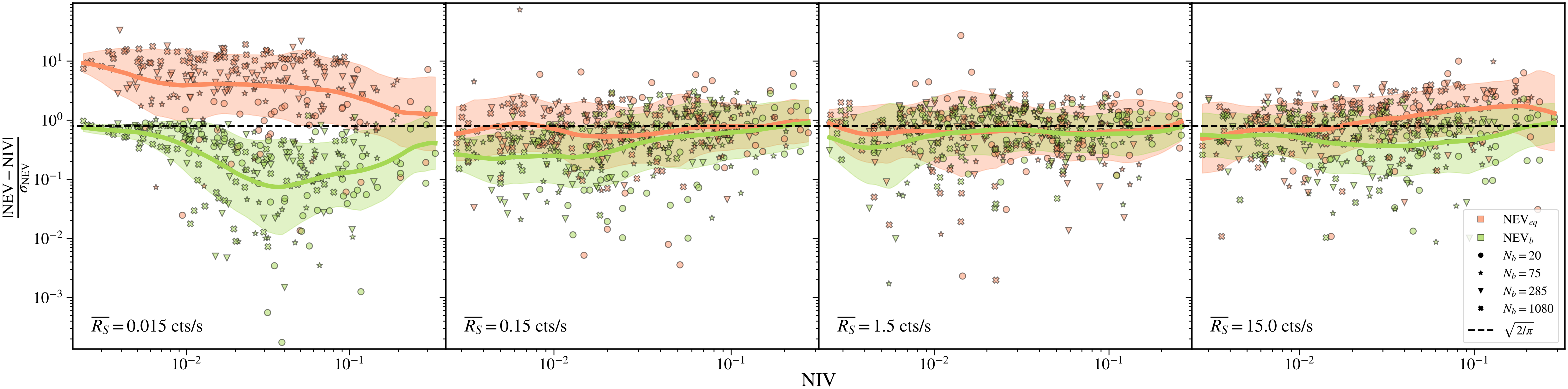}}
    \caption{For light curves simulated with white noise PSDs of $P\propto \nu^0$}
    \label{MetCompNEVerr0}
\end{subfigure}
\begin{subfigure}{\textwidth}
    \resizebox{\hsize}{!}{\includegraphics{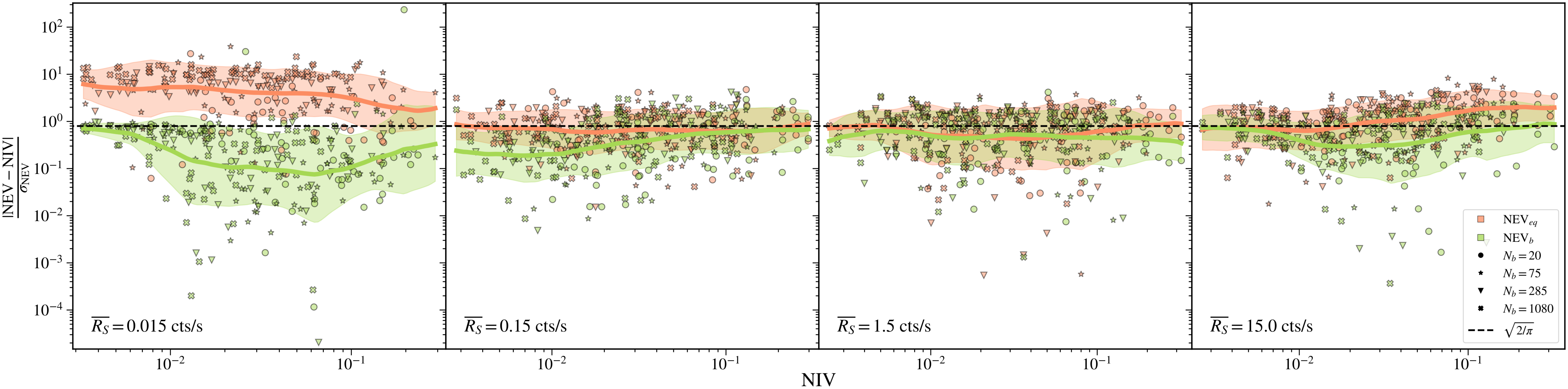}}
    \caption{For light curves simulated with pink PSDs of $P\propto \nu^{-1}$}
    \label{MetCompNEVerr-1}
\end{subfigure}
\begin{subfigure}{\textwidth}
    \resizebox{\hsize}{!}{\includegraphics{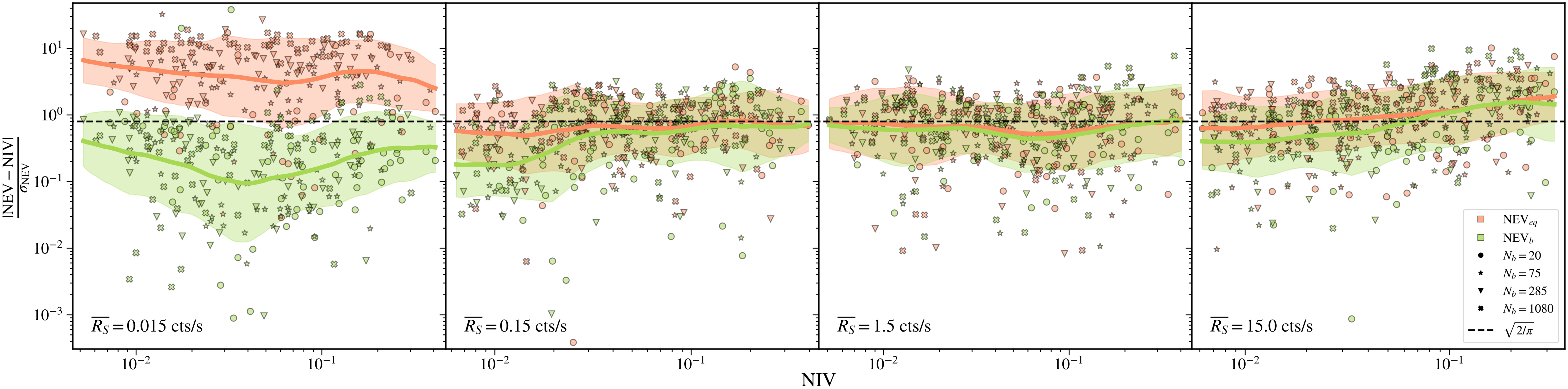}}
    \caption{For light curves simulated with red PSDs of $P\propto \nu^{-2}$}
    \label{MetCompNEVerr-2}
\end{subfigure}
\caption{The accuracy of the uncertainties of $\mathrm{NEV}_{\mathrm{eq}}$ and $\mathrm{NEV}_{\mathrm{b}}$, for a range of different count rates, number of bins, and types of PSDs. This figure used the results of the same simulated light curves as are displayed in Fig. \ref{MetCompNEV}. The parameter $\mathrm{NEV}_{\mathrm m}$ denotes a NIV estimate, which is either $\mathrm{NEV}_{\mathrm{b}}$, or $\mathrm{NEV}_{\mathrm{eq}}$. The parameter $\sigma_{\mathrm m}$ denotes the uncertainty in the measurement. The black dashed line at a value of $\sqrt{2/\pi}$, indicates the level that $\left|\mathrm{NEV}_{\mathrm m} - \mathrm{NIV}\right| / \sigma_{NEV,m} = \left|Z_{\mathrm m}\right|$ should have. This figure excludes points from simulated light curves that have such negative values of $\mathrm{NEV}_{\mathrm{eq}}$, that the error in those measurements cannot be determined. Other features are similar to those of Fig. \ref{MetCompNEV}.}
\label{MetCompNEVerr}
\end{figure*}

The $\mathrm{NEV}_{\mathrm{i}}$ and $\mathrm{NEV}_{\mathrm{eq}}$ methods estimate the NIV with similar accuracy. The $\left|D_{\mathrm m}\right|$ parameter follows a similar anti-correlation as a function of the NIV, for both $\mathrm{NEV}_{\mathrm{i}}$ and $\mathrm{NEV}_{\mathrm{eq}}$. In contrast, the mean difference between the $\mathrm{NEV}_{\mathrm{b}}$ estimate, and the NIV, are maintained at an almost constant ratio of the NIV. The anti-correlation is predominantly caused by the normalisation of $\left|D_{\mathrm m}\right|$. The existence of such an anti-correlation indicates that the average difference between the estimate, and the true value of the NIV is mainly independent of the NIV. Therefore, this also shows that the $\mathrm{NEV}_{\mathrm{b}}$ is a more reliable estimate of the NIV.

Another relevant feature to consider when comparing different methods of estimating the NIV in a light curve is the ability to estimate the uncertainty in the estimate accurately. To determine how accurate the $\mathrm{NEV}_{\mathrm{eq}}$, and $\mathrm{NEV}_{\mathrm{b}}$ errors are, we investigated the ratio of the absolute value of the difference between the estimate, and the real value of the NIV, to the estimated error:  $\left|\mathrm{NEV}_{\mathrm m} - \mathrm{NIV}\right| / \sigma_{\mathrm{NEV},m} = \left|Z_{\mathrm m}\right|$. Here, $\sigma_{\mathrm{NEV},m}$ represents the error in $\mathrm{NEV}_{\mathrm{eq}}$, or $\mathrm{NEV}_{\mathrm{b}}$. Fig. \ref{MetCompNEVerr} shows how this parameter varies as a function of the NIV, the mean count rate, the number of bins, and the type of variability (white noise, pink noise, and red noise). The figure display $\left|Z_{\mathrm m}\right|$ for the two NIV estimators $\mathrm{NEV}_{\mathrm{eq}}$, and $\mathrm{NEV}_{\mathrm{b}}$. The $\mathrm{NEV}_{\mathrm{i}}$ is not considered here, as we found the computation of its error to be unreliable. 

Assuming that the individual NIV estimates are distributed normally, with a mean equal to the NIV, and a standard deviation equal to the measurement uncertainty, $\left|Z_{\mathrm m}\right| = \sqrt{2/\pi}$ \citep{10.1093/biomet/27.3-4.310}. This is the value we compare the two methods against, indicated in Fig. \ref{MetCompNEVerr} as a dashed black line. If the mean value of $\left|Z_{\mathrm m}\right|$ for either method is found to lie below $\sqrt{2/\pi}$, then the uncertainties in the measurement are overestimated. If, instead, the value of this parameter lies above $\sqrt{2/\pi}$, then either the mean NIV estimate differs significantly from its actual value, or the uncertainties are underestimated. The closer the average value of $\left|Z_{\mathrm m}\right|$ lies to $\sqrt{2/\pi}$, the better the method is at correctly estimating the measurement uncertainties. 

At low count rates, the errors in $\mathrm{NEV}_{\mathrm{b}}$ are slightly overestimated, which can be seen in the panels on the far left of Fig. \ref{MetCompNEVerr}. This may be due to the decision to increase the size of the lower bound errors on $\mathrm{NEV}_{\mathrm{b}}$, that extended below the lower limit of variability detectable by bexvar (see Section \ref{NEVest}).

In contrast, the $\mathrm{NEV}_{\mathrm{eq}}$ uncertainties are significantly underestimated at $\overline{R_{\mathrm S}}=0.015~\mathrm{cts/s}$. There are even some instances, when $\sigma_{\mathrm{obs}}^2 < \overline{\sigma_{\mathrm{err}}}^2 / 2$, such that $\mathrm{NEV}_{\mathrm{eq}} = -\overline{\sigma_{\mathrm{err}}}^2 / 2$. When this happens, the error in $\mathrm{NEV}_{\mathrm{eq}}$ cannot be calculated. The figure does not include light curves with these properties as data points. The $\mathrm{NEV}_{\mathrm{eq}}$ errors based on \citet{2003MNRAS.345.1271V} are inaccurate whenever $\mathrm{NEV}_{\mathrm{eq}} < 0$, and that happens very frequently at low count rates.

The accuracy of the errors on both NIV estimators improves with an increasing average source count rate, and they are reasonably reliable for $\overline{R_{\mathrm S}} \gtrsim 0.15~\mathrm{cts/s}$, for both methods. This can be seen in the second, third, and fourth panels from the left in Fig. \ref{MetCompNEVerr}. At $\overline{R_{\mathrm S}}=0.15~\mathrm{cts/s}$, the $\mathrm{NEV}_{\mathrm{eq}}$ uncertainties more accurately represent the error in the measurement than the $\mathrm{NEV}_{\mathrm{b}}$ ones. At $\overline{R_{\mathrm S}}=1.5~\mathrm{cts/s}$, and $\overline{R_{\mathrm S}}=15~\mathrm{cts/s}$, both sets of errors perform about equally well, for all three different types of variability investigated here. The $\mathrm{NEV}_{\mathrm{b}}$ errors are, however still more likely to be slightly overestimated, and the $\mathrm{NEV}_{\mathrm{eq}}$ are still more likely to be slightly underestimated. The $\mathrm{NEV}_{\mathrm{b}}$ errors are also reasonably reliable for red and white noise variability, even though they are based on the assumption of pink noise variability. 

These errors only depict the measurement errors of estimating the NIV for the selected set of observations. They do not contain the intrinsic scatter in the NIV, which is discussed in Sections \ref{SecSysErr}, and \ref{SecRedSysErr}.

We note that $\mathrm{NEV}_{\mathrm{b}}$ is the best method for estimating the NIV of \emph{eROSITA}-like light curves. Its estimates are reliable across a wide range of degrees of variability, count rates, number of bins, and types of variability. It never estimates the NIV to be negative, as $\mathrm{NEV}_{\mathrm{i}}$, and $\mathrm{NEV}_{\mathrm{eq}}$ do, so it can always be used to estimate $\mathrm{NIV}_{\infty}$, by computing its geometric mean, $\overline{\mathrm{NEV}_{\mathrm{b}}}$. It can identify when an estimate of the NIV is merely an upper limit. The uncertainties of $\mathrm{NEV}_{\mathrm{b}}$ are well-defined and close to the size they should be, but are overestimated at low count rates. The downside to $\mathrm{NEV}_{\mathrm{b}}$ is that it is more computationally expensive than $\mathrm{NEV}_{\mathrm{i}}$, and $\mathrm{NEV}_{\mathrm{eq}}$. We recommend using $\mathrm{NEV}_{\mathrm{b}}$ to estimate the NIV in all light curves, except when a source has $\gtrsim 20$ source counts in every bin, or when it shows non-power law variability. 

The methods for computing $\mathrm{NEV}_{\mathrm{i}}$ and $\mathrm{NEV}_{\mathrm{eq}}$, as they are described in Sections \ref{SecNEVi} and \ref{SecNEVeq} can be used for any other instrument, as they were not explicitly developed for \emph{eROSITA}. The modifications to the standard version of these estimates of the NIV are necessary whenever a light curve consists of varying fractional exposures. 

\section{Corner plots}

\subsection{Corner plot for $\mathrm{NEV}_{\mathrm{b}}$ as a function of $\overline{\sigma_{\mathrm{b}}}$}\label{AppCorPlot}

\begin{figure*}[h]
\resizebox{\hsize}{!}{\includegraphics{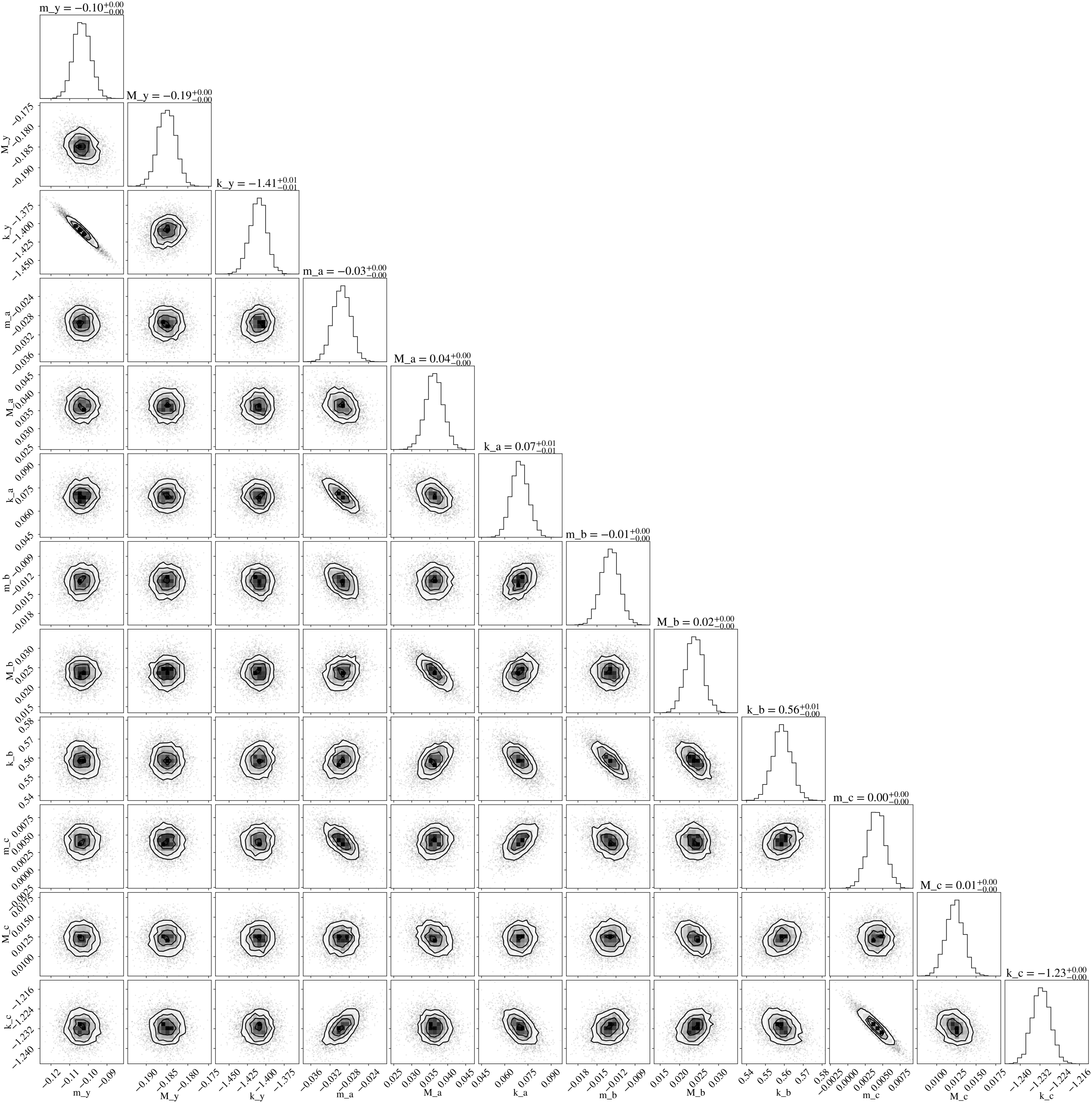}}
\caption{The corner plot of the best fitting parameters of Eq. \ref{NEVtoBVfit_quadabc}, for fitting $\log(\overline{\sigma_{\mathrm{b}}})$ as a function of the rescaled true input $\log(\mathrm{NIV}')$, across all simulations spanning count rates from 0.001 cts/s to 30 cts/s, and number of bins from 50 to 1000. The values of the best fit parameters shown here differ from those displayed in Table \ref{TBVNEVquad(nbcr)}, as the fit was performed for $\log(\overline{\sigma_{\mathrm{b}}})$ as a function of the rescaled $\log(\mathrm{NIV}') = \log(\mathrm{NIV}) - \overline{\log(\mathrm{NIV})}$. Table \ref{TBVNEVquad(nbcr)} instead shows the parameter values for the fit of $\log(\overline{\sigma_{\mathrm{b}}})$ as a function of $\log(\mathrm{NIV})$, and those are the ones that should be used for estimating $\mathrm{NEV}_{\mathrm{b}}$ from $\overline{\sigma_{\mathrm{b}}}$.
 \label{BV(NEV)quadabc}}
\end{figure*}

\subsection{Corner plot for the best fit of $\Delta_{sys}$ as a function of $N_{\mathrm b}$, and $\log(\mathrm{NIV}_{\infty}')$} \label{ACpSysErr}

\begin{figure}[h]
\resizebox{\hsize}{!}{\includegraphics{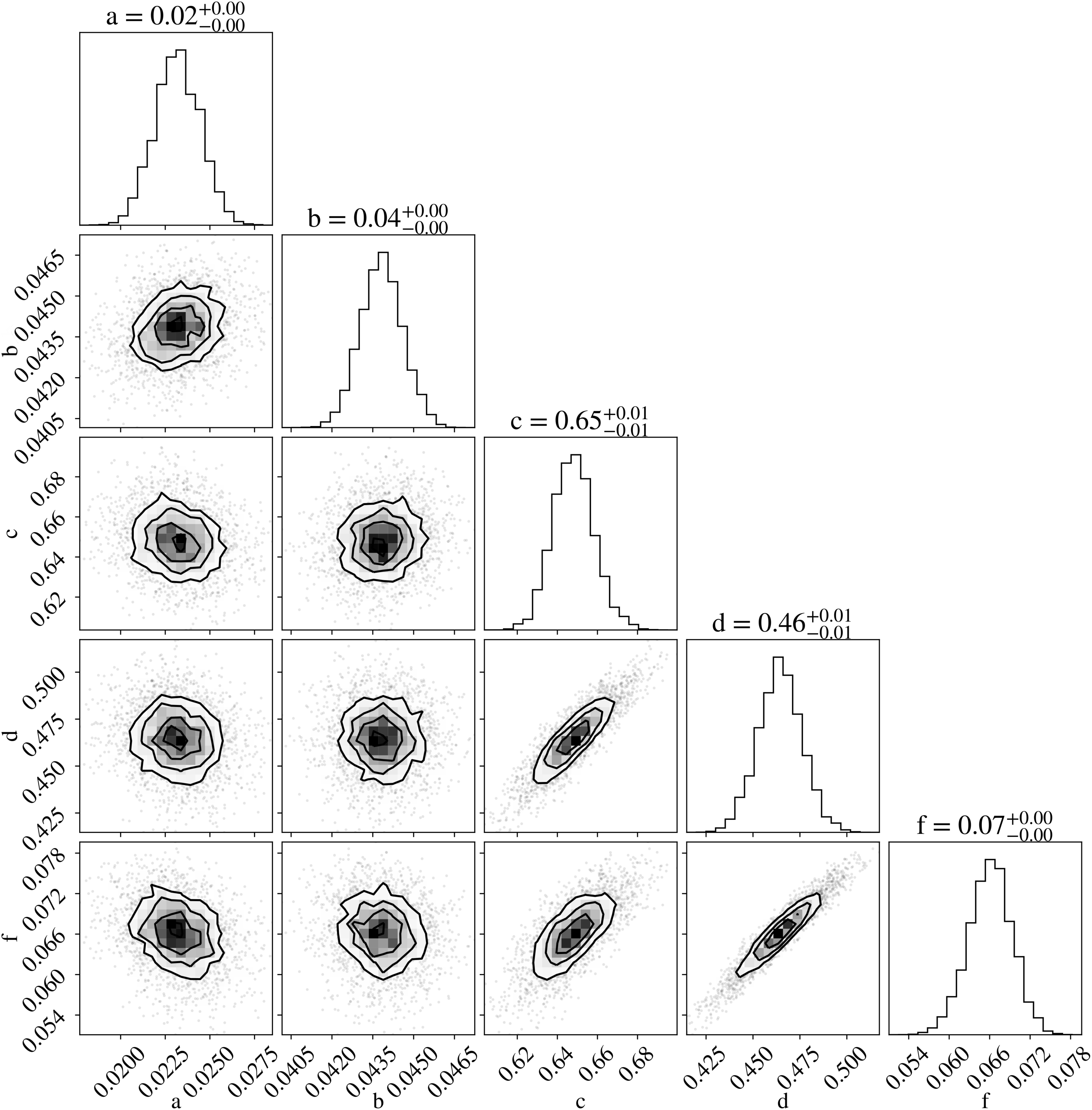}}
\caption{Corner plot for the best fitting parameters of $a$, $b'$, $c$, $d$, and $f'$, for fitting $\Delta_{\mathrm s}$, as a function of the $N_{\mathrm b}$, and $\log(\mathrm{NIV}_{\infty}')$. This figure is the result of $3.2\times10^5$ simulated pink noise light curves.
 \label{cplotsyserr}}
\end{figure}

\subsection{Corner plot for the excess noise in a periodogram as a function of $\overline{\epsilon}$, and $\sigma_\epsilon^2$}\label{ACpExcessNoise}

\begin{figure}[h]
\resizebox{\hsize}{!}{\includegraphics{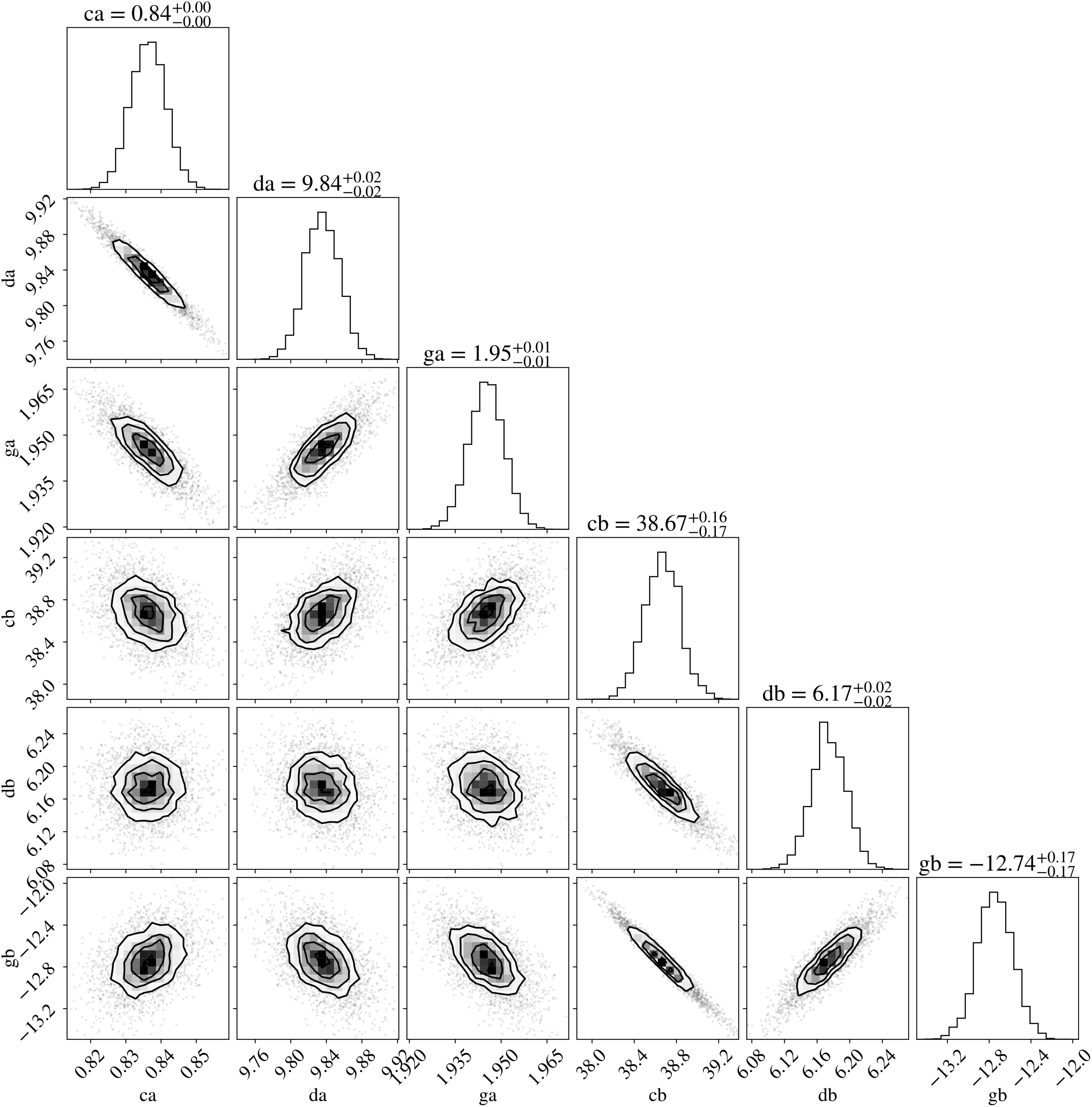}}
\caption{Corner plot for the best fitting parameters of $c_a$, $d_a$, $g_a$, $c_{\mathrm{b}}$, $d_{\mathrm{b}}$, and $g_{\mathrm{b}}$ for fitting the excess periodogram noise due to variable fractional exposures with Eq. \ref{fplexeq}.
 \label{exPLfitcorner}}
\end{figure}

\end{document}